\documentclass[%
 reprint,
superscriptaddress,
 amsmath,amssymb,
 aps,
twocolumn]{revtex4-2}

\usepackage{graphicx}% Include figure files
\usepackage{dcolumn}% Align table columns on decimal point
\usepackage{bm}% bold math
\usepackage{stackrel}

\usepackage{hyperref}
\usepackage[shortlabels]{enumitem}

\usepackage{multirow}
\newcommand{\on}{q_{\rm on}}
\newcommand{\off}{q_{\rm off}}

\begin{document}

\preprint{APS/123-QED}

\title{The Nanocaterpillar's Random Walk: Diffusion With Ligand-Receptor Contacts}% Force line breaks with \\

\author{Sophie Marbach}
\affiliation{%
 Courant Institute of Mathematical Sciences, New York University,
NY, 10012, U.S.A.
}%
\affiliation{CNRS, Sorbonne Universit\'{e}, Physicochimie des Electrolytes et Nanosyst\`{e}mes Interfaciaux, F-75005 Paris, France }%Lines break automatically or can be forced with \\
\email{sophie@marbach.fr}
\author{Jeana Aojie Zheng}%
\affiliation{%
 Department of Physics, New York University, 
NY, 10012, U.S.A.
}%
\author{Miranda Holmes-Cerfon}%
\affiliation{%
 Courant Institute of Mathematical Sciences, New York University, 
NY, 10012, U.S.A.
}%


\begin{abstract}
Particles with ligand-receptor contacts  
 bind and unbind fluctuating ``legs" to surfaces, whose fluctuations cause the particle to diffuse. 
Quantifying the diffusion of such ``nanoscale caterpillars" 
is a challenge, since binding events often occur on very short time and length scales. 
Here we derive an analytical formula, validated by  simulations, for the long time translational diffusion coefficient of an overdamped nanocaterpillar, under a range of modeling assumptions. 
%Here we present a robust analytical framework, validated by simulations, to coarse-grain fast dynamics and obtain 
%the long time translational diffusion coefficient of a nanocaterpillar. 
%In contrast with previous perturbative approaches, we find that 
We demonstrate that the effective diffusion coefficient, which depends on the microscopic parameters governing the legs, can be orders of magnitude smaller than the background diffusion coefficient. Furthermore it varies rapidly with temperature, and reproduces 
%quantitatively predicts 
the striking variations seen in existing data and our own measurements of the diffusion of DNA-coated colloids.
%, and in particular can be nearly unrelated to the background diffusion coefficient. 
%Our analytical expression for the diffusion coefficient can vary over orders of magnitude, and accurately reproduces existing and additional data on diffusion of DNA-coated colloids. 
%We furthermore compare our model to a range of other models and assumptions found in the literature, and find ours is the most general, encapsulating others as special limits. 
%We verify our theory experimentally, by measuring diffusion coefficients of DNA-coated colloids on DNA-coated surfaces. % and find close agreement with our prediction.  
Our model gives insight into the mechanism of motion, and allows us to ask: when does a nanocaterpillar prefer to move by \textit{sliding}, where one leg is always linked to the surface, and when does it prefer to move by \textit{hopping}, which requires all legs to unbind simultaneously? 
We compare %whether a nanocaterpillar prefers to hop or slide for 
a range of systems (viruses, molecular motors, white blood cells, protein cargos in the nuclear pore complex, bacteria such as Escherichia coli, and DNA-coated colloids)  and present guidelines to control the mode of motion for materials design. 
\end{abstract}

\maketitle

%\tableofcontents

%\sophie{S to do: update labels in SI when main is final}\\
%\mhc{there is a package to do this automatically, have you used it before?}\\
%\sophie{no lol. Happy to learn this trick!}
%\mhc{it's called "xr". you'd probably have to put both .tex files in the same folder, to make it work together. with this package you just reference an equaion/figure in the other file, the same way you normally would (but you can add an "S" or "T" or whatever else you want in front, to distinguish.) }
% good to know for later!

% +33 6 74 55 57 41

\begin{figure}[h!]
\includegraphics[width = 0.99\linewidth]{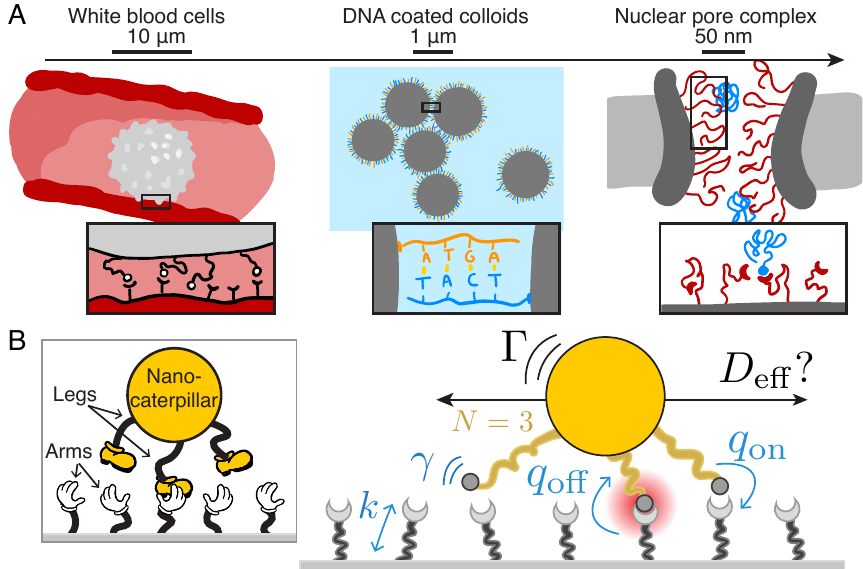}% Here is how to import EPS art
\caption{\label{fig:fig1} \textbf{Overview of nanocaterpillars.} (A) Multivalent ligand-receptor systems span the micro to nanoscales. White blood cells stick to vessel walls through selectin mediated bonds (inspired from Ref.~\citenum{korn2008dynamic}); DNA-coated colloids self-assemble through hybridization of complementary DNA strands; Protein cargos translocate through the polymer mesh of the nuclear pore complex (inspired from Ref.~\citenum{fogelson2018enhanced}). (B) Ligand-receptor systems are modeled here with an arbitrary number of legs $N$ (ligands) and/or arms (receptors). 
The stochastic model includes binding and unbinding rates $\on$ and $\off$, spring constant $k$, and leg friction $\gamma$ (all fast, in blue); and the bare friction coefficient $\Gamma$ of the nanocaterpillar (slow, in black). We seek the long-time effective longitudinal diffusion coefficient $D_{\rm eff}$. }
\end{figure}

Particles with ligand-receptor contacts -- or \textit{nanocaterpillars} -- harvest binding and unbinding dynamics of their fluctuating \textit{legs} at the nanoscale to move, target, stick, or assemble into large structures~\cite{mammen1998polyvalent,bressloff2013stochastic,hammer2014adhesive,rogers2016using}. Nanocaterpillars are found across multiple scales, spanning a great variety of systems in biology and biomimetic assays -- see Fig.~\ref{fig:fig1}-A. To name but a few, microscale white blood cells with protein linkers stick and roll on blood vessel walls until they reach a healing target~\cite{alon2002rolling,ley2007getting,korn2008dynamic}. Microscale droplets with protein linkers are used to study cellular-like adhesion~\cite{zhang2017sequential,pontani2016cis,merminod2021avidity}.  Microscale to nanoscale colloids coated with complementary deoxyribonucleic acid (DNA) strands self-assemble into macroscopic crystals~\cite{macfarlane2011nanoparticle,rogers2016using,lewis2020single} with novel optical or selectivity properties~\cite{park2014full,he2020colloidal,merindol2021fast,bilchak2020tuning}. Nanoscale viruses transiently adhere with spike proteins to the respiratory mucus to find vulnerable host cells~\cite{mammen1998polyvalent,sakai2017influenza,sakai2018unique,muller2019mobility}. At even smaller scales, protein cargos bind to receptors in the nuclear pore complex for selective transport to a cell's nucleus~\cite{allen2000nuclear,aramburu2017floppy}. %\sophie{It should be clear at this stage that these are some of possible examples, but there are much more, because molecular motor people might feel akward. } %The past two decades have also seen tremendous advances in the artificial design of sticky surfaces. Further, 
%enable to probe small forces in soft media~\cite{zhang2017sequential,lin2016evidence,merindol2019modular} and 
%ho1990existence

For all these systems to function,  a nanocaterpillar must \emph{move} relative to the surface to which its legs are attracted. An important question therefore is to characterize \emph{how} it moves, over scales much larger than individual legs. %, such by an average mobility or equivalently diffusivity tensor. %-- via an average diffusion coefficient, average velocity, or other characterization. 
Since legs constantly bind and unbind to the surface, imparting force each time they do so, the particle's macroscopic mobility depends on the microscopic details of its legs. 
%details of how they do so can strongly affect the particle's macroscopic motion. 
%, while the manner in which it moves macroscopically depends on the microscopic details of its legs.  
%The microscopic details of leg fluctuations and adhesive interactions with surfaces determines how nanocaterpillars move. 
For example, leg flexibility and bond lifetimes control the average mobility of the particle~\cite{alon1997kinetics,shrivastava2019stiffness,muller2019mobility}, and differences in both parameters can be harvested to detect infected cells~\cite{dasanna2018adhesion,yehl2016high,karnik2008nanomechanical} or prevent viral infections~\cite{wang2014igg}. 
%Bond lifetime and leg flexibility control the velocity of the particle~\cite{alon1997kinetics,shrivastava2019stiffness,muller2019mobility}, and differences can be harvested to detect infected cells~\cite{dasanna2018adhesion,yehl2016high,karnik2008nanomechanical} or prevent viral infections~\cite{wang2014igg}. 
As another example, leg density affects how DNA-coated colloids nucleate and grow into  crystals~\cite{hensley2021classical,lewis2018dna} and governs the long-range alignment of crystals~\cite{wang2015crystallization,holmes2016stochastic,jana2019translational}. 
Overall, microscopic details underlie a variety of large-scale modes of motion, such as hopping~\cite{hammer2014adhesive,sakai2017influenza,loverdo2009quantifying,hamming2019influenza}, cohesive motion including rolling and crawling~\cite{sakai2017influenza,vahey2019influenza}, %\mhc{memory effects -- in influenza paths} %\sophie{do you have a specific influenza ref in mind ? again though, I'm not so keen on insisting on these behaviors, similarly as for molecular motors, especially because they are mediated by active forces (that cut the bonds) and therefore correspond to an out-of-equilibrium process.}, 
and also transient or firm arrest~\cite{alon2002rolling,hammer2014adhesive,ramesh2015significance}, resulting in large differences in macroscopic mobility.   %\sophie{diffusion vs directed motion, with linear response theory we may expect that "mobility" covers these regimes. I think we can say that we focus on diffusion only at the point where we start to discuss what we do, otherwise diffusion only papers are quite limited (a lot of works study out of equilibrium systems yet some of the effects (e.g. dependence on leg parameters) are expected to translate in equilibrium.}
%\mhc{something about diffusion, and directed motion} %\sophie{do you want this because the next paragraph discusses macroscopic mobility, so diffusion basically? A way of leading to that could be something else such as "Even macroscopic mobility on surfaces is already intricate, with a diversity of large-scale modes of motion, such as hopping etc" (use the previous sentence, because all these examples are all about mobility). Otherwise I think discussing more directed motion vs diffusion would maybe add text in a direction that is not necessarily critical to the reader.}

%Yet, the microscopic details of the legs are usually not the quantity of interest in most of these systems -- rather, one wants to understand the overall mobility of the nanocaterpillar, on scales much larger than individual legs. 

Investigating how microscopic binding details lead to macroscopic mobility is challenging, as it requires probing time and length scales that can often be quite different~\cite{muller2019mobility,fan2021microscopic} -- legs can be much smaller than the nanocaterpillar they are attached to, while leg dynamics can be orders of magnitude faster than the timescales of macroscopic motion. Furthermore, many systems have a valency of thousands of leg contacts~\cite{xu2011subdiffusion,wang2015crystallization,fan2021microscopic}, too many degrees of freedom to resolve experimentally or computationally~\cite{fogelson2018enhanced,etchegaray2019stochastic}. 
%\cite{cohen2015notable,hamming2019influenza}
To make progress, numerical and analytical models often rely on simplified assumptions, \textit{e.g.} excluding stochastic relaxation of the legs~\cite{ziebert2021influenza,licata2007colloids}, limiting the analysis to a small number of legs~\cite{bose2010semianalytical,kowalewski2021multivalent,ziebert2021influenza}, or assuming small perturbations~\cite{fogelson2018enhanced}. 
Such models have given insight into a variety of phenomena, such as how specific parameters could favor rolling over sliding~\cite{caputo2005effect,korn2008dynamic,bose2010semianalytical,grec20181d,ziebert2021influenza} or how specific mechanisms could increase overall mobility (with coupling effects such as binding dynamics depending on bond number~\cite{klumpp2005cooperative,fenz2017membrane,miles2018analysis} %, a specific combination of modes of motion~\cite{loverdo2009quantifying} 
or when numerous adhesive sites are available for a single ligand~\cite{goodrich2018enhanced,fogelson2018enhanced,fogelson2019transport}). 
%\mhc{refs from Hammer to add here? nancy forde? molecular motor refs? other modeling refs?} \sophie{molecular motor refs are in. I did skim modeling refs quite a lot but maybe you're thinking of others. I have some Daniel Hammer refs and he's more a bio person, I'm thinking that you might be thinking of another one. Nancy forde I'm not sure what she did, I can look it up if you don't have a specific paper in mind.}
Nevertheless, such modeling assumptions are not always justified; for example
stochasticity plays a critical role for mobility, facilitating rolling~\cite{ramesh2015significance}, targeted arrest~\cite{etchegaray2019stochastic}, or other walking modes~\cite{korosec2021substrate}. Furthermore, such models can also not reproduce the order of magnitude decrease of diffusion of DNA-coated colloids ~\cite{xu2011subdiffusion,wang2015crystallization}. Hence, a systematic derivation of macroscopic mobility from microscopic details that is valid under a broad range of parameters is needed.

In this paper we derive an analytical expression for the effective mobility of a nanocaterpillar in an overdamped system, %-- through the calculation of its diffusion coefficient --  
by systematically coarse-graining over the microscopic details of its legs. %We consider a nanocaterpillar in an overdamped system, moving in one horizontal direction parallel to a sticky surface, with legs that bind and unbind to it passively, so the system is reversible. 
Starting from a model that includes the detailed spatial fluctuations of the legs, we use homogenization techniques~\cite{pavliotis2008multiscale,lee2018modeling,fogelson2018enhanced} to average over these fluctuations. We obtain an analytical expression for the effective long-time translational diffusion coefficient of the particle, $D_{\rm eff}(N,\Gamma,\gamma,k,q_{\rm off},q_{\rm on})$, as a function of the microscopic parameters governing the legs (Eq.~\eqref{eq:gammaN}; see also Fig.~\ref{fig:fig1}-B and Sec.~\ref{sec:sec1}.)
The expression depends in a non-trivial way on the friction coefficients of the individual components of the system (legs and particle), with the frictions either adding up arithmetically (like springs in parallel) or harmonically (like springs in series) according to the mechanistic details.  
%We highlight this analogy throughout the paper as we give intuitive interpretations of our results. 
We validate our analytical calculations with numerical simulations, which show the expression is accurate over a wide range of parameter values.

Our model gives insight into the mechanism of nanocaterpillar motion, as it allows us to distinguish between two long term modes of motion: \textit{sliding}, where at least one bond is always attached to the surface, and \textit{hopping}, where the particle detaches completely, moves in free space and reattaches. These regimes are controlled by physical properties of the legs, such as stiffness and adhesive strength, allowing us to investigate existing biological and biomimetic systems in a so-called Ashby chart for nanocaterpillars (Sec.~\ref{sec:slideorhop}). We identify how critical design parameters (such as the coating density for DNA-coated colloids) controls the preferential mode of motion and reconcile disparate experimental observations on similar systems~\cite{xu2011subdiffusion,wang2015crystallization}.

Importantly, the effective diffusion can sometimes be orders of magnitude smaller than the background diffusion coefficient, showing the critical effect of the legs on the particle's mobility. 
This analytical prediction of a dramatically decreased diffusivity is borne out with experimental measurements of the diffusion of DNA-coated colloids, both from existing data~\cite{xu2011subdiffusion,wang2015crystallization} and additionally measured in this study. Our model agrees with the data within experimental accuracy over a range of temperatures and for different DNA coating densities on the colloids (Sec.~\ref{sec:slideorhop}). %This agreement was surprising to us, since our model does not yet capture two-dimensional modes of motion such as rolling, which are thought to lead to faster diffusion for DNA-coated colloids~\cite{jana2019translational,lee2018modeling} %\sophie{does mognetti really say that ? he doesn't compare it to sliding, I think ?! Dave Pine doesn't ever really see or measure rolling diffusion. I'm also a bit unsettled by this comment at this stage since we didn't talk so much about rolling for DNA before but maybe it's ok. }.   

%Although our model is limited to motion in one dimension,
%ur results lay the grounds to tune mobility features in artificial designs.

Finally, we derive the effective diffusion coefficient for several variations of the model with varying assumptions, and show that our model incorporates these assumptions as special limits~\cite{lee2018modeling,fogelson2018enhanced}, but is accurate over a broader range of parameters and system designs (Sec.~\ref{sec:Models}). In particular, previous approaches can not describe the observed orders of magnitude decrease in diffusion~\cite{fogelson2018enhanced}. 
Overall, our results lay the ground to tune mobility features in artificial designs, and provide methodological tools to study more complex motion mediated through ligand-receptors, including rolling or self-avoiding walks due to active cutting of bonds. %\sophie{I removed 2D/3D because that's not the main point, rolling seems more to the point.  }

\section{Deriving an analytical formula for the effective diffusion coefficient}
\label{sec:sec1}

In Sections \ref{eq:1legEqn}-\ref{sec:1legnterpret} 
we illustrate our homogenization technique pedagogically by considering a 1-legged caterpillar. Our main result for the effective diffusion coefficient of an $N$-legged caterpillar, Eq.~\eqref{eq:gammaN}, is presented in Section \ref{sec:Nlegs}. 

%We begin by presenting our homogenization technique to obtain the long time diffusion coefficient for a simple model of nanocaterpillar. We expose first the method on a 1-legged caterpillar model to facilitate notations and understanding then generalize our results to an N-legged caterpillar. 

\subsection{1-legged caterpillar: constitutive equations}\label{eq:1legEqn}

\begin{figure}[h!]
\includegraphics[width = 0.99\linewidth]{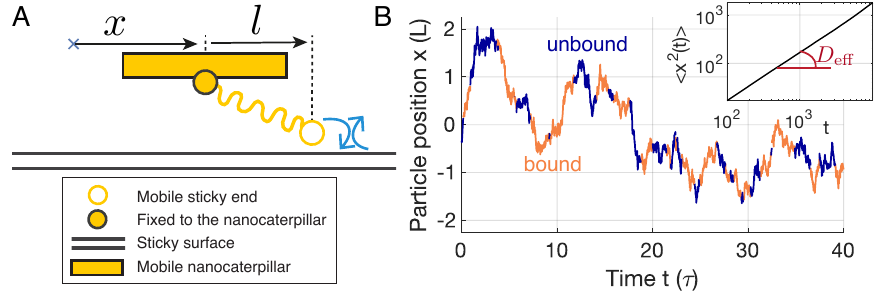}% Here is how to import EPS art
\caption{\label{fig:fig1model} \textbf{1-legged nanocaterpillar model.} (A) The longitudinal extension of the single leg ($l$) is monitored and feeds back into the longitudinal position ($x$) of the particle. (B) Simulation trace of the position of a 1-legged particle  with time. (inset) The effective long time diffusion $D_{\rm eff}$ is half the slope of the mean squared displacement over long times.}
\end{figure}

We begin with the simplest possible model: a nanocaterpillar with a single leg (Fig.~\ref{fig:fig1model}). The leg is permanently fixed to the caterpillar while its other end is mobile, and can attach anywhere on the binding surface. We consider for now a one-dimensional model, where leg fluctuations and particle motion occur on a line, longitudinal to the surface. 

The dynamics of the particle position $x(t)$ and leg length $l(t)$ occur over nano to microscales, mostly in dense fluids such as water. In this context, dynamics are well captured by overdamped Langevin equations~\cite{bian2016111}, where inertia plays a negligible role. This is in contrast to previous modeling efforts which used the Langevin equation (with inertia)~\cite{lee2018modeling}, a point we return to in Sec.~\ref{sec:Models}, where we show that the two approaches can give predictions that are orders of magnitude different in certain parameter regimes. 

When the legs are unbound they evolve as
\begin{equation}
    \frac{dl}{dt} = - \frac{k}{\gamma} (l(t) - l_0) + \sqrt{\frac{2k_B T}{\gamma}} \eta_l(t)\,.
    \label{eq:dlUnbound}
\end{equation}
Here $k$ is a spring constant describing the recoil force of the leg material, $\gamma$ is its friction coefficient, $l_0$ its rest length, $k_B$ is Boltzmann's constant, $T$ is temperature and $\eta_l$ is a Gaussian white noise satisfying $\overline{\eta_l(t)} = 0$ and $\overline{\eta_l(t) \eta_l(t')} = \delta(t - t')$ where $\overline{\cdot}$ is the average over realizations of the noise. In most systems we consider, legs are made of polymers or proteins, where small leg deformations around equilibrium are well captured by a constant spring constant $k$~\cite{rubinstein2003polymer,miller2006mechanical,lim2006flexible}. 

The particle's position $x$ when the leg is unbound obeys
\begin{equation}
\frac{dx}{dt} = \sqrt{\frac{2k_B T}{\Gamma}} \eta_x(t)  
\label{eq:dxUnbound}
\end{equation}
where $\Gamma$ is the friction coefficient of the particle and $\eta_x(t)$ is a Gaussian white noise uncorrelated with $\eta_l(t)$. The diffusion coefficient for the unbound particle is $D_0 = \frac{k_B T}{\Gamma}$. 
%Note that this model for the simultaneous fluctuations of the particle + unbound leg is valid if the leg is fixed by its center of mass to the particle, with $\Gamma$ representing $\Gamma' + \gamma$, the arithmetic sum of the bare particle friction $\Gamma'$ and the leg friction $\gamma$ (Supplementary~4.3). A model where the leg is not fixed by its center of mass results in minor modifications. Importantly, if the particle is 1-armed instead of 1-legged, with the fixed point on the surface (see Fig.~\ref{fig:fig1}-B) the equations presented here are exact. We ignore such details in the following and use 1-legged particles to account for particles with 1 adhesive point -- discussing possible extensions in Sec.~\ref{sec:handsFeet}.

%A model where the leg is not fixed by its center of mass can also be treated within this framework and results in minor modifications. Importantly, if the particle is 1-armed instead of 1-legged, with the fixed point on the surface (see Fig.~\ref{fig:fig1}-B) the equations presented here are exact. We ignore such details in the following and use 1-legged particles to account for particles with 1 adhesive point. 
%Whether the spring is always bound to the particle or to the opposing surface results in no significant difference -- see Sec.~\ref{sec:handsFeet} -- and is ignored here. We consider in the following that the spring is indeed a leg, always connected to the particle, that may bind to the opposing surface -- alike Fig.~\ref{fig:fig1}-B.

We consider for now that %the particle is always close to the surface,} and that 
the surface is uniformly coated with receptors. The leg can thus bind at any location on the surface with a constant binding rate $q_{\rm on}$ and constant unbinding rate $q_{\rm off}$. Detailed balance requires $\frac{q_{\rm on}}{q_{\rm off}} = \frac{\pi_b}{\pi_u}$ where $\pi_{b/u}$ is the equilibrium probability of the system to be bound or unbound. Typically $\frac{\pi_b}{\pi_u} = e^{-\beta \Delta G}$, where $\beta^{-1} = k_B T$ and $\Delta G< 0$ is the free energy change when the leg binds to the surface~\cite{fan2021microscopic,varilly2012general}.  % -- it corresponds to the energy well in the surface vicinity associated with complementary ligand-receptor pairs compared to a reference energy state with uncomplementary pairs

We now seek to describe motion of the system when the leg is bound. In this case, variables are constrained as $x(t) + l(t) - x_{\rm r} = 0$ where $x_{\rm r}$ is the location of the receptor where the leg tip is attached, which is constant until the leg detaches and reattaches to another location. 
The stochastic dynamics Eqns.~\eqref{eq:dlUnbound} and \eqref{eq:dxUnbound} must be projected~\cite{ciccotti2008projection,holmes2016stochastic} onto the constraint surface, see Appendix~A. We obtain
\begin{equation}
    \frac{dx}{dt} = - \frac{dl}{dt} = \frac{k}{\Gamma + \gamma} (l(t) - l_0) + \sqrt{\frac{2k_B T}{\Gamma + \gamma}} \eta(t)
    \label{eq:dxProjected}
\end{equation}
where $\eta(t)$ is a Gaussian white noise. Here we see that the projected dynamics have a natural expression where the effective friction in the bound state is the arithmetic sum of the friction coefficients in the unbound states, $\Gamma + \gamma$. Note that this projection is a crucial step that is often ignored in such derivations~\cite{fogelson2018enhanced,lee2018modeling,holmes2016stochastic}, and modifies the dynamics in non trivial ways especially with a large number of legs.

The dynamics are now specified through the set of  Eqns.~\eqref{eq:dlUnbound}-\eqref{eq:dxProjected}, together with the binding and unbinding dynamics. To see what happens over long times, we simulate trajectories for 1 leg -- see Fig.~\ref{fig:fig1model}-B (and simulation details in Appendix B). 
Over long times, the particle's mean-squared displacement grows linearly with time, and we may extract an effective long time diffusion coefficient $D_{\rm eff}$  -- see inset of Fig.~\ref{fig:fig1model}-B.

\subsection{Homogenization to coarse-grain the fast dynamics}
\label{sec:homo}

The computational cost of simulating Eqns.~\eqref{eq:dlUnbound}-\eqref{eq:dxProjected} is high, since small time steps are required to resolve the fast relaxation and binding events. We therefore seek an analytical method to coarse-grain over these fast timescales. To apply this method we identify a non-dimensional separation of scales, which is novel compared to other approaches~\cite{lee2018modeling,fogelson2018enhanced,fogelson2019transport} and will allow us to find a result valid over a broad range of parameters. We use homogenization theory to average over the fast scales, eventually obtaining an effective diffusion equation, Eq.~\eqref{eq:dxa}, with effective diffusivity  (Eq.~\eqref{Deff1}) and related effective friction (Eq.~\eqref{eq:gammaeff1}), which is one of the main results of this paper for the special case of a 1-legged caterpillar. A reader interested in the results and physical implications may skip to Section~\ref{sec:1legnterpret}.

\subsubsection{Set up: partial differential equations to be coarse-grained}

The set of stochastic Eqns.~\eqref{eq:dlUnbound}-\eqref{eq:dxProjected} 
defines a Markov process that
is conveniently studied via the Fokker-Planck equation and its adjoint, the Kolmogorov backward  equation~\cite{gardiner1985handbook,pavliotis2008multiscale}. 
Let $p(x,l,t) = \left( p_u(x,l,t), p_b(x,l,t) \right)^T$ be the probability density function of finding the system at time $t$ and positions $x,l$ in the unbound or bound states. We obtain from Eqns.~\eqref{eq:dlUnbound}-\eqref{eq:dxProjected} the Fokker-Plank equation
\begin{equation}\label{eq:FPE}
    \partial_t p = \mathcal{L}^{\star} p\,,
\end{equation}
with $\mathcal{L}^{\star} = \mathcal{V}^{\star} + \mathcal{Q}^{\star}$ where
\begin{align*}  
    \mathcal{V}^{\star} &= \mathrm{diag} \begin{pmatrix} \partial_l \left( \frac{k}{\gamma} (l-l_0) + \frac{k_B T}{\gamma} \partial_l\right) + \frac{k_B T}{\Gamma} \partial_{xx} \\
    (\partial_l -  \partial_x) \left( \frac{k}{\Gamma + \gamma} (l-l_0) + \frac{k_B T}{\Gamma + \gamma} (\partial_l -  \partial_x)  \right)
    \end{pmatrix}, \nonumber \\
    \mathcal{Q}^{\star} &= \begin{pmatrix}  - q_{\rm on} & q_{\rm off} \\  q_{\rm on} & - q_{\rm off} \end{pmatrix},
\end{align*}
with an appropriate initial condition. 
% \begin{eqnarray}  
%     & \partial_t p = \mathcal{L}^{\star} p \,\, \mathrm{with} \,\,\,  \mathcal{L}^{\star} = \mathcal{U}^{\star} + \mathcal{Q}^{\star},   \nonumber \\
%     \mathcal{U}^{\star} &= \mathrm{diag} \begin{pmatrix} \partial_l \left( \frac{k}{\gamma} (l-l_0) + \frac{k_B T}{\gamma} \partial_l\right) + \frac{k_B T}{\Gamma} \partial_{xx} \\
%     (\partial_l -  \partial_x) \left( \frac{k}{\Gamma + \gamma} (l-l_0) + \frac{k_B T}{\Gamma + \gamma} (\partial_l -  \partial_x)  \right)
%     \end{pmatrix}, \nonumber \\
%     & \mathrm{and} \,\, \mathcal{Q}^{\star} = \begin{pmatrix}  - q_{\rm on} & q_{\rm off} \\  q_{\rm on} & - q_{\rm off} \end{pmatrix}.
%   \label{eq:FPE}
% \end{eqnarray}
Additionally we require the flux in either state to vanish at infinity, to conserve total probability. The stationary solution of Eq.~\eqref{eq:FPE} is $\pi = \frac{e^{-\beta k (l-l_0)^2/2}}{Z} \left( q_{\rm off} , q_{\rm on} \right)^T $ where $Z$ is a normalization constant. This is therefore the equilibrium probability density of the system; it satisfies detailed balance. 
 
While probability densities have an intuitive physical meaning, in the following it will be easier -- and mathematically better posed -- to consider the adjoint of the Fokker-Planck equation and the corresponding dual functions. These are functions $f(x,l,t) = \int p(x',l',t|x,l) g(x',l') dl' dx'$ that give the expectation of any scalar function $g(x(t),l(t))$, given an initial condition $x(0)=x,l(0)=l$. Once we know how such functions $f$ evolve, we may calculate any statistic $g$ of our stochastic process. %,  and hence the dynamics of $f$ fully specify the dynamics of our process. %, in the same way that these dynamics are equivalently described by the Fokker-Planck equation.  
%\mhc{sohpie-please check/edit, or we can discuss. maybe this sentence is unnecessary?} \sophie{I think we can stop after ``our process". also do we have a ref for mathematically better posed?} \mhc{not really... any SDE textbook?} 
%The dynamics of the collection of such functions $f(x,l,t)$ fully determines a Markov process by its statistics. 
% we reference the Gardiner later so I believe this is fine. 
Writing $f(x,l,t) = \left( f_u(x,l,t) , f_b(x,l,t) \right)^T$, we have that $f$ satisfies the Kolmogorov backward equation~\cite{gardiner1985handbook}
\begin{equation}\label{eq:backward}
    \partial_t f = \mathcal{L} f\,, \quad f(x,l,0) = g(x,l)\,.
\end{equation}
Here $\mathcal{L}$ is the adjoint operator of $\mathcal{L}^{\star}$, defined by the operator that satisfies $\langle f , \mathcal{L}^{\star} p \rangle = \langle \mathcal{L} f , p \rangle$ for any probability density $p$ and statistic $f$, where $\langle f , p \rangle = \iint (f_u p_u + f_b p_b) dl dx$ is the inner product.

\subsubsection{Non-dimensionalization and assumptions on scales.}

%, both in time and space,We  expect  a  scaleseparation both in space (spring length fluctuations arebounded compared to particle motion that is unbounded)and in time (binding and unbinding dynamics, as well asspring relaxation dynamics, are fast compared to parti-cle motion that can be observed at infinitely long times)
%To use averaging techniques we now need to make the different scales at play explicit, both in time and space. We write the nondimensionalization

We now seek to coarse-grain the fast dynamics,  by applying homogenization techniques to the backward equation, Eq.~\eqref{eq:backward}. 
%We now seek to coarse-grain the backward equation Eq.~\eqref{eq:backward}, in order to average over the fast relaxation and binding terms and obtain the effective dynamics of the particle.
To start, we non-dimensionalize the equation using 
\begin{equation*}
 x \rightarrow L_x \tilde{x}, \,\, l - l_0 \rightarrow L \tilde{l}, \,\,  t \rightarrow \tau \tilde{t}, 
\end{equation*}
where $L = \sqrt{k_B T/k}$ is the reference length of the leg fluctuations,  $L_x$ is the scale for the long-time average motion of $x$, and $\tau$ is the timescale associated with this average motion. 
The latter two scales are not determined \textit{a priori} by any intrinsic scales in the system, but rather are chosen large enough that averaging will be appropriate over such scales; hence we choose $L_x = L/\epsilon$ where $\epsilon\ll 1$ is a small non-dimensional number. 
We are interested in long time scales corresponding to the diffusion of the particle, hence we expect $\tau = L_x^2/D_0$, which corresponds to $\tau = \frac{1}{\epsilon^2}\frac{\Gamma}{k}$. 
%As we may observe the dynamics over \textit{arbitrarily} long times $\tau$, this also corresponds to $L_x$ reaching \textit{arbitrarily} broad scales. As a result we may define the small parameter $\epsilon = L/L_x$ and hence $\tau = \frac{1}{\epsilon^2}\frac{\Gamma}{k}$. 
Importantly, and in contrast with other works~\cite{fogelson2018enhanced,fogelson2019transport}, here $\epsilon$ does not measure the value of physical parameters, but rather, it measures the large observation time scale over which the coarse-grained model is valid. Such long observation times are quite likely in experiments, as typical binding rates and leg dynamics occur at most over $1~\mathrm{ms} - 1~\mathrm{s}$ while observation (or other biophysical processes such as internalisation for viruses~\cite{sakai2017influenza}) happens over the course of $10~\mathrm{min}$ at least~\cite{fan2021microscopic}. This non-dimensionalization step is crucial as it will allow us to find order of magnitude changes in the diffusion coefficient according to the physical parameters, something that was not captured by previous perturbative approaches~\cite{fogelson2018enhanced,fogelson2019transport}.

We now assume that the observation time scale is long enough, such that binding and unbinding events, as well as relaxation dynamics, will both occur on comparably short time scales. We can therefore write $ \tilde{q}_i = q_{i} \Gamma/k = O_\epsilon(1)$ and $\gamma/\Gamma = O_\epsilon(1)$. In Sec.~\ref{sec:Models} we  will see that taking different limits for these physical parameters (such as $\gamma/\Gamma \ll 1$) yields the same result as applying these limits to the final result. Our choices of scalings are therefore quite general and can be easily adapted to more detailed systems. 

Using non-dimensional variables (and dropping the $\tilde{.}$ for simplicity) we obtain from the backward equation Eq.~\eqref{eq:backward} a separation in orders of $\epsilon$ as
\begin{equation}
    \partial_t f = \mathcal{L} f = \left( \frac{1}{\epsilon^2} \mathcal{L}_0 + \frac{1}{\epsilon} \mathcal{L}_1 + \mathcal{L}_2 \right) f
    \label{eq:Generator1}
\end{equation}
where
\begin{equation*}
\begin{split}
    &\mathcal{L}_0 = \begin{pmatrix}  - q_{\rm on} + \frac{\Gamma}{\gamma} ( - l \partial_ l + \partial_{ll} ) & q_{\rm on} \\
 q_{\rm off} & - q_{\rm off} + \frac{\Gamma}{\Gamma + \gamma} ( - l \partial_ l + \partial_{ll} ) 
    \end{pmatrix}, \\
   & \displaystyle \mathcal{L}_1 = \mathrm{diag} \left( 0, \frac{\Gamma}{\Gamma + \gamma} \left( l \partial_x - 2 \partial_{lx} \right) \right), \\
   &\displaystyle  \mathcal{L}_2 =  \mathrm{diag} \left( \partial_{xx}  ,\frac{\Gamma}{\Gamma + \gamma}\partial_{xx}  \right) .
    \end{split}
\end{equation*}

\subsubsection{Homogenization method.}

%It is now possible to use homogenization to obtain long time effective dynamics for the particle position $x$. One of these techniques, homogenization~\cite{pavliotis2008multiscale}, is quite relevant as it allows to properly \textit{average} over all the short time scale dynamics to obtain dynamics valid at long times (for $\epsilon \ll 1$). We explain the methodology below (see also Supp. Mat. 1.2). 

We seek a solution to Eq.~\eqref{eq:Generator1}  of the form $f = f_0 + \epsilon f_1 + \epsilon^2 f_2 + ...$. We obtain a hierarchy of equations at different orders in $\epsilon$:
\begin{eqnarray}
    & O_{\epsilon}\left( \frac{1}{\epsilon^2} \right): \quad & \,\, \mathcal{L}_0 f_0 = 0,  \label{eq:order0} \\
    & O_{\epsilon}\left( \frac{1}{\epsilon} \right): \quad & \,\,  \mathcal{L}_0 f_1 = - \mathcal{L}_1 f_0,  \label{eq:order1}  \\ 
    & O_{\epsilon}\left( 1 \right): \quad & \,\, \mathcal{L}_0 f_2 = \partial_t f_0 - \mathcal{L}_1 f_1 - \mathcal{L}_2 f_0 , \label{eq:order2}  \\
   & \vdots & \,\, \phantom{\mathcal{L}_0 f_2 }  \;  \vdots \nonumber
\end{eqnarray}
and we solve these iteratively for $f$ at each order in $\epsilon$. 
At lowest order we obtain from Eq.~\eqref{eq:order0} and the vanishing flux at boundaries, $f_0 = a(x,t) \begin{pmatrix} 1 \\ 1\end{pmatrix}$, where $a(x,t)$ is an unknown function of the slow variable $x$, whose dynamics we seek to determine. The associated equilibrium distribution at lowest order, $\mathcal{L}^{\star}_0 \pi_0 = 0$ is simply the full one $\pi_0 = \pi$.
%$\mathcal{L}_0 f_0 = 0$. The general solution is $f_0 = a(x,t) \begin{pmatrix} 1 \\ 1\end{pmatrix} + b(x,t) \int_0^l e^{y^2}dy.$ We expect the flux in probability space to vanish far from the center, $\left[ l p_0 + \partial_l p_0 \right]_{l = \pm \infty} = 0$ translating into $\left[ \partial_l f_0 \right]_{l = \pm \infty} = 0$, yielding $b(x,t) = 0$. 

At the next order, one can check that
\begin{equation*}
    f_1 = \begin{pmatrix} \gamma q_{\rm on} \\ \Gamma + \gamma q_{\rm on}  \end{pmatrix} \frac{l \partial_x a}{\Gamma (1 + q_{\rm off}) + \gamma (q_{\rm on} + q_{\rm off}) } 
\end{equation*}
is a particular integral of Eq.~\eqref{eq:order1}, and is the unique solution since we impose that $f_1$ does not contain terms in the nullspace of $\mathcal{L}_0$. 

Finally Eq.~\eqref{eq:order2} possesses a solution if and only if it satisfies the Fredholm alternative~\cite{pavliotis2008multiscale}
\begin{equation*}
    \langle (\partial_t f_0 - \mathcal{L}_1 f_1 - \mathcal{L}_2 f_0) , \pi_0 \rangle = 0.
\end{equation*}
Standard algebra yields an effective long time diffusion equation for $a$ (in dimensional variables) 
\begin{equation}
   % \partial_t a = D_{\rm eff} \partial_{xx} a 
   % \quad \Leftrightarrow \quad  
    \partial_t a = D_{\rm eff} \partial_{xx} a, 
    \label{eq:dxa} 
\end{equation}
% where
% \begin{equation}\label{eq:gammaeff1}
%      \frac{1}{\Gamma_{\rm eff}} = \frac{p_0}{\Gamma_0} + \frac{p_1}{\Gamma_1} ,
% \end{equation}
% with
% \begin{equation*}
% \Gamma_0 = \Gamma, \;\;\Gamma_1 = \Gamma + \gamma_{\rm eff},
%  \,\, \gamma_{\rm eff} =  \gamma +  k\left( \frac{1}{q_{\rm off}}  + \frac{\gamma}{k} \frac{\on}{\off} \right).
% \end{equation*}
where
\begin{equation}\label{Deff1}
     D_{\rm eff} = \frac{k_B T }{\Gamma_{\rm eff}},
\end{equation}
with 
\begin{equation}
\begin{split}
    \frac{1}{\Gamma_{\rm eff}} &= \frac{p_0}{\Gamma_0} + \frac{p_1}{\Gamma_1} , \quad\text{with }\;\Gamma_0 = \Gamma, \;\;\Gamma_1 = \Gamma + \gamma_{\rm eff} \\
    &\mathrm{and} \,\, \gamma_{\rm eff} =  \gamma +  k\left( \frac{1}{q_{\rm off}}  + \frac{\gamma}{k} \frac{\on}{\off} \right).
\end{split}
 \label{eq:gammaeff1}
\end{equation}
%The effective diffusion coefficient is $D_{\rm eff} = \frac{k_B T }{\Gamma_{\rm eff}}$.
In the above expressions, $p_0 = \frac{q_{\rm off}}{q_{\rm off} + q_{\rm on}}$ is the equilibrium probability to have no bond, and $p_1 = 1 - p_0$ the equilibrium probability to have one bond. $\Gamma_0 = \Gamma$ is the friction in the unbound state and $\Gamma_1$ is the effective friction contributing to the bound state.

Eq.~\eqref{eq:dxa}, which is the backward equation for the particle+leg over long times, is one of the main results of this paper, in the case of a 1-legged caterpillar. It is the backward equation for a particle that evolves as 
\begin{equation}
    \frac{dx}{dt} = \sqrt{2D_{\rm eff}} \eta_x(t).
    \label{eq:dxeff}
\end{equation}
That is, the particle diffuses, with effective diffusion coefficient $D_{\rm eff}$ and effective friction $\Gamma_{\rm eff}$.
The effective diffusivity and friction have the usual interpretation. In particular, if a potential $\mathcal{U}(x)$ were added to the particle Eqns.~\eqref{eq:dxUnbound} and \eqref{eq:dxProjected}, one would recover in Eq.~\eqref{eq:dxeff}, following the same coarse-graining procedure,  a term $- \frac{1}{\Gamma_{\rm eff}} \partial_x \mathcal{U}$.

In Fig.~\ref{fig:fig2} we compare the analytical result obtained in Eq.~\eqref{eq:gammaeff1} (gray line) to numerical simulations of the full stochastic Eqns.~\eqref{eq:dlUnbound}-\eqref{eq:dxProjected} (gray dots). We show the results for a number of system parameters and find perfect agreement over several orders of magnitude of physical parameters. We also predict order of magnitude changes in the diffusion coefficient as the microscopic parameters change.

%\medskip
%((Note that Eq.~\eqref{eq:gammaeff1} has been properly arranged to highlight that when the leg has zero resistance $k = 0$ (starting from Eq.~\eqref{eq:dlUnbound}) homogenization would simply yield $\gamma_{\rm eff} =  \gamma$. \mhc{this is wrong. when $k=0$ you get $\gamma+\gamma  \frac{\on}{\off}$.})) \sophie{not when you start from Eq.~() and redo the math. But maybe I didn't take the limit properly from the start. but if you have no spring constant, one should really get just $\gamma$. I think the problem is to define well $k \rightarrow 0$}
% actually the limit k -> 0 is just ill-defined because then you can not define the non-dimensional length L. So let's just forget about this comment. 

\subsection{Microscopic parameters determine long term diffusion}\label{sec:1legnterpret}

% To understand the long term Eq.~\eqref{eq:dxa} on the statistics, we revert to it's equivalent stochastic formulation for the particle's position $x$
% \begin{equation}
%     \frac{dx}{dt} = \sqrt{\frac{k_B T}{\Gamma_{\rm eff}}} \eta_x(t).
%     \label{eq:dxeff}
% \end{equation}
% At long times the particle diffuses with the diffusion coefficient $D_{\rm eff} = \frac{k_B T}{\Gamma_{\rm eff}}$. The effective diffusion and friction have the usual interpretation. In particular, if a potential $\mathcal{U}(x)$ were added to the particle Eqns.~\eqref{eq:dxUnbound} and \eqref{eq:dxProjected}, one would recover in Eq.~\eqref{eq:dxeff} a $- \frac{1}{\Gamma_{\rm eff}} \partial_x \mathcal{U}$ term.

How shall we interpret the expressions for the effective diffusivity Eq.~\eqref{Deff1} and the effective friction Eq.~\eqref{eq:gammaeff1}? 
The effective diffusivity is a weighted sum of the diffusivity in each state, $D_{\rm eff}  = p_0 D_0 + p_1 D_1$ where the weights correspond to the probability to be in either state, and $D_i=k_BT/\Gamma_i$. The effective friction, on the other hand, is a harmonic weighted sum of the friction coefficients. That the diffusivity averages arithmetically is to be expected, since the mean squared displacement is an extensive quantity in a system with multiple states. Over a time $t$ we can write \begin{equation*}
\begin{split}
    \overline{x^2(t)} = 2 D_{\rm eff} t  &= 2   D_0 p_0 t + 2  D_1 p_1 t \\ &= 2 D_0 t_0 + 2 D_1 t_1 = \overline{x^2(t)}|_{0} + \overline{x^2(t)}|_{1},
    \end{split}
\end{equation*} 
where $t_0$ and $t_1$ refer to the time spent in either state. The novelty here is that the diffusivity in the bound state, 
\[
D_1 = k_BT(\Gamma +\gamma_{\rm eff})^{-1} \neq k_BT(\Gamma + \gamma)^{-1},
\]
is obtained not just from the friction in the bound state, see Eq.~\eqref{eq:gammaeff1},
%that $\Gamma_1 = \Gamma +\gamma_{\rm eff} \neq \Gamma + \gamma$ is not just the friction in the bound state, see Eq.~\eqref{eq:gammaeff1}, 
but is modified by spring resistance during binding events by an additional term $\gamma_{\rm eff}-\gamma$.

We can interpret this additional term  by writing it as  
\[
\gamma_{\rm eff}-\gamma =  k \tau_{\rm eff}, \quad \text{where}\;\; \tau_{\rm eff} = \tau_{\rm b} + \tau^{\rm relax}_{\rm u}
\]
is the typical time over which the leg's spring resistance acts, with
$\tau_{\rm b} = 1/q_{\rm off}$ representing the average bound time, and 
$\tau^{\rm relax}_{\rm u} = \frac{\gamma}{k} \frac{\on }{\off} = \frac{\gamma}{k} \frac{\tau_{\rm b}}{\tau_{\rm u}}$ representing the bare relaxation time $\gamma/k$ increased by the ratio of average bound time to average unbound time. This is coherent as the leg fluctuations may only relax in the unbound state. 
% \begin{itemize}[nosep]
%     \item $\tau_{\rm b} = 1/q_{\rm off}$ the average bound time and
%     \item $\tau^{\rm relax}_{\rm u} = \frac{\gamma}{k} \frac{\on }{\off} = \frac{\gamma}{k} \frac{\tau_{\rm b}}{\tau_{\rm u}}$ is the bare relaxation time $\gamma/k$ increased by the ratio of average bound time to average unbound time. This is coherent as the spring may only relax when it is unbound. 
% \end{itemize}
The interpretation of $\tau_{\rm eff}$ is comparable to that in Ref.~\citenum{lee2018modeling} although the results of Ref.~\citenum{lee2018modeling} were obtained from underdamped dynamics.

\begin{figure}[h!]
\includegraphics[width = 0.99\linewidth]{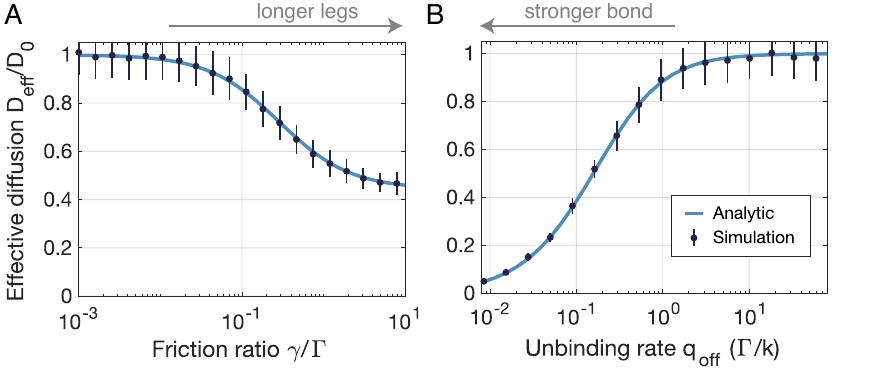}% Here is how to import EPS art
\caption{\label{fig:fig2} \textbf{Effective diffusion $D_{\rm eff}$ of a 1-legged particle.} Simulation and analytical result Eq.~\eqref{eq:gammaeff1} for a 1D system with 1 leg, with respect to (A) friction ratio $\gamma/\Gamma$ and (B) unbinding rate $q_{\rm off}$. (A) and (B) share the same y-axis. The other numerical parameters are $q_{\rm on}\Gamma/k = 1.0$,  and for (A) $q_{\rm off}\Gamma/k = 0.8$ while for (B) $\gamma/\Gamma = 0.1$. Error bars represent one standard deviation for 100 independent runs. }
\end{figure}

Fig.~\ref{fig:fig2} shows how the effective diffusion coefficient depends on microscopic parameters such as the leg friction and binding rates. 
As the leg friction $\gamma$ increases, the effective diffusion of the particle decreases (Fig.~\ref{fig:fig2}-A). When the leg friction $\gamma$ is large compared to all other contributions to friction, diffusion in the bound state is frozen $D_1 = 0$, and the effective diffusion corresponds only to mobility in the unbound state $D_{\rm eff} = p_0 D_0$ ($p_0 = 0.8/1.8 \simeq 0.44$ in Fig.~\ref{fig:fig2}-A).
%\mhc{but not arbitrarily -- it is bounded below by something} % Good remark!
As leg friction is typically proportional to the size of the legs, it is thus expected that the bigger the legs, the slower the particle. As the  unbinding rate $q_{\rm off}$ decreases,  $D_{\rm eff}$ decreases to arbitrarily small values (Fig.~\ref{fig:fig2}-B). This slow down is due to spring recoil forces acting over longer times, eventually freezing the particle in a given location. Note that similar qualitative dependencies of the diffusion coefficient on the unbinding rate ($D_{\rm eff} \sim k_B T \off/k$) were noted in a numerical model of multivalent transport on discrete sites~\cite{kowalewski2021multivalent}, in a scaling law investigation of sticky reptation in polymers~\cite{leibler1991dynamics}, and experimentally in Influenza A viruses~\cite{muller2019mobility}. 
%This suggests that slow down due to recoil forces is a universal phenomenon across nanocaterpillar systems. 

As a test of modeling choice, the analytical expression may also be plotted against numerical simulations of the non-dimensional equations with any value of $\epsilon$. We find perfect agreement up to $\epsilon \lesssim 10$ (Supplementary Fig.~S1), regardless of the choice of physical parameters. This highlights that the natural choice  $\epsilon = L/L_x$ for coarse-graining purposes, corresponding to bound leg length scales versus unbound particle long range motion, is especially well suited for these types of problems. In the following $\epsilon$ is not incorporated in numerical simulations.

% according to the value of the small parameter $\epislon = L/L_x$ 
% for the \epsilon graph

\subsection{Diffusion of N-legged caterpillar spans orders of magnitude}
\label{sec:Nlegs}

\begin{figure}[h]
\includegraphics[width = 0.99\linewidth]{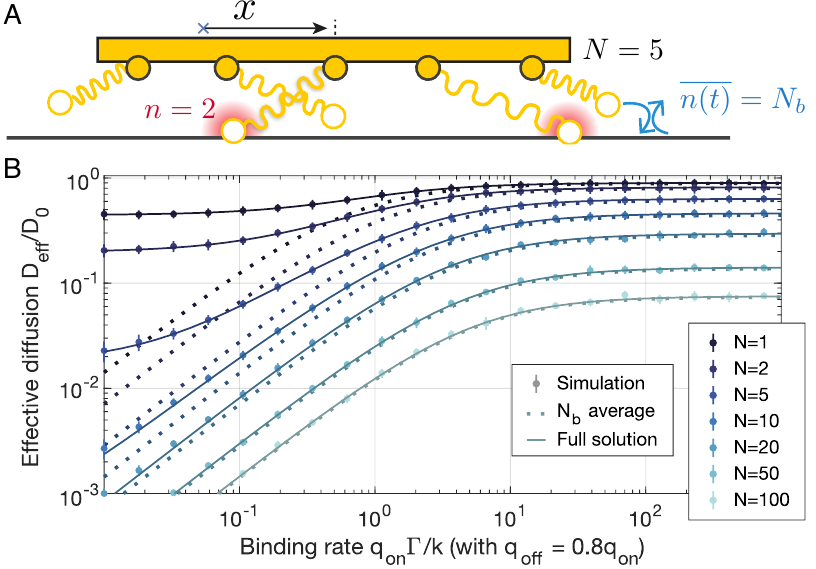}% Here is how to import EPS art
\caption{\label{fig:figNmodel} \textbf{N-legged nanocaterpillar model.} (A) The longitudinal extension of $N$ legs are monitored (here $N=5$) with binding and unbinding. The number of bonds $n(t)$ changes in time, here $n(t) = 2$. The average number of bonds $\overline{n(t)} = N_b$ depends on the binding and unbinding rates. (B) Simulations and analytical results of the effective diffusion coefficient for $N$-legs according to the binding rate $q_{\rm on}\Gamma/k$. ``$N_b$ average" corresponds to Eq.~\eqref{eq:gammaNb} and ``full solution" to Eq.~\eqref{eq:gammaN}. The other numerical parameters are $\gamma/\Gamma = 0.1$ and $q_{\rm off} = 0.8 q_{\rm on}$.}
\end{figure}

We extend our framework to probe nanocaterpillar dynamics with an arbitrary number of legs $N$ (see Fig.~\ref{fig:figNmodel}-A). Eq.~\eqref{eq:dlUnbound} is repeated for each unbound leg, and each leg binds to the surface with rates $\on, \off$ independently. Eq.~\eqref{eq:dxUnbound} gives the particle dynamics when no legs are bound.  When $n$ legs are bound, indexed by $i= 1,\ldots,n$, the dynamics of the particle and bound legs are constrained as (Supplementary~1.2)
\begin{equation}
    \frac{dx}{dt} = - \frac{dl_i}{dt} = \frac{k}{\Gamma + n \gamma} \sum_{i=1}^n (l_i - l_0) + \sqrt{\frac{2k_B T}{\Gamma + n\gamma}} \eta.
    \label{eq:dxProjectedN}
\end{equation}
Note here that the projection step yields a friction coefficient scaling linearly with the number of bonds $n$, and hence is not a perturbative effect~\cite{fogelson2018enhanced}. The set of stochastic equations is now fully determined and can be simulated for any $N$, see Fig.~\ref{fig:figNmodel}-B. 

Similarly as in Sec.~\ref{sec:homo}, coarse-graining predicts a long time effective diffusion with $N$ legs as (Supplementary~1.2)
\begin{equation}
    D_{\rm eff}^{N\, \rm legs} =  \frac{k_B T}{\Gamma^{N\, \rm legs}_{\rm eff}} = k_B T\sum_{n=0}^N \frac{p_n}{\Gamma_n}
    \label{eq:gammaN}
\end{equation}
where $p_n = \binom{N}{n} \frac{q_{\rm off}^{N-n} q_{\rm on}^n}{(\off + \on)^N} $ is the equilibrium probability to have $n$ bonds and $\Gamma_n$ is the friction coefficient in a state with $n$ bonds. The frictions $\{\Gamma_n\}$ solve a linear system of equations that does not have a simple analytical solution (see Eqns.~(S1.20-22)), but can be solved using numerical linear algebra for given parameters as
reported in Supplementary~1.2. 

Eq.~\eqref{eq:gammaN} is one of the main results of this paper. It predicts the long-term diffusion coefficient of a nanocaterpillar, as a non-trivial function of the microscopic parameters of the legs. 
We compare the numerically solved Eq.~\eqref{eq:gammaN}  (full lines) to numerical stochastic simulations with $N$ legs (dots) in Fig.~\ref{fig:figNmodel}-B and find excellent agreement. %Eq.~\eqref{eq:gammaN} is the second main result of our work. 

The coefficients $\Gamma_n$ contributing to each bound state can be further investigated to yield an analytical approximation for $\Gamma^{N\, \rm legs}_{\rm eff}$. When a large number of legs $N$ is involved in the process, the dominant term in the sum of Eq.~\eqref{eq:gammaN} corresponds to the average number of bonds $N_b = \sum_{n=0}^N n p_n = \frac{\on}{\off + \on} N$. Furthermore,  one expects that the coefficients vary weakly around $n = N_b$, simplifying the linear system for the  $\{\Gamma_n\}$, yielding
%In the limit where the average number of bonds $N_b = \sum_{n=0}^N n p_n = \frac{\on}{\off + \on} N$ is small compared to the total number of legs, $N_b \ll N$, we find that $\Gamma_n$ grows linearly with the number of bonds as
%\begin{equation}
%     \Gamma_n \stackrel[N_b \ll N]{}{\simeq} \Gamma + n \gamma_{\rm eff} \,,
%\end{equation}
%where $\gamma_{\rm eff} = \gamma + k\left( \frac{1}{q_{\rm off}}  + \frac{\gamma q_{\rm on}}{k q_{\rm off}} \right)$ as in Eq.~\eqref{eq:gammaeff1} (Fig. S4). 
% and is coherent with Eq.~\eqref{eq:gammaeff1} (Fig. S4). 
% \begin{equation}
%     \Gamma_n \stackrel[N_b \ll N]{}{\simeq} \Gamma + n \gamma_{\rm eff} \,\, \mathrm{with}  \,\, \gamma_{\rm eff} = \gamma + k\left( \frac{1}{q_{\rm off}}  + \frac{\gamma q_{\rm on}}{k q_{\rm off}} \right)
% \end{equation}
% and is coherent with Eq.~\eqref{eq:gammaeff1} (Fig. S4). 
\begin{equation}
    \frac{1}{\Gamma^{N\, \rm legs}_{\rm eff}} \stackrel[N \gg 1]{}{\simeq} \frac{1}{\Gamma_{N_b}} = \frac{1}{\Gamma + N_b \gamma_{\rm eff}}.
    \label{eq:gammaNb}
\end{equation}
The right hand side of Eq.~\eqref{eq:gammaNb} is valid regardless of parameter values (Fig. S3) and provides a good approximation for $\Gamma^{N\, \rm legs}_{\rm eff}$ for large values of $N$ (Fig. S2). For example, close agreement with Eq.~\eqref{eq:gammaN} is obtained as early as $N=20$, while good qualitative agreement is obtained  for $N=5$ (see Fig.~\ref{fig:figNmodel}-B, dotted line). Eq.~\eqref{eq:gammaNb} shows that the effective friction with $N$ legs decays linearly with the \textit{average} number of bonds $N_b$. For systems with a large number of legs (and hence potentially a large average number of bonds)~\cite{xu2011subdiffusion,wang2015crystallization,fan2021microscopic}, we therefore expect a strong diffusion decrease, covering potentially several orders of magnitude, due to enhanced friction with the surface. 

\section{Do nanocaterpillars hop or slide?}
\label{sec:slideorhop}

Our model and analytical formula Eq.~\eqref{eq:gammaN} are useful not only for quantitatively predicting the diffusion coefficients of existing nanocaterpillar systems, but also to obtain insight into the \emph{mechanism} by which particles diffuse. 
%We will now use our model to investigate the diffusive motion of existing nanocaterpillar systems. Beyond quantitatively predicting the diffusion coefficient, our model  gives insight into the mechanism by which particles move on the surface.
%For example, in our experiments, we can see colloids diffusing cohesively on the surface, except slightly above the melting temperature where colloids lift-off for short times and return to the surface. 
Different experiments with DNA-coated colloids made puzzling and seemingly contradictory observations, whereby similar systems appear to diffuse in different ways. For example, some DNA-coated colloids appear to diffuse through a succession of uncohesive moves, namely hops above the surface~\cite{xu2011subdiffusion}, while others move cohesively along the surface~\cite{wang2015crystallization}. The difference between cohesive and uncohesive modes of motion has been noted in a variety of other systems, ranging from virus mobility on surfaces~\cite{muller2019mobility,sakai2017influenza} to sticky polymer reptation~\cite{leibler1991dynamics}. 
%A systematic derivation of mobility for both these modes of motion from microscopic details  has not, to the best of our knowledge, been carried out. 
Yet the parameters that characterize and quantify these different modes of motion remain to be elucidated.
Our model gives insight into this question -- 
 do nanocaterpillars prefer to diffuse by ``sliding'' along the surface, or by ``hopping'' along it (see Fig.~\ref{fig:fig4}-A)?

\subsection{What are hopping and sliding?}

\begin{figure}[h]
\includegraphics[width = 0.99\linewidth]{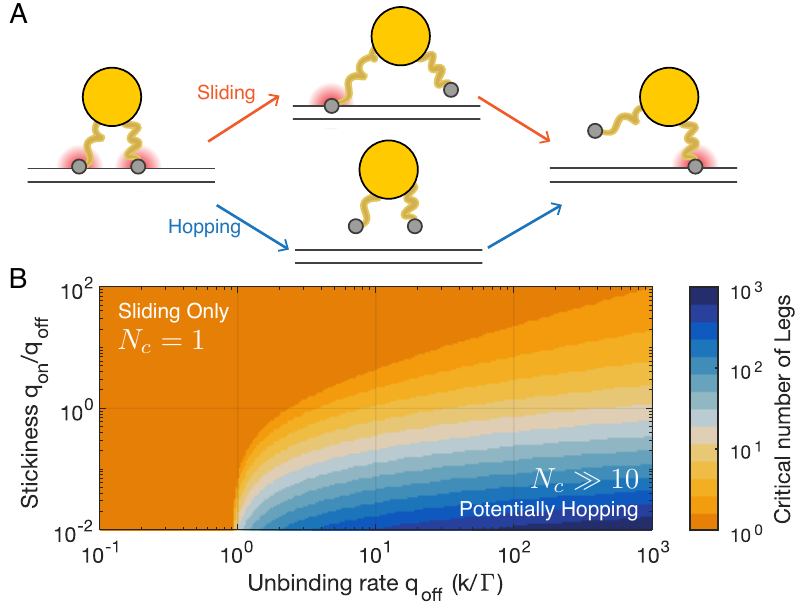}% Here is how to import EPS art
\caption{\label{fig:fig4} \textbf{Nanocaterpillar diffusion modes with $N$ legs.} (A) Typical modes of motion with $N$ bonds: the nanocaterpillar may either slide (at least one bond remains attached to the surface) or hop (all bonds detach for the particle to move). (B) Critical number of legs $N_c$ required for sliding to be more effective than hopping as a function of stickiness $\on/\off$ and unbinding rate. %\mhc{is stickiness defined somewhere?} \sophie{yep was missing in this section, added it now. Do I need to put $\kappa$ in the label and/or caption? I try just to use it to simplify notations in some spots of the text. It's used for the first time in paragraph 2 right below.}\mhc{I would just not say kappa at all here}
%Given a number of legs of the system, say $N = 10$, the sliding region corresponds to stickiness and unbinding rate such that $N \geq N_c$, therefore to the orange region above the dashed black line $N_c = 10$. Hopping corresponds to $N \leq N_c$ and therefore parameter regions in the shades of blue.
%\mhc{I'm a bit confused about N vs Nc. Is this plot at fixed N? why is N=Nc?}
% \sophie{is this clearer ? $N_c (\on, \off, k/\gamma)$ so it all depends on whether $N \geq N_c$ or not.}\mhc{I actually find it clearer without the explanation! And possibly without the dashed line. The dashed line makes me wonder what is special about it. Adding "N" in the caption makes me wonder where "N" is in the figure. I would just delete everything after "... qon/qoff and unbinding rate." And possibly delete the black line.}\mhc{I am also confused about why some regions are labelled "sliding" and "hopping". I thought this was just a heatmap of Nc.} 
}
\end{figure}

We start by quantifying the diffusion associated with either hopping or sliding. 
The mean squared displacement of a particle whose diffusion coefficient is determined from Eq.~\eqref{eq:gammaN} can be split into two contributions, as 
\begin{align*}
\langle x^2 \rangle = 2 D_{\rm eff} t &= 2 p_0 \frac{k_B T}{\Gamma_0} t + 2 \sum_{n=1}^N p_n \frac{k_BT}{\Gamma_n} t \\
%& \equiv \langle x^2\rangle_{\rm hop} + \langle x^2\rangle_{\rm slide}\\
& \equiv 2 D_{\rm hop} t + 2 D_{\rm slide} t.
\end{align*}
We identify (a) a \textit{hopping} mode (in accordance with Refs.~\citenum{loverdo2009quantifying} and \citenum{xu2011subdiffusion}) where the particle detaches \textit{all bonds} with the surface and moves in free space (see Fig.~\ref{fig:fig4}-A), until it forms another bond. In this hopping mode
    \begin{equation}
        D_{\rm hop} = p_0 \frac{k_B T}{\Gamma} = \left( \frac{\off}{\off + \on} \right)^N \frac{k_B T}{\Gamma}.
        \label{eq:hopping}
    \end{equation}
    %Eq.~\eqref{eq:hopping} is interpreted in the following way. 
    %The mean squared displacement for hopping is obtained as $\langle x^2\rangle_{\rm hop} = 2 D_{\rm hop} t = 2 D_0 p_0 t = 2 D_0 t_{\rm unbound}$ where $t_{\rm unbound}= p_0 t$ is the time spent unbound. 
    We also isolate (b) a \textit{sliding} mode (see Fig.~\ref{fig:fig4}-A) where the particle keeps at least one bond with the surface, a form of walking with no preferred direction,
    \begin{equation}
            D_{\rm slide} = \sum_{n=1}^N\frac{p_n}{\Gamma_n} \simeq \frac{k_B T}{\Gamma_{N_b}} = \frac{k_BT}{\Gamma + N \frac{\on}{\off + \on}  \gamma_{\rm eff}}.
            \label{eq:slide}
    \end{equation}
     %The mean squared displacement for sliding is simply $\langle x^2\rangle_{\rm slide} = 2 D_{\rm slide} t$, \textit{i.e.} it is the displacement of the particle while it is bound to the surface by at least one bond. \mhc{why not t(bound)?} \sophie{It's not t(bound) right, you would have to expand $ = 2 D_i p_i t = 2 (D_i p_i)/(p_i) t_{bound}$ and that wouldn't be nice??}
    % \mhc{ok -- but then this is a bit different from the unbound case...}
The total mean-squared displacement can be broken up into the sum of the mean-squared displacement when hopping, and the mean-squared displacement when sliding, as 
$\langle x^2 \rangle = 2 D_{\rm hop} t + 2 D_{\rm slide} t = \langle x^2\rangle_{\rm hop} + \langle x^2\rangle_{\rm slide}$.

%Remarkably, the sliding diffusion coefficient $D_{\rm slide}$ decays as $1/N$ where $N$ is the total number of legs. This is interesting because it simply illustrates why diffusion is greatly stalled (typically by 1 to 2 orders of magnitude) in systems where a large number of legs are at play, see Refs.~\cite{xu2011subdiffusion,sakai2017influenza,muller2019mobility}. We will come back to typical orders of magnitude relative to actual systems in Sec.~\ref{sec:realsystems}. 

An important observation is that $D_{\rm slide}$ decays with the number of legs roughly as  $1/N$, while $D_{\rm hop}$ decays exponentially with $N$, \textit{i.e.} much faster. As soon as a few legs are involved, we may therefore expect that \textit{sliding} dominates \textit{hopping}. This interpretation is natural, since when a system has just a few legs ($N \simeq 1-2$), the odds that the legs all \textit{detach at once} are quite high, therefore favoring hopping. In contrast, in a system with a large number of legs, the odds that all legs \textit{simultaneously} detach are simply too small, and the system walks randomly, remaining close to the surface. In a sense, nanocaterpillars truly are caterpillars walking with nanoscale legs. The scaling quantifying both modes of motion is another essential analytical result of our work. 
%\mhc{should you reference Fig 6 somewhere in this section?}
% yep it's already the case

In general, the critical number of legs $N_c(\on,\off,k,\gamma,\Gamma)$ required to favor sliding ($N \geq N_c$) over hopping ($N \leq N_c$) satisfies 
\begin{equation}
    \frac{\langle x^2\rangle_{\rm hop}}{\langle x^2\rangle_{\rm slide}} =  \frac{D_{\rm hop}}{D_{\rm slide}} = \left( \frac{\off}{\off + \on} \right)^{N_c} \left( 1 + N_c \frac{\on}{\off + \on}  \frac{\gamma_{\rm eff}}{\Gamma} \right) = 1.
    \label{eq:Nc}
\end{equation}
The critical number of legs is controlled by the ratio $\on/\off$, termed henceforth \textit{stickiness}, and by the magnitude of the effective friction in the bound states $\gamma_{\rm eff}$, itself dominated in most systems by the unbinding rate $\off$. We can therefore investigate $N_c$ as a function of stickiness $\on/\off$ and unbinding rate $\off$ (Fig.~\ref{fig:fig4}-B). Overall, a system with say $N = 10$ legs is typically dominated by sliding motion. Yet hopping may still occur \textit{e.g.} with large unbinding rate $\off$. In fact $\off$ increases the friction $\gamma_{\rm eff}$ in the bound states and reduces $D_{\rm slide}$. %This naturally calls for a proper investigation of the values of physical parameters for other nanocaterpillars.
The number of legs is thus a critical parameter for nanocaterpillar diffusion: controlling both the magnitude of the diffusion decrease and the mode of motion. 

\subsection{Distinguishing the diversity of biophysical nanocaterpillars}
\label{sec:realsystems}

Whether a nanocaterpillar slides or hops, 
as predicted by Eq.~\eqref{eq:Nc}, depends on numerous system parameters. 
%is not only determined by the number of legs $N$ but also by other system parameters. 
Existing biological and biomimetic systems cover a broad range of parameters that we now explore, to ask which systems prefer to move by sliding and which by hopping, within the framework of our model.  

%While DNA-coated colloids possess similar physical parameters (for $k, \gamma, \off, \on, \Gamma$) amongst one another, a broad range is covered by the diversity of biological and biomimetic systems that corresponds with nanocaterpillars. We thus use our model to investigate these systems, and ask whether each tends to prefer sliding or hopping. %We now investigate the expected regimes, between sliding and hopping, in a variety of biological and biomimetic systems.  A central point is to efficiently grasp the diversity of behaviors. 

Our model relies on 6 physical parameters $k, \gamma, \off, \on, \Gamma, N$ that can be estimated from the literature for many systems : viruses, molecular motors, white blood cells, protein cargos in the nuclear pore complex, bacteria such as Escherichia coli, and DNA-coated colloids (Supplementary~3). 
Typically, stickiness values are similar across systems with $\on/\off \sim 0.05 - 0.8 \geq 1$ -- when the system is not thermally manipulated as will be explored in Sec.~\ref{sec:DNAccs}. Therefore we consider $\on/\off \simeq 0.1$. Additionally, as legs are generally small compared to particles, $\gamma/\Gamma \simeq 10^{-3} - 10^{-1}$ and therefore the dominant factor in $\gamma_{\rm eff}/\Gamma$ is usually controlled by spring recoil force and unbinding times, as $k / \Gamma \off$. We find $ k / \Gamma \off \simeq 10^{-2} - 10^{8}$ in the range of systems studied, confirming that this is a critical factor to discriminate nanocaterpillars. Additionally, as systems have a varied number of legs $N$, we define an effective relaxation rate 
$$\frac{k^{(N)}}{\Gamma} = \frac{k}{\Gamma} N \frac{\on}{\off + \on}  \left[\left(\frac{\off + \on}{\off} \right)^N - 1 \right]^{-1}$$ that will allow us to predict either sliding or hopping. 

\begin{figure}[t]
\includegraphics[width = 0.99\linewidth]{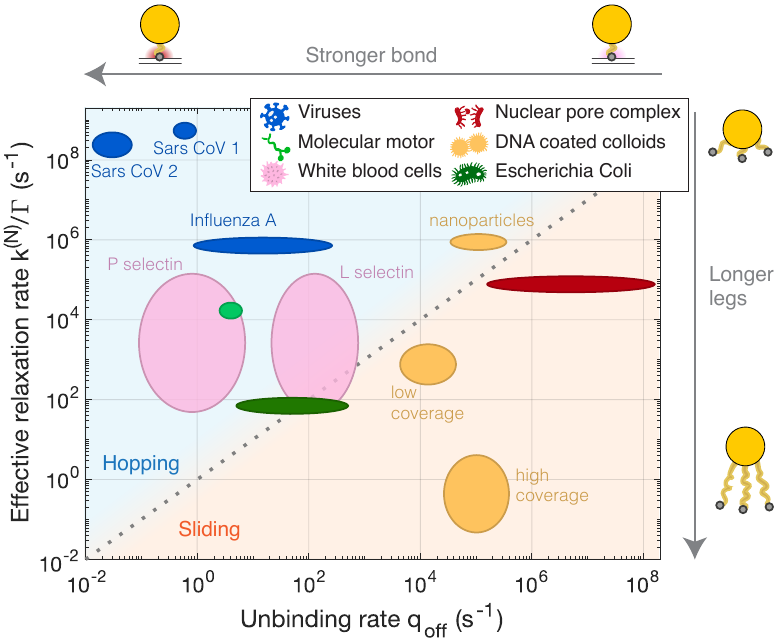}% Here is how to import EPS art
\caption{\label{fig:fig6} \textbf{Sorting biophysical systems.} Expected regimes of sliding or hopping according to the effective relaxation rate $k^{(N)}/\Gamma$ and unbinding rate $\off$. The gray line corresponds to $k^{(N)}/\Gamma = \off$ and separates the sliding and the hopping regions. Circles represent the range of values found in the literature for parameters of each system. Systems are color coded according to their category in the legend. When multiple systems belong to a category, details are indicated next to the circles. %Note that the light green circle (molecular motors) overlaps with the circle corresponding to white blood cell motion mediated by P-selectin adhesion. 
Low and high coverage DNA-coated colloids refer to $1~\mathrm{\mu m}$ size colloids and nanoparticles to $15~\mathrm{nm}$ size.}
\end{figure}

We sort systems in a so-to-speak \textit{Ashby} chart, according to the effective relaxation rate $k^{(N)}/\Gamma$ and unbinding rate $\off$ (Fig.~\ref{fig:fig6}). 
This chart summarizes parameter ranges for different systems, and predicts which systems move by sliding and which move by hopping, within the assumptions of our model. % (note that another mode of motion is rolling, which the model does not capture.). 
If $k^{(N)}/ \Gamma \off \leq 1$, according to Eq.~\eqref{eq:Nc}, sliding (orange region) is favored over hopping (blue region). While other modes of motion could occur for such complex systems, our aim here is to observe these systems in the ``projected'' sub-space where only sliding and hopping is considered. Interestingly, we find that different groups of systems emerge according to this classification, that we review below. 

\subsubsection{Sticky hoppers}

We predict that viruses, %(in blue in Fig.~\ref{fig:fig6}), 
white blood cells, and molecular motors cannot slide. These systems show very long bond lifetimes, with $\tau_{\rm off} = \off^{-1} \simeq 1 - 100 ~\mathrm{s}$. This is characteristic of strong bonds, for which the interaction energy $|\Delta G| \gg k_B T$. Since for the protein ligands in these systems, $k \simeq 10^{-4}~\mathrm{N/m}$ and $\Gamma \simeq 10^{-9}~\mathrm{N.s/m}$ for $1\mathrm{\mu m}$ particles, we expect $k/\Gamma \simeq 10^5 \gg \off$ and $\gamma_{\rm eff} \gg \Gamma$. Therefore such systems simply can not slide. Sliding is even more disfavored for coronaviruses (Sars CoV 1 and 2), since the legs are made of very rigid proteins, with $k \simeq 0.5~\mathrm{N/m}$~\cite{cao2021biomechanical,ponga2020quantifying}. Hopping is therefore a probable mode of motion for these systems.

%\mhc{I don't understand the point of this paragraph. Didn't you just say, in the paragraph above, that you predict hopping, for viruses? What does this paragraph add? You should say explicitly what it adds, in the first sentence. Eg., ``...''} Fair point !

These predictions are qualitatively consistent with experimental measurements. The diffusion coefficient of an influenza A virus on protein-coated surfaces was measured as $D_0/D_{\rm eff} \simeq 4-190$~\cite{muller2019mobility,sakai2017influenza}. %\mhc{was this including the active cutting proteins?} \sophie{so both of them basically study experimentally the activity of the cutting proteins, and when there's no-no cutting proteins the particle does not move, while the more they are active, the faster it moves. But there are more problems here. One of them is that actually the receptors seem to be embedded in bilayers in which case they are mobile. Another of them is that the on/off rates are typically measured in specific conditions including these so called cutting proteins. In fact it's not clear also how the cutting proteins work because in fact they are just say a type (2) of spike protein that seems to accelerate the cutting of type (1) being bound, but generally type 1 is not bound. I think we shouldn't go into the details, just acknowledge that there is more complexity, and say that hopping is probable -- but it doesn't mean that it's the right explanation and that there's more to it. I changed the text slightly, let me know if that's ok. }
%Although the virus particles in these experiments were more complex than our model, including \textit{e.g.} cleaving bonds, and possibly mobile receptors, we may nevertheless compare the measurements to the model's predictions.  
Estimating the typical number of available legs $N \simeq 10$~\cite{harris2006influenza,reiter2019force} and the bound probability $\on/(\on + \off) = 20\%$~\cite{reiter2019force} yields $D_0/D_{\rm hop} = \left[ \off/(\on + \off) \right]^N   \simeq 10$, in the range of measured values. Our model predicts that hopping is therefore more probable than sliding for influenza A, at least when considering its translational motion under passive binding and unbinding. %\mhc{qualify? when moving passively?} \sophie{so i can't really answer your question. Level 0 of active motion would be taking qon/qoff with the active cleaving, which is more or less what is done here, but of course there's also the fact that the particle can't go back on its tracks. Again, I would add caveats after, and start just by saying we made some attempt and it qualitatively seems to work. Notice that also to do this estimate we take a very crude estimation, that really just depends on the relative probability to be bound/unbound, not really the underlying details since this is just the estimation for hopping.} 
This is consistent with Ref.~\citenum{sakai2017influenza}, which observed infrequent yet very long spatial steps, termed \textit{gliding} moves. We note that the influenza A virus has also been observed to move via cohesive short spatial steps, 
%Additionally, cohesive short spatial steps were observed in these systems
that have been attributed to rolling motion~\cite{sakai2017influenza,muller2019mobility,alon2002rolling,ziebert2021influenza}, which may be due in this context to active bond cleaving~\cite{muller2019mobility,sakai2017influenza,ziebert2021influenza} that is beyond the scope of passive binding as presented here. %However, this serves as a reminder that other modes of motion may be observed for these systems, especially out-of-equilibrium. 
%\mhc{please don't end a sentence with ``are observed''. The most interesting / important information in a sentence should come at the beginning and end, since this is what readers pay the most attention to / what readers give the most emphasis to in the heads. The fact that something was observed is neither interesting nor important. You should end with ``very long spatial steps'' or something on that level of interest.} OK !!

Turning to DNA-coated colloids,
while the binding kinetics are roughly independent of colloid size, the effective relaxation rate can vary strongly. Nanometre-sized DNA-coated colloids (yellow nanoparticles) have fast relaxation rates as they are small (and therefore $\Gamma$ is smaller), and are thus sticky hoppers. In contrast, micronscale colloids have slower relaxation rates $k^{(N)}/\Gamma$, all the more as usually a great number of bonds $N\simeq 100$ are involved in the binding process, and thus are prone to slide. We will turn in more detail to DNA-coated colloids in Sec.~\ref{sec:DNAccs}. 
%The framework presented in Sec.~\ref{sec:sec1} could be extended to calculate rolling friction -- albeit for a few non-trivial mathematical subtleties that will be the purpose of future work. Nonetheless, as similar ingredients would enter (such as spring relaxation times and unbinding rates) we may expect similar parameters to discriminate between rolling and other modes of motion.

% \mhc{similar comments in the above paragraph. for example, I changed "Additionally, cohesive short spatial steps were observed in these systems" to ``The influenza A virus has additionally been observed to move by cohesive short spatial steps'' -- again, putting interesting / relevant information at the beginning and end (influenza A virus, a description of how it moves). It is NEVER interesting that something was observed -- this is boring, useless words that are best avoided if possible, (unless the manner of observation is relevant), but if not, put these words in the middle of the sentence where they won't take up brainwaves. I also changed ``these systems'' to ``influenza A'', since this is what you are talking about -- might as well highlight that.
% EDIT: on second thought, we can get rid of the useless observation statement, by just saying directly how the influenza A virus moves -- ``The influenza A virus additionally moves by cohesive short spatial steps''. We have references -- obviously these are records of observations -- so we can just say what happens, directly, which is much more interesting to read and much easier to follow. }

\begin{table*}[t]
\small
\caption{Method used to calculate model parameters for the DNA-coated colloids studied experimentally in this work. Parameter values are reported only at the melting temperature $T_m$. Their dependence on temperature is indicated in the ``Comments and References'' column.}
\label{tab:Params}
\begin{tabular*}{\textwidth}{@{\extracolsep{\fill}}p{1.5cm}p{3.0cm}p{2.5cm}p{10cm}}
\textrm{Parameter}& \textrm{Formula used} &
\textrm{Value at $T_m$}& 
\textrm{Comments and References}\\
\hline
$\Gamma$ & $\Gamma = 2\times 6\pi\eta(T) R$ & $1.6\times 10^{-8}~\mathrm{N.s/m}$ & hydrodynamic friction near a surface~\cite{brenner1961slow}; colloid radius $R=500~\mathrm{nm}$; $\eta(T)$ water viscosity with temperature. \\
$\gamma$& $\gamma = 6 \pi \eta(T) h$ & $1.8\times10^{-10}~\mathrm{N.s/m}$ & with brush height $h \simeq 22~\mathrm{nm}$, calculated with Milner-Witten-Cates theory~\cite{milner1988theory}, and accounting for increased brush density due to Pluronic F127 (see Ref.~\citenum{fan2021microscopic}). \\
$k$ & $\displaystyle k = 3 k_B T / 2 L \ell$ & $0.16~\mathrm{mN/m}$ & spring constant for polymers~\cite{rubinstein2003polymer}; extended brush length $L \simeq 84~\mathrm{nm}$ (6500 g/mol PEO + 20 single stranded DNA (ssDNA) bases); persistence length $\ell = 0.5~\mathrm{nm}$ (average of PEO + ssDNA at $140~\mathrm{mM}$ salt concentration ~\cite{chen2012ionic}) \\
$\on $ & $q_{\rm on} = k_{\rm on} \bar{\sigma} / h \mathcal{N}_A$ & $ 4~\mathrm{kHz}$ & where $k_{\rm on} = 1.6\times 10^6 ~\mathrm{M^{-1}.s^{-1}}$ from Ref.~\citenum{zhang2018predicting}, using the exact sequence as in our experiments; $\bar{\sigma} = \sqrt{\sigma \sigma_g}$ where $\sigma = 1/ (3.27~\mathrm{nm})^{2}$ is the particle coating density and $\sigma_g = 1/ (10.8~\mathrm{nm})^{2}$ is the glass substrate coating density; Avogadro's number $\mathcal{N}_A$; Independent of $T$.  \\
$\off$ & $\off = \on \frac{N(T) - N_b(T)}{N_b(T)}$ & $ 18~\mathrm{kHz}$ & $N_b$ average number of bound legs and $N$ total number of legs available for binding in the interaction region; Dependent on $T$. \\
\end{tabular*}
\end{table*} 

\subsubsection{Slippery sliders}

Reciprocally, we predict that systems with weak adhesion (equivalent to short bond lifetimes, \textit{i.e.} large $\off$) may move by sliding. Such systems include 
%Sliding motion is therefore accessible to systems with weak adhesion (equivalent to short bond lifetimes) such as 
proteins translocating through the nuclear pore complex, or white blood cells adhering through L-selectin linkers, which are notably weaker than P-selectin~\cite{alon1997kinetics}. Sliding may also be accessible to systems with short effective relaxation rate, for which the sticky friction mediated by $k/\Gamma$ is low. This corresponds to large particles with long legs, as is the case for Escherichia Coli~\cite{miller2006mechanical} (dark green). DNA-coated colloids with high DNA coverage are prone to slide due to their large number of legs.

\subsection{DNA-coated colloids hop \textit{and} slide, with order of magnitude decrease in their diffusion coefficient}
\label{sec:DNAccs}

We now turn to probe in more detail the predicted modes of motion and strong decrease in diffusion of DNA-coated colloids by comparing our model's predictions with experimental measurements of DNA-coated colloids. DNA-coated colloids provide a well-controlled model system for testing our analytical results, especially their dependence on $N$, since the number of DNA legs involved in the sticking process may be easily tuned by changing the temperature~\cite{fan2021microscopic}. %We will also compare predicted and observed modes of motion.
Our aim here is not to build a detailed model to describe all the possible modes of motion of DNA-coated colloids. Rather, we seek potential key parameters that control the magnitude of the diffusion and the mode of motion. To do so, we test whether the predicted strong decrease is coherent with experimental observations over a range of temperatures and for three different experimental designs. %We show that we find excellent agreement with experimental data of our current model, and discuss potential improvements. 
%

%We predict the diffusion coefficients for existing  DNA-coated colloid and compare them to experimental data.  %For the purpose of this work, note that additional diffusion data was acquired using common fabrication~\cite{fan2021microscopic}  and acquisition techniques~\cite{xu2011subdiffusion,wang2015crystallization}, as existing data was rather scarce~\cite{xu2011subdiffusion,wang2015crystallization} (Supplementary 2). 

%To probe the predicted strong decrease in diffusion with increasing numbers of bound legs, and to further validate our model, we conduct experiments with DNA-coated colloids,
 %Their interaction strength and mobility may be controlled via the DNA sequence and the surface density of legs. 
%The number of legs $N$ in the contact region can exceed 100~\cite{fan2021microscopic,oh2015high}, suggesting they could cause a significant decrease in diffusion. The number of legs may be tuned by changing the temperature of the bath, as temperature controls the relative distance between surfaces~\cite{fan2021microscopic}. 

%the effective global diffusivity $D_{\rm eff}$ using Eq.~\eqref{eq:gammaN}, that for sliding $D_{\slide}$ and that for hopping $D_{\rm hop}$ for existing DNA-coated colloids.

\subsubsection{Model parameters can be directly established from experimental data.}

We predict the diffusion coefficients $D_{\rm eff}$ (and $D_{\rm slide}$ and $D_{\rm hop}$) for three different experimental systems, by determining the parameters involved in Eq.~\eqref{eq:gammaN} from the literature or from independent measurements, with no fitting parameters (apart from calibrating to the melting temperature, as discussed below).
The diffusion coefficients for DNA-coated colloids on flat DNA-coated surfaces have been measured in two different experimental systems reported in the literature~\cite{xu2011subdiffusion,wang2015crystallization}. These studies report only very few data points around the melting temperature where motion is diffusive, since in these experimental systems diffusive motion is only observed in a narrow range of temperatures, so the studies focused mainly on the low temperature regime where motion is subdiffusive. We complemented the scarce existing data by performing our own experiments, using recently-developed fabrication~\cite{fan2021microscopic}  and acquisition techniques~\cite{xu2011subdiffusion,wang2015crystallization}, and we observe diffusive motion over a wider range of temperatures (Supplementary 2). For each of the three experimental datasets, we map reported experimental parameters to the parameters of the model, and detail our process below.

%To evaluate $D_{\rm eff}$ from Eq.~\eqref{eq:gammaN} %(and similarly for $D_{\rm slide}$ and $D_{\rm hop}$), we must evaluate the parameters of the nanocaterpillar model. 
Some parameters are easily estimated using standard results, see Table~\ref{tab:tab1}. The friction coefficient $\Gamma$ is taken as the hindered lateral hydrodynamic friction near a wall~\cite{brenner1961slow}; $\gamma$ and $k$ correspond to hydrodynamic friction and spring resistance of the polymer linker (that links the surface and the complementary DNA strand) and are directly established from polymer dynamics~\cite{rubinstein2003polymer}. The binding rate $\on$ depends on the exact -- known -- DNA sequence used for the complementary stickers and the density of coated DNA strands on surfaces~\cite{zhang2018predicting}. 

Other parameters, such as $N$ and $N_b$ (or equivalently $N$ and the ratio $\on/\off$) require more extensive modeling of the detailed leg-arm interactions to be evaluated. Recently Refs.~\citenum{fan2021microscopic} and \citenum{varilly2012general} have shown how to establish $N$ and $N_b$ with no fitting parameters, taking as input parameters the DNA sequence used, the coating densities, and the properties of the DNA linker (see Fig.~S5), and we employ the method we have developed in Ref.~\citenum{fan2021microscopic}. 

Finally, since measurements include colloid vertical motion beyond the binding range\footnote{The binding range is about $20~\mathrm{nm}$, but this is not optically removable as the vertical resolution is about $200~\mathrm{nm}$}, we further include vertical motion and hence particle buoyancy through a 2$\times$1D model. Such vertical motion is generally slow and only affects the effective probabilities $p_n$, not the friction coefficients $\Gamma_n$. Motion in two lateral dimensions can be straightforwardly extended from our 1D model (see Supplementary 2 for more details). 

All parameters are thus readily expressed from detailed experimental system design. The diffusion coefficient $D_{\rm eff}$ is decreased by orders of magnitude at low temperatures. It progressively increases to its ``bare'' value -- corresponding to non-sticky DNA -- at high temperatures, with a sharp transition. This sharp transition from the bound to unbound state occurs at a melting temperature $T_m$ specific to each experimental design. The predicted $T_m$ is always close to the experimentally measured $T_m$ (less than $1^{\circ}$C difference) with no fitting parameters.

Nonetheless, intrinsic variations remain in experimental parameters. In particular, different \textit{e.g.} humidity conditions can affect the coating process and exact coating density obtained, and hence the experimental $T_m$, over about $2^{\circ}$C. To investigate data over the relevant short temperature range where diffusion can be measured, one option could be to fit \textit{e.g.} the value of the coating density on colloids, to obtain the exact experimental $T_m$ -- effectively fitting the location of the sharp transition. Instead, we choose to align all data (theoretical or experimental) with respect to its own melting point $T_m$ (predicted or measured). This has the advantage of avoiding fitting and allowing us to easily compare similar experimental systems with slightly different $T_m$ (Supplementary 2).

\begin{figure}[h!]
\centering
\includegraphics[width = 0.99\linewidth]{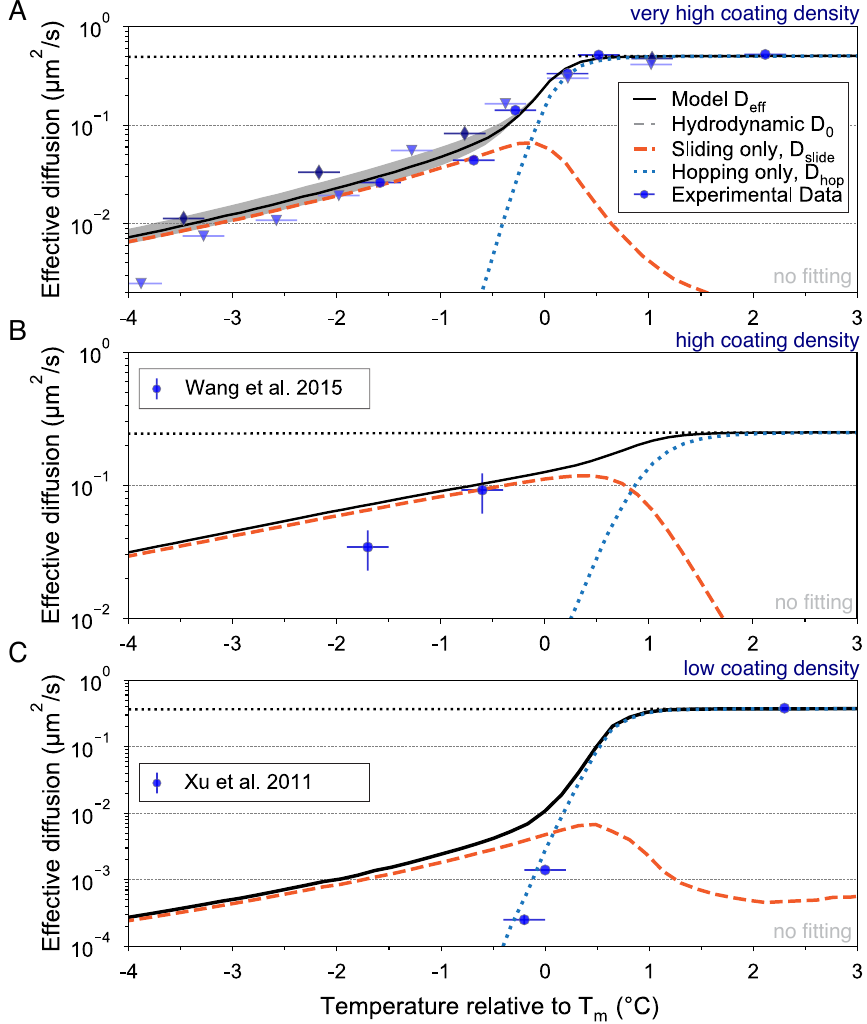}% Here is how to import EPS art
\caption{\label{fig:fig5} \textbf{Diffusion coefficients of DNA-coated colloids.} Comparison between experimentally measured diffusion coefficients of DNA-coated colloids on DNA-coated surfaces and analytical predictions of $D_{\rm eff}$, $D_{\rm slide}$, and $D_{\rm hop}$ (Eqns.~\eqref{eq:gammaN},~\eqref{eq:slide} and~\eqref{eq:hopping}). The DNA-coated colloids have (A) highly dense coatings (1 DNA per $10~\mathrm{nm^2}$, Supplementary~2) (B) dense coatings (1 DNA per $27~\mathrm{nm^2}$) from Ref.~\citenum{wang2015crystallization} and (C) sparse coatings (1 DNA per $144~\mathrm{nm^2}$) from Ref.~\citenum{xu2011subdiffusion}. 
%All calculated diffusion coefficients include corrections to account for the average height of the colloids with respect to the surface, obtained using vertical jumping rates $Q_{\rm on}$ and $Q_{\rm off}$ (see  Supplementary~2). 
In (A) the gray region corresponds to uncertainties on the coating density of the substrate, and the different symbols correspond to repeated experiments repeated. The hydrodynamic diffusion $D_0 = k_BT / 12 \pi \eta R$ corresponds to lateral diffusion near a flat rigid wall, where $R$ is the radius of the colloid and $
\eta$ the solution viscosity. Horizontal error bars correspond to uncertainties on imposed temperature and vertical error bars correspond to uncertainties in determining the diffusion coefficient from data (Supplementary 2). }
\end{figure}

\subsubsection{The coating density controls the mode of motion and the magnitude of the diffusion coefficient decrease.}

The number of legs implied in the sticking process $N$ changes significantly with temperature. At low temperatures $N  \gtrsim 100$; the colloids are strongly bound. With increasing temperatures $N$ decreases until the particles are completely unbound and $N=0$ (see Fig.~S5), with a sharp transition at the melting temperature $T_m$. Importantly, the number of legs is the parameter that changes the most with temperature and controls therefore the magnitude of the long time diffusion $D_{\rm eff}$. 

The three experimental systems differ mainly in the DNA coating density, which implicitly controls the number of legs $N$ involved in the binding process. 
For densely coated colloids (Fig.~\ref{fig:fig5}, A and B), we find excellent agreement between our model calculation for $D_{\rm eff}$ and experimental data, predicting a fast diffusion decrease over 2 orders of magnitude in barely a few temperature degrees. Further, we predict that sliding, or some form of cohesive motion with the surface, is the dominant mode of motion below the melting temperature $T_m$. %Hopping is unfavorable as it requires coordination of too many legs. 
In fact the high number of available legs, $N \simeq 100$, due to high coverage, prevents hopping  below the melting temperature and colloids primarily slide, consistent with the observed cohesive motion~\cite{wang2015crystallization}.
Hopping emerges as a favorable mode above the melting point, where the average number of available and bound legs significantly decreases due to particle lift-off from the surface. This prediction is consistent with our qualitative observations above the melting point: particles perform long moves over short time intervals, accompanied by more frequent and longer excursions far from the surface. The transition between motion modes occurs for about $N = 40$ legs in contact (Fig. S5). 

For DNA-coated colloids with low coverage densities, as in Ref.~\citenum{xu2011subdiffusion} (Fig.~\ref{fig:fig5} C), our model predicts a diffusion coefficient that is far too large. Yet, $D_{\rm hop}$ is in remarkable agreement with experimental data. In fact, $D_{\rm eff}$ contains sliding motion yet the spacing between legs in Ref.~\citenum{xu2011subdiffusion} is too large and geometrically prevents sliding. Hence only hopping, or uncohesive motion with the surface, is possible. In fact, for such systems only hopping is observed, resulting in a much stronger slow down of diffusion with decreasing temperature~\cite{xu2011subdiffusion}. 
The DNA coating density therefore appears to be a significant factor in determining how DNA-coated colloids move, allowing it to vary from sliding to hopping.

%This faster increase could be due to various factors; one possibility is the polymer brush is softer than modelled~\cite{fan2021microscopic}. 

%The diversity of modes of motion for DNA-coated colloids extends beyond hopping and sliding in specific experimental regimes, and could be probed with the analytical tools set forth here. 
%Comparing rolling diffusivity to either sliding or hopping diffusivity would require building a higher-dimensional model, a challenge we discuss further in the conclusion. 
%Nevertheless, because our model agrees with our experimental data within an order of magnitude at all temperatures, we expect rolling mobility to depend in a similar way on the microscopic DNA parameters.  
\subsubsection{Other possible modes of motion.}

There are other ways that DNA-coated colloids could move in specific experimental regimes, that could be probed with the analytical tools set forth here, yet that we have not yet explored. 
At lower temperatures, particles don't diffuse, they rather subdiffuse~\cite{xu2011subdiffusion,wang2015crystallization}, potentially due to inhomogeneities in the coated surfaces~\cite{wang2015crystallization,licata2007colloids,xu2011subdiffusion}. Such spatial dependencies are not accounted for in our model but could be studied through spatially dependent attachment rates $\on(x)$ or leg number $N(x)$. 

Particles may also move by rolling instead of by sliding~\cite{wang2015crystallization}, a motion that could also be investigated with homogenization techniques. Rolling may have a higher mobility at some temperatures~\cite{lee2018modeling,jana2019translational}, since the strands closest to the contact point on the surface do not resist rolling, for geometrical reasons. Yet when a large number of bonds are implied in the binding process, numerous bonds are actually far from the contact point and hence resist rolling. It is possible that rolling is thus favorable only over a small range of temperatures.

%The sharper experimentally measured transition is likely attributable to the same factors as mentioned for the melting transition (softer brush tips than modelled). \mhc{what about other factors, like rolling, which may be easier than sliding?} 
Although our model lacks these more complex ingredients and geometries, it is in surprisingly good agreement with our experimental measurements. This suggests we have identified some critical parameters controlling the observed effective diffusion, precisely the coating density and working temperature as they set the number of legs $N$. Even in a more complex model, containing \emph{e.g.} inhomogeneous coating density, or rotational degrees of freedom, we therefore expect these parameters to play an important role in mobility.

%still quantitatively predicts the hindered diffusion of DNA-coated colloids on surfaces near the melting temperature within experimental accuracy. 
%and it explains why this diffusion changes over orders of magnitude.  

\subsection{Design rules for sliding versus hopping}

Herewith we can draw simple design rules for sliding or hopping. Numerous, long wobbly legs with weak adhesive bonds are well adapted for sliding. Short and stiff legs with strong adhesive bonds facilitate hopping. 
DNA-coated colloids offer various design features to control their mobility: for example, larger particle size, higher DNA coverage, and lower temperature all favor sliding. Further control can be achieved by tuning the microscopic features of the legs, such as their spring constants $k$, for example by choosing the length of the ligand leg~\cite{fan2021microscopic}. However, such control is especially hard to achieve experimentally \textit{without} changing other experimental features at the same time. For example, current coating processes generally result in less dense coatings for longer legs~\cite{fan2021microscopic}.

Overall, these design rules allow one to tune artificial systems to control their mobility. This could have consequences in particular in the field of self-assembly of artificial structures, where facilitated cohesive motion is believed to be essential for long-range alignment~\cite{wang2015crystallization,holmes2016stochastic,jana2019translational}. 

\section{Coarse-graining under different models and assumptions}\label{sec:Models}

In the physical and biological systems we explored, the range of physical parameters was quite broad, suggesting that other scaling ans\"{a}tze might be appropriate to study long term dynamics. We review alternative approximations and modeling assumptions and compare them to the predictions of the model presented in Section~\ref{sec:sec1}. We find that our model is the most general, encapsulating perturbative results obtained with other approximations, and that it is naturally modified to account for additional features (such as arms as well as legs). To make the argument simpler, we mainly focus on a 1-legged caterpillar; the comparisons should be similar for a multi-legged caterpillar. Detailed coarse-graining steps are reported in Supplementary~4. All results are summarized in Table~\ref{tab:tab1} (displayed in the Appendix) and compared  in Fig.~\ref{fig:fig3}.

\subsection{Dynamics with inertia}

One may include particle inertia with a small yet finite mass $m \neq 0$, by starting with the underdamped Langevin equations for the particle (rather than the overdamped as we have done) -- see Ref.~\citenum{lee2018modeling}. 
%In a broader context, particle inertia may be considered with a small yet finite mass $m \neq 0$ -- see Ref.~\citenum{lee2018modeling}. 
To understand the scales associated with mass, one can compare the correlation time of the particle's velocity when spring recoil forces are at play, $\tau_v \simeq \frac{m (L_x/\tau)}{L k}$, to the time scale of observation $\tau$~\cite{lee2018modeling}. Coarse-grained dynamics require $\frac{\tau_v}{\tau} = \frac{mL_x}{L k \tau^2} = O(\epsilon)$, which is apparently coherent with a small mass. %, typical for microscopic systems. 

\begin{figure}[h!]
\includegraphics[width = 0.99\linewidth]{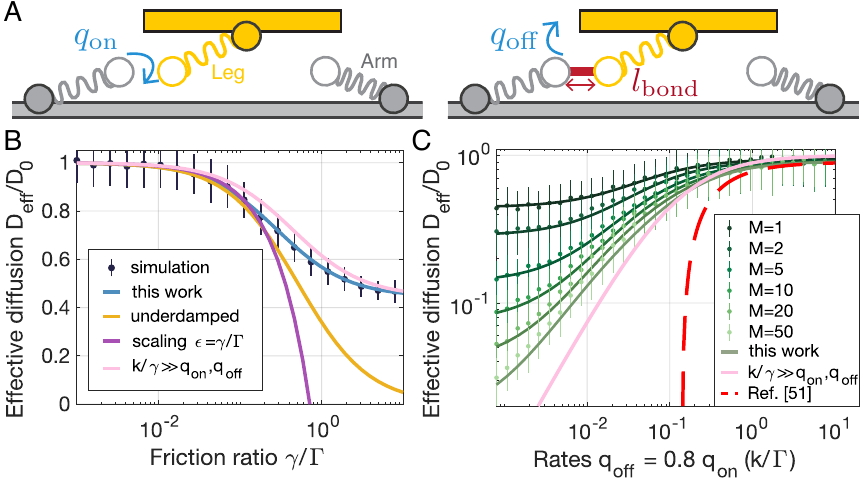}% Here is how to import EPS art
\caption{\label{fig:fig3} \textbf{Comparing with other coarse-grained models and assumptions.} (A) Schematic for arm and leg dynamics considered in this work. (B) Effective diffusion with respect to friction ratio $\gamma/\Gamma$:  calculated with Eq.~\eqref{eq:gammaeff1} (``This work''), Eq.~\eqref{eq:nonOverdamped} (``underdamped''), Eq.~\eqref{eq:scaling} (``scaling $\epsilon = \gamma/\Gamma$'') and Eq.~\eqref{eq:pre-averaged} (``$k/\gamma \gg \on,\off$''). (C) Effective diffusion with respect to binding and unbinding rates (keeping $q_{\rm on}/q_{\rm off}$ constant), for a particle with 1 leg facing $M = 1- 50$ arms: calculated with Eq.~\eqref{eq:Narms} (``This work'') and  Eq.~\eqref{eq:pre-averaged} (``$k/\gamma \gg q_{\rm on}, q_{\rm off}$''), taking $p_0 = 0$ and $p_1 = 1$ to match the limits in $M \rightarrow \infty$.  Ref.~\citenum{fogelson2019transport} corresponds both to $k/\gamma \gg q_{\rm on}, q_{\rm off}$ and $\gamma/\Gamma = \epsilon$ and was plotted for consistency.  For (A) and (B), shared numerical parameters are $q_{\rm on}\Gamma/k = 1.0$, $q_{\rm off}\Gamma/k = 0.8$ and $\gamma/\Gamma = 0.1$. }
\end{figure}

Coarse-graining steps (Supplementary~4.1) lead to an effective friction
\begin{equation}
    \Gamma^{m}_{\rm eff} = p_0 \Gamma_0 + p_1 \Gamma_1.
    \label{eq:nonOverdamped}
\end{equation}
Notice that the effective friction is the \textit{arithmetic} sum of the frictions in each state -- not the \textit{harmonic} sum obtained in Eq.~\eqref{eq:gammaeff1}
~\footnote{Eq.~\eqref{eq:nonOverdamped} corresponds to the result derived in Ref.~\citenum{lee2018modeling}, with in addition projected dynamics for the bound state, and base friction of the particle ($\Gamma \neq 0$)}. Eq.~\eqref{eq:nonOverdamped} is equivalent to Eq.~\eqref{eq:gammaeff1} in the limit where the friction correction is small, $\gamma_{\rm eff} \ll \Gamma$  -- see Fig.~\ref{fig:fig3}-B (yellow). 

However, differences arise beyond this regime. For stiff legs ($\gamma/\Gamma \gg 1$, $k / q_{\rm off} \Gamma \gg 1$) one finds $\Gamma^{m}_{\rm eff} \sim 0$ while $\Gamma_{\rm eff} \sim \Gamma$. This stark difference has an intuitive explanation: the particle may not move when it is attached with the stiff leg, but it can still move when it is unbound, and therefore the effective friction should remain finite. This is true \textit{unless} the particle has significant inertia and therefore does not have the time to accelerate within the unbound periods. In fact, in the non-dimensionalization we implicitely assumed that $m/\Gamma = \epsilon L k \tau^2/ \Gamma L_x = \Gamma/k \epsilon^2$, such that the inertial relaxation time was in fact assumed to be large compared to the time scale of velocity fluctuations. 

This drives the general question of how to account for inertia in such systems, and whether inertia plays a role in the macroscopic diffusion of nanocaterpillars. We will address this question thoroughly in another paper~\cite{marbach2021inertial}, in which we reconcile Eq.~\eqref{eq:nonOverdamped}  and Eq.~\eqref{eq:gammaeff1}.% through a model which includes inertia in all the degrees of freedom. 

\subsection{Choice of time-scale hierarchy}

There are other choices for the ordering of time scales. We review these below: 
 we describe their experimental relevance, then briefly examine the effective friction under these different approximations and compare it to our main result Eq.~\eqref{eq:gammaeff1}.% We find that our choice of time-scale hierarchy is fairly general, as it encompasses other approximations as special limits of ours. 

\subsubsection{Fast leg dynamics compared to particle dynamics}
\label{sec:Fastlegs}

%The choice of the small parameter $\epsilon$ is not unique. 
One common approximation is to assume rapid leg dynamics compared to particle dynamics, with $\epsilon = \gamma/\Gamma$~\cite{fogelson2019transport}. Such an approximation is consistent with numerous experiments, as legs are typically short, hence fast because of Stokes relation, compared to the large particles investigated (such as white blood cells~\cite{korn2008dynamic} or DNA-coated colloids~\cite{oh2015high}).  %

%To perform coarse-graining, nondimensionalization is slightly different.

With this assumption one typically relaxes the restriction on lengthscales, as $L \sim L_x$. The observation time-scale is $\tau = L^2/D_0 = \Gamma/k$ and binding and unbinding are taken to be fast compared to this time scale, $q_{\rm on} \sim q_{\rm off} \sim 1/\tau \epsilon$. One obtains (Supplementary~4.2.1) 
\begin{equation}
    \frac{1}{\Gamma^{\epsilon=\gamma/\Gamma}_{\rm eff}} = \frac{p_0}{\Gamma} + \frac{p_1}{\Gamma} \left( 1 - \frac{\gamma_{\rm eff}}{\Gamma} \right).
    \label{eq:scaling}
\end{equation}
Eq.~\eqref{eq:scaling} results in a small correction to the effective friction, of order $\epsilon$.  It is equivalent to Eq.~\eqref{eq:gammaeff1} in the limit where $\gamma_{\rm eff} \ll \Gamma$ is small. The assumption $\epsilon = \gamma/\Gamma$ appears thus quite restrictive as it implicitly also requires to observe the system at long time scales compared to the other time scales in the system. Furthermore, contrary to Eq.~\eqref{eq:gammaeff1} where the small parameter $\epsilon$ disappears, here $1/\Gamma^{\epsilon=\gamma/\Gamma}_{\rm eff}$ is a first order expansion in $\epsilon \sim \gamma_{\rm eff}/\Gamma$. We present Eq.~\eqref{eq:scaling} against Eq.~\eqref{eq:gammaeff1} in Fig.~\ref{fig:fig3}-B (purple vs black) and find that Eq.~\eqref{eq:scaling} is indeed only valid for small values of $\gamma/\Gamma$. Our choice of scaling $\epsilon = L/L_x$ can thus account for a broad range of bare friction values. Additionally, such an approach can only account for small perturbations to the background mobility, while we find perturbations over several orders of magnitude. 

\subsubsection{Fast leg dynamics compared to binding dynamics}

Another approximation assumes fast leg relaxation dynamics compared to binding dynamics, $k/\gamma \gg q_{\rm on}, \off$ (and both are fast compared to particle dynamics). In this case leg lengths are sampled from their equilibrium distribution when they bind, corresponding to a ``pre-averaging" approximation. Leg lengths are not tracked when they are unbound, allowing to speed up simulations~\cite{fogelson2018enhanced,fogelson2019transport,jana2019translational,noe2018}. This limit is relevant to describe stiff legs, \textit{e.g.} rigid polymers such as double stranded DNA -- see Table S1. 

Coarse-graining gives (Supplementary~4.2.2)
\begin{equation}
    \frac{1}{\Gamma_{\rm eff}^{\rm k/\gamma \gg q}} = \frac{p_0}{\Gamma} + \frac{p_1}{\Gamma + \gamma +  \frac{k}{q_{\rm off}}}.
    \label{eq:pre-averaged}
\end{equation}
%Compared to Eq.~\eqref{eq:gammaeff1}, pre-averaging removes the term involving $\tau_{\rm u}^{\rm relax}$ in $\gamma_{\rm eff}$. This confirms that $\tau_{\rm u}^{\rm relax}$ originates from unbound relaxation dynamics. 
%How can we relate pre-averaging dynamics to our theory for a leg binding to uniformly sticky surface described in Sec.~\ref{sec:sec1}?
The pre-averaged result Eq.~\eqref{eq:pre-averaged}  is comparable to 
%the coarse-grained result for a leg binding to uniformly sticky surface of Sec.~\ref{sec:sec1}, 
Eq.~\eqref{eq:gammaeff1}, yet misses the relaxation term involving $\tau_{\rm u}^{\rm relax}$ in $\gamma_{\rm eff}$. This confirms that $\tau_{\rm u}^{\rm relax}$ originates from unbound relaxation dynamics. This difference results in some differences in $D_{\rm eff}$, depending on the microscopic parameters (Fig.~\ref{fig:fig3}-B). Additionally, the pre-averaged limit may be viewed as the limit regime for a nanocaterpillar with a large number of legs, say $N \gg 1$, where on average 1 or 0 leg is bound to the surface, $N_b \lesssim 1$. This typically requires $q_{\rm on} \ll q_{\rm off} \ll k/\gamma$, and indeed Eq.~\eqref{eq:gammaN} converges to the pre-averaged result in that limit (Supplementary Fig~S4). 

The validity of pre-averaging is limited to the domain $q_{\rm on}, q_{\rm off} \ll k/\gamma$. In systems such as DNA-coated colloids, binding rates $\on$ and $\off$ may be manipulated over orders of magnitude~\cite{xu2017real}, by choosing the DNA sequence or by adjusting temperature, potentially accessing $q_{\rm on} \gg q_{\rm off} \gg k/\gamma$ at low temperatures. In this regime, Eq.~\eqref{eq:gammaeff1} predicts that the nanocaterpillar is frozen in the bound state, while pre-averaged dynamics still predict a non zero mobility.  In these situations pre-averaged dynamics are therefore not suitable. We show later however that introducing numerous arms -- more generally a lot of binding partners -- can extend the validity range of pre-averaging.  
%\mhc{I still don't understand -- see comments} \sophie{Ok now hopefully it's clear -- and also coherent with what comes after (end of C3) }

\subsubsection{Fast binding dynamics compared to leg dynamics}

Finally, one can consider fast binding dynamics compared to leg dynamics, $q_{\rm on}, q_{\rm off} \gg k/\gamma$. Although this limit is not often considered in simulations, it is relevant for dense arrangements of receptor sites~\cite{oh2015high}. In fact as the binding rate $q_{\rm on}$ scales linearly with the concentration of receptors, it can increase by orders of magnitude for a leg potentially in contact with a dense array of arms -- see Table S1. 

Coarse-graining yields (Supplementary~4.2.3)
\begin{equation}
    \frac{1}{\Gamma^{q\, \rm fast}_{\rm eff}} = \frac{p_0}{\Gamma} + \frac{p_1}{\Gamma + \gamma + k\left( \frac{\gamma}{k} \frac{\on}{\off} \right)}
\end{equation}
which is exactly what is expected in the limit $q_{\rm on}, q_{\rm off} \gg k/\gamma$ in Eq.~\eqref{eq:gammaeff1}. Again, this highlights the physical mechanisms yielding the different contributions in $\gamma_{\rm eff}$. Here the average bound time of the leg is small, $\tau_{\rm b} \ll \gamma/k$, and therefore does not contribute to $\gamma_{\rm eff}$.

\subsection{Arms and/or legs}
\label{sec:handsFeet}

The diversity of nanocaterpillars resides also in their geometry: some particles have legs that attach to a surface~\cite{chang2000state}, some have no legs (or infinitesimally small legs), with binding sites directly on the particle that attach to outstretched receptors on the surface that we refer to as arms~\cite{fogelson2018enhanced,fogelson2019transport} (1 arm case in Table~\ref{tab:tab1}) and some have both outstretched legs connecting to outstretched arms~\cite{jana2019translational} (arms and legs in Table~\ref{tab:tab1}). 

\subsubsection{Arms or legs}

A particle with a leg or a bare particle attaching to an arm (1-legged and 1-armed respectively, see Table~\ref{tab:tab1}) have nearly equivalent effective dynamics. The only difference resides in the interpretation of $\Gamma$ in the unbound leg dynamics Eq.~\eqref{eq:dxUnbound} -- see Supplementary~4.3.1. For the 1-legged case, if the leg's center of mass corresponds to the point grafted to the particle, the unbound friction coefficient is simply increased by the leg as $\Gamma \rightarrow \Gamma + \gamma$, where $\Gamma$ is the bare particle friction coefficient and $\gamma$ the leg's. If the leg's center of mass is offset from the grafting point on the surface, minor modifications have to be made to Eq.~\eqref{eq:dxUnbound} yet lead to very similar dynamics overall. For the 1-armed case, we simply have the unbound friction coefficient $\Gamma$ to be the bare friction coefficient of the particle. This justifies our approach in Sec.~I, where we ignore the details of the leg or arm location and simply treat them as mathematically equivalent. 

\subsubsection{Arm and leg}

A 1-legged particle attaching to 1 arm has slightly more interesting dynamics. To investigate this case, we simplify the problem and consider that the leg can bind to the arm regardless of their relative location, with a rigid rod of length $l_{\rm bond}$ that bridges the gap between the sticky points (see Fig.~\ref{fig:fig3}-A). In the bound state the constraint is thus $x + l_{\rm leg} - l_{\rm arm} =l_{\rm bond}$. The relative distance $l_{\rm bond}$ is unimportant and can be assumed to be zero, and therefore this model effectively creates an arm with the correct length each time the leg binds. 

Although the model is simplistic, it is realistic in the presence of a dense periodic array of arms %(where the leg can only bind to 1 arm at a time), 
and allows us to compare the mechanical properties of this geometry compared to a single leg or arm. 
We find using similar coarse-graining techniques (Supplementary~4.3.2) 
\begin{equation}
    \frac{1}{\Gamma_{\rm eff}^{leg+arm}} = \frac{p_0}{\Gamma} + \frac{p_1}{\Gamma + \gamma_{\rm eff,1}(1,1)} \, \text{where} \,\, \gamma_{\rm eff,1}(1,1) = \frac{\gamma_{\rm eff}}{2}.
     \label{eq:gammaeffarmleg}
\end{equation}
The added friction in the bound state is only half that with a single leg or a single arm: friction is distributed harmonically, like the effective spring constant of two springs in series~\footnote{Note however that attaching springs with different spring constants would not lead to a similar harmonic sum of effective frictions, as the effective friction contains more contributions than those originating from the spring recoil force (analytical results not shown here).}. Slightly improved mobility is therefore achieved with both an arm and a leg, while the qualitative behavior of the original model is preserved.

\subsubsection{Leg facing numerous arms}

We now consider a leg that can bind to \textit{multiple} arms at the same time. As in the previous section, the $M$ arms do not have particular locations but rather appear with the correct lengths when needed. In that case, the binding rate depends on the number of bound legs. For a given leg, the effective binding rate is $(M-n)q_{\rm on}$, where $n$ is the current number of bound legs, such that $M-n$ corresponds to the number of available binding sites. The effective unbinding rate of each leg remains $q_{\rm off}$.  
Following the formalism of arm and leg dynamics detailed above (Supplementary~4.3.3) one finds that with $M$ arms, 
\begin{equation}
        \frac{1}{\Gamma_{\rm eff}^{leg+M\,arms}} = \frac{p_{M,0}}{\Gamma} + \frac{p_{M,1}}{\Gamma + \gamma_{\rm eff, 1}(M,1)}
        \label{eq:Narms}
\end{equation}
where $p_{M,0} = \off/(\off + M \on)$ and $p_{M,1} = 1 - p_{M,0}$ are the probabilities to have 0 or 1 bond. The added friction $\gamma_{\rm eff, 1}$ is a harmonic average when $M$ is large
\begin{equation}
        \frac{1}{\gamma_{\rm eff, 1}(M,1)} \stackrel[M \gg 1]{}{\simeq}  \frac{1}{ \gamma_{\rm eff, M, 1}} + \frac{1}{ \gamma_{\rm eff, 1, 1}} , \,\,
        \label{eq:inverseSums}
\end{equation}
with $\gamma_{\rm eff, M, 1} =  k \left( \frac{1}{\off} + \frac{\gamma}{k} \frac{(M-1)\on + \off}{\off} \right)$ the effective friction due to the leg $\gamma_{\rm eff,1,1} =  k \left(\frac{1}{\off} + \frac{\gamma}{k} \right)$ due to arms. We see that the factors implying the unbound relaxation time $\tau_{\rm u}^{\rm relax}$ are modified in each case. We give the following interpretation: the average unbound time for the leg is $\tau_{\rm u} = 1/(M-1)\on $, due to $M-1$ other available arms to bind to. For the arms, $\tau_{\rm u} = \infty$ as there are no other legs to bind to once the only leg is bound. 
The harmonic average in Eq.~\eqref{eq:inverseSums} highlights again that the leg-arm configuration is mathematically similar to the effective force of springs in series. % Hence, $\gamma_{\rm eff}$ is divided by 2 in the 1 arm-1 leg scenario compared to only 1 leg. 
%k \left( \frac{1}{\off} + \frac{\gamma \on}{k \off} \frac{N + 1}{2} \frac{1}{1 + \frac{N-1}{2} \frac{\on}{k/\gamma + \off}} \right)

In the limit of a large number of arms $M$, the leg is always bound to the surface ($p_1 = 1$) and the correction to the bound state friction converges to 
\begin{equation}
    \gamma_{\rm eff, 1}(M,1) \xrightarrow[M \rightarrow \infty]{}  \gamma_{\rm eff, 1}(1,1)  = \gamma + \frac{k}{\off},
    \label{eq:limitarm}
\end{equation}
which is the correction to the effective friction for the pre-averaged result, Eq~\eqref{eq:pre-averaged}.

This limit is surprising. Sec~\ref{sec:sec1}, Eq.~\eqref{eq:gammaeff1} showed that for a leg binding to a uniformly sticky surface, in the limit where the leg is always bound ($p_1 = 1$), the nanocaterpillar is frozen and $D_{\rm eff} = 0$. When the leg is bound to a great many arms this is no longer the case: we recover the diffusion coefficient associated with pre-averaging. We interpret this discrepancy as follows. 
With many arms binding to a leg, the particle may still move, even in a parameter regime where the leg is always bound. In fact, the leg rapidly swaps between different arms, which have different random lengths and hence apply different random forces, causing the particle's position to fluctuate. 
%While the leg fluctuations may not relax and move the particle, the leg still swaps between different arms. The particle is then carried over because arm lengths fluctuate and relax. 
Indeed, in Eq.~\eqref{eq:limitarm} it is apparent that the remaining friction is due to arms and not to the leg. Swapping the particle upside down, this is equivalent to a particle with a large number $M$ of legs binding to a uniformly sticky surface, but where on average only 0 or 1 leg is bound to the surface at a time. Therefore, this limit is equivalent to the pre-averaged result: each time a new arm is bound it is sampled from its equilibrium distribution -- as so many arms are within reach. 

%In fact, if a leg can bind to numerous arms, this is similar to sampling the arm length from its equilibrium unbound distribution. More interestingly, 
%\mhc{there is still something inconcisstent with beofre -- we should discusss}
%\mhc{this result seems inconsistent with what you said before, that preaveraging is not relevant when there are lots of arms to bind to.} \sophie{I haven't changed the wording here but now there should be no contradiction.} \mhc{I'm still not sure -- in one case, you say having lots of places to bind makes preaveraging not correct, now here you say the opposite} \sophie{I removed the first statement that was saying that qon was large because of that. I believe qon can be large in many systems regardless of having lots of places to bind to, for example at low temperatures it's just always large. -- we should discuss if not clear.}
%Notice how it's a bit more subtle. I indeed say the ``correction" converges, because the effective total friction is not the same, since for the preaveraging there's still the $p_0$, $p_1$, while here $p_0 = 0$ and $p_1 = 1$. So the results are equivalent only in the limit $\on \rightarrow \off$ but still both of them smaller than $k/\gamma$ which is not trivial 

Simulations with M arms are presented in Fig.~\ref{fig:fig3}-C with analytical solutions Eq.~\eqref{eq:Narms} (green colors). They indeed converge to the pre-averaged result (pink). For consistency, we also record the result of Ref.~\citenum{fogelson2019transport} (Eq. (2.48)) that corresponds to pre-averaging \textit{and} assumes $\epsilon = \gamma/\Gamma$. It is plotted in Fig.~\ref{fig:fig3}-C (red) and agrees with our result only over a limited range of parameters, corresponding to the validity range of Ref.~\citenum{fogelson2019transport}. 

\subsubsection{Numerous legs facing numerous arms}

$N$ legs binding to $M$ arms induce a long time effective friction that encapsulates the previous result for $M$ arms and that for $N$ legs in Sec.~\ref{sec:Nlegs} (Supplementary~4.3.4). Eq.~\eqref{eq:gammaN} still holds with adapted bond probabilities $p_n$, and $\gamma_{\rm eff}$ in Eq.~\eqref{eq:gammaNb} is the harmonic average between arm and leg contributions, $(\gamma_{\rm eff,n} (M,N))^{-1} = \gamma_{\rm eff, M, n}^{-1} + \gamma_{\rm eff, N, n}^{-1}$.

\vspace{4mm}

Overall, spanning different limits shows that our methodology to investigate long time dynamics is robust, as it accounts for a broad range of physical parameters and a variety of geometries. It also justifies the use of ``pre-averaging'' approximations (sampling leg lengths from equilibrium distributions upon binding) to accelerate simulations in specific situations. It also highlights that taking limits of various parameters is subtle, and care must be taken when doing so as the limits do not commute in general. %It also shows that limits can be subtle, and care must be taken when interchanging the limits 

\section*{Conclusion}

When a particle is coated with ligands that bind and unbind stochastically to receptors on a surface, the ligands impart a random force to the particle each time they bind, causing the particle to undergo a random walk on long timescales.
%Ligands on particles bind and unbind to surface receptors, imparting a random force to the particle each time they bind, and thereby causing the particle to undergo a random walk on long timescales.
%Binding and unbinding ligands to surface receptors transmits random forces to the ligand-bearing particle thereby causing the particle to undergo a random walk on long timescales.
We constructed a model for the coupled dynamics of such a nanocaterpillar and its leg-like ligands, and derived an analytical expression for the 
 nanocaterpillar's long-term effective diffusion coefficient as a function of the microscopic leg parameters. Our simulations showed this expression is valid over a broad range of parameters. %, and our experiments using DNA-coated colloids showed this expression accurately predicts the observed diffusion in a complex physical system. In particular, 
 Our expression predicts a dramatic decrease in the diffusion coefficient, by several orders of magnitude, as temperature decreases by a few degrees, a prediction that is borne out in our experimental measurements. 
 
 Our model allows us to distinguish between two modes of motion, sliding and hopping, and to identify parameters that govern which mode of motion is preferred, across a wide range of biophysical systems. Typically, systems with a large number of legs will slide, since the mean-squared displacement due to hopping decreases exponentially with the number of bound legs. % compared to bulk mobility, with measurements and analytical predictions $D_{\rm hop}/D_{0} \sim 1/100$. 
%In contrast, sliding can be much more efficient, with $D_{\rm slide}/D_{0} \sim 1/10$. 
%A similar conclusion was obtained in a different context, namely sticky reptation of entangled polymers~\cite{leibler1991dynamics}. 
Hopping is favored for systems with short, stiff legs, and/or strong bonds. 
%Our analytical framework also allows us to infer how microscopic features of the legs and particle affect overall diffusion. For example, systems with short stiff (resp. long wobbly) legs and strong (resp. weak) bonds will favor hopping (resp. sliding). 
%Overall, our calculations demonstrate how nanocaterpillar motion may be harvested for specific transport beyond standard Einstein diffusion~\cite{bian2016111}.
Regardless of the mode of motion, the fast binding and relaxation dynamics at the microscale result in an overall slow diffusion of the nanocaterpillar, sometimes many times smaller than the background hydrodynamic diffusion. %, sometimes orders of magnitude smaller than the background hydrodynamic diffusion. 

We derived the effective diffusivity for a range of other models and scaling assumptions, which allowed us to tease out \textit{e.g.} the effect of having arms (flexible receptors) as well as legs, having significantly more arms than legs or \textit{vice versa}, having significant inertia, \textit{etc}. 
In particular, we explored the validity range of specific approximations used to accelerate simulations, such as that upon binding, leg lengths are sampled from their equilibrium distributions~\cite{jana2019translational,fogelson2018enhanced,fogelson2019transport}. We showed this approximation is valid for fast leg dynamics $\gamma/k \ll q_{\rm on},q_{\rm off}$ in 1D, or when binding to a great number of binding partners, such as many arms, $M\gg 1$, yet its validity should be reassessed in more complex geometries.

There are numerous ways to build upon our model to address additional complexities within the same coarse-graining framework. 
An important step would be to incorporate particle rotational degrees of freedom, and to ask how rolling compares to hopping and sliding. Rolling has been predicted to lead to a low effective friction in systems with stiff legs, because it doesn't require stretching legs at the contact point~\cite{lee2018modeling,jana2019translational}. While rolling has been modeled in special situations, none of these account for the full stochastic nature of the motion, nor do they systematically derive a coarse-grained equation from microscopic parameters~\cite{ziebert2021influenza}. A systematic derivation of a rolling diffusion coefficient would involve a few additional mathematical subtleties beyond those that occur here, such as including binding rates with spatial dependencies to account for the variable separation between surfaces~\cite{korn2007mean,schwarz2004selectin}, but we may nevertheless expect similar parameters  (such as spring relaxation times and unbinding rates) to discriminate between rolling and other modes of motion.
%Such a systematic derivation would involve handling a few additional mathematical subtleties beyond those that occur here, such as coupling rotational and translational degrees of freedom, and including binding rates with spatial dependencies, to account for the variable separation between surfaces~\cite{korn2007mean,schwarz2004selectin}. 
%Nevertheless, we may expect similar parameters  (such as spring relaxation times and unbinding rates) to discriminate between rolling and other modes of motion.

%There are other ways that nanocaterpillars may move macroscopically, beyond hopping and sliding. 
%For example, they may roll, a type of motion that has been predicted to lead to a low effective friction in systems with stiff legs, because it doesn't require stretching legs at the contact point~\cite{lee2018modeling,jana2019translational}. While rolling has been modelled in special situations, none of these account for the full stochastic nature of the motion, nor do they systematically derive a coarse-grained equation from microscopic parameters~\cite{ziebert2021influenza}.
%Such a systematic derivation would involve handling a few additional mathematical subtleties beyond those that occur here, such as coupling rotational and translational degrees of freedom, and including binding rates with spatial dependencies, to account for the variable separation between surfaces~\cite{korn2007mean,schwarz2004selectin}. 
%Nevertheless, we may expect similar parameters  (such as spring relaxation times and unbinding rates) to discriminate between rolling and other modes of motion.

Going further, other effects that could be studied include the details of binding kinetics, \textit{e.g.} non-exponential kinetics in DNA hybridization~\cite{wallace2001non,rogers2013kinetics,wu2013kinetics}, which could also impact the long time response~\cite{licata2007colloids};  mobility of the leg roots, corresponding to fluidity of the bilayer~\cite{sarpangala2021cargo,merminod2021avidity}; and out-of-equilibrium effects, such as white blood cells streaming in blood flow~\cite{alon2002rolling,schwarz2004selectin}, active stepping of molecular motors~\cite{miles2018analysis,mckinley2012asymptotic,peskin2000role}, or proteins that actively cleave bonds on influenza A~\cite{vahey2019influenza,sakai2017influenza}. 
Accounting for such effects would require adapting bond dynamics to include increased bond rigidity or bond lifetime in flow~\cite{bell1978models,doyle2000dynamics,chen2001selectin,e2011novel,hammer2014adhesive,rakshit2014biomechanics}; binding kinetics coupled to the number of bonds~\cite{klumpp2005cooperative,miles2018analysis}; or memory effects associated with dead zones created by cleaved bonds~\cite{yehl2016high,vahey2019influenza,korosec2021substrate}. Importantly, such improvements require carefully adapting binding rates to preserve detailed balance and physical constraints~\cite{korn2007mean,holmes2016stochastic}.

Furthermore, detailed hydrodynamic effects may be important to describe certain kinds of nanocaterpillar dynamics. We have accounted for hydrodynamics via the bare friction coefficients ($\Gamma, \gamma$), but these coefficients themselves are coarse-grained, and in reality depend on the distance of a nanocaterpillar to a surface~\cite{brenner1961slow} and are coupled to the details of the polymer leg mesh. 
%Further hydrodynamic effects could be at play in nanocaterpillar dynamics. 
%Bare friction coefficients ($\Gamma, \gamma$) account for a coarse-grained description of the fluid. % surrounding the caterpillar and its legs. %Yet the legs, particle, and wall all interact hydrodynamically in a complex manner. 
%A more detailed model could incorporate how fluid flows within and around the polymer leg mesh. 
Indeed, elasticity of the polymer mesh could modify the particle's mobility near the interface, as was predicted for elastic membranes~\cite{daddi2016long,bertin2021soft}. %Furthermore, hydrodynamic friction between a particle and a surface depends strongly on the distance between them%, 
%an effect known as lubrication or hydrodynamic hindrance, and that is well established for flat rigid walls~\cite{brenner1961slow}. 
%The vertical diffusion coefficient of a nanocaterpillar should therefore also include hydrodynamic corrections%, taking into account the polymer mesh. 
%The vertical motion of a nanocaterpillar near a surface is also modified by stochastic binding and unbinding of the legs~\cite{Mani:2012did}, and it is not even clear that a diffusion coefficient would be sufficient to accurately capture this vertical motion. 
A more detailed description of the hydrodynamic flow near a nanocaterpillar
could help shed light on other systems where mobility through fluid is mediated by slender legs, such as for the Vampire amoeba~\cite{hess2012shedding}. 

%Building upon the model to consider and compare these other modes of motion is an important step for future work. 

%The diversity of modes of motion extends beyond hopping and sliding. Accounting for other modes, for example for rolling, may also be done within the framework presented, provided suitable extensions. Specifically, binding rates including spatial dependencies~\cite{korn2007mean,schwarz2004selectin} could probe rolling motion. Rolling has been predicted to lead to a low effective friction, at least in systems with stiff legs, because it doesn't require stretching legs at the point on a particle which is closest to a sticky surface~\cite{lee2018modeling,jana2019translational}. While rolling has been modelled in special situations, none of these account for the full stochastic nature of the motion, nor do they systematically derive a coarse-grained equation involving rolling, from microscopic binding and unbinding dynamics~\cite{ziebert2021influenza}. Such a systematic derivation involve extra mathematical subtleties beyond what occur here, and will be the purpose of future work; nevertheless we may expect similar parameters  (such as spring relaxation times and unbinding rates) to discriminate between rolling and other modes of motion.

Beyond its biophysical details, nanocaterpillar motion resonates with other fields where mobility is determined through adhesive contacts. %Questions raised in other fields therefore translate to nanocaterpillar motion. 
For example, solid state sliding friction is created by bonds breaking between atoms. Close neighbor interactions between bonds, originating from mechanical interactions, can result in dramatic avalanches of bond breaking that change the sliding motion~\cite{de2019collective,ji2021geometry}. Similar correlations between nearby bonds could be at play in some nanocaterpillars. For example, in white blood cells, membrane tension mediates bond-bond interactions~\cite{klumpp2005cooperative,fenz2017membrane}. It is therefore interesting to speculate whether avalanches of bond unbinding could also occur for nanocaterpillar systems.
%Reciprocally, similar coarse-grained approaches could be applied to related fields and shed light on underlying mechanisms, for example understanding how rapid binding of ions to specific proteins controls their selective and rapid translocation through nanopores~\cite{dutta2019dynamic,knowles2021current}.% This suggests that s
Overall, the mathematical framework of coarse-graining is well suited to explore how microscopic features determine  macroscopic modes of motion for nanocaterpillars and could facilitate predictive capacity for materials design and biophysical systems.

% \begin{table*}
% \small
%   \caption{\ An example of a caption to accompany a table \textendash\ table captions do not end in a full point}
%   \label{tbl:example2}
%   \begin{tabular*}{\textwidth}{@{\extracolsep{\fill}}lllllll}
%     \hline
%     Header one & Header two & Header three & Header four & Header five & Header six  & Header seven\\
%     \hline
%     1 & 2 & 3 & 4 & 5 & 6  & 7\\
%     8 & 9 & 10 & 11 & 12 & 13 & 14 \\
%     15 & 16 & 17 & 18 & 19 & 20 & 21\\
%     \hline
%   \end{tabular*}
% \end{table*}

\section*{Author Contributions}
S.M. derived the mathematical framework, and solved it in all cases; acquired biological data for the Ashby chart; designed and analyzed the numerical simulations; found predictions for the diffusion of DNA-coated colloids. J.A.Z. synthesized DNA-coated colloids, conducted the experiments, and analyzed the experimental data to find diffusion coefficients. 
M.H.C. supervised the project. S.M. and M.H.C. wrote the paper. 

\section*{Conflicts of interest}
There are no conflicts to declare.

\section*{Acknowledgements}
The authors are grateful for fruitful discussions with Fan Cui, Aleksandar Donev, Christopher E. Miles, and David J. Pine. 
S.M. acknowledges funding from the MolecularControl project, European Union’s Horizon 2020 research and innovation programme under the Marie Skłodowska-Curie grant award number 839225. All authors were supported in part by the MRSEC Program of the National Science Foundation under Award Number DMR-1420073. M.H.-C. was partially supported by the US Department of Energy under Award No. DE-SC0012296, and acknowledges support from the Alfred P. Sloan Foundation.

\begin{table*}
\small
\caption{Summary of different models and their effective long time friction. The 1-leg case corresponds to a system where the leg's center of mass is fixed on the particle. Apart from the 1-leg case, we ignore differences between $\Gamma$ and $\tilde{\Gamma}$ to simplify notations. }
\label{tab:tab1}
\begin{tabular*}{\textwidth}{@{\extracolsep{\fill}}lll}
\hline
\textrm{Model}&
\textrm{Sketch}&
\textrm{Result}\\
\hline
\multicolumn{3}{l}{\textit{Main geometries}} \\
\hline  \\  [-0.3cm]
1-arm  & \raisebox{-0.3cm}{\includegraphics[width=2cm]{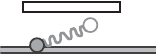}} & $\displaystyle \frac{1}{\Gamma_{\rm eff}}  = \frac{p_0}{\Gamma_0} + \frac{p_1}{\Gamma_1}$, $\Gamma_0 = \Gamma$, $\Gamma_1 = \Gamma + \gamma_{\rm eff}$, $\gamma_{\rm eff} =   k\left( \frac{1}{q_{\rm off}}  + \frac{\gamma}{k} \frac{\on}{\off} \right)$ \\ [-0.2cm] \\
%\colrule
1-leg  & \raisebox{-0.3cm}{\includegraphics[width=2cm]{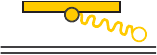}} & $\displaystyle \frac{1}{\Gamma_{\rm eff}}  = \frac{p_0}{\Gamma_0} + \frac{p_1}{\Gamma_1}$, $\Gamma_0 = \tilde{\Gamma}$, $\Gamma_1 = \tilde{\Gamma} + \gamma_{\rm eff}$, $\tilde{\Gamma} = \Gamma + \gamma$ \\[-0.2cm] \\
%\colrule
N-legs  & \raisebox{-0.3cm}{\includegraphics[width=2cm]{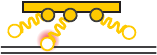}} & $\displaystyle \frac{1}{\Gamma_{\rm eff}}  = \sum_{n=0}^N \frac{p_n}{\Gamma_n}$, $p_n  = \binom{N}{n} \frac{\off^{N-n} \on^n}{(\off + \on)^N} $, $\Gamma_n \stackrel[N \gg 1]{}{\simeq} \Gamma + n \gamma_{\rm eff} $ \\ [-0.2cm] \\
\hline
\multicolumn{3}{l}{\textit{Inertial dynamics}} \\
\hline \\  [-0.3cm]
1-leg, inertia  & \raisebox{-0.3cm}{\includegraphics[width=2cm]{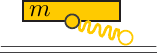}} & $\displaystyle \Gamma_{\rm eff}  = p_0 \Gamma_0 + p_1\Gamma_1$ , $\Gamma_0 = \Gamma$, $\Gamma_1 = \Gamma + \gamma_{\rm eff}$ \\ 
\hline
\multicolumn{3}{l}{\textit{Limit regimes}} \\
\hline \\  [-0.3cm]
Small legs & \raisebox{-0.3cm}{\includegraphics[width=2cm]{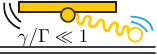}} & $\displaystyle \frac{1}{\Gamma_{\rm eff}}  = \frac{p_0}{\Gamma} + \frac{p_1}{\Gamma} \left(1 -  \frac{\gamma_{\rm eff}}{\Gamma}\right)$ \\[-0.2cm] \\ 
Fast legs &  \raisebox{-0.3cm}{\includegraphics[width=2cm]{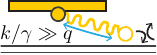}} & $\displaystyle \frac{1}{\Gamma_{\rm eff}}  = \frac{p_0}{\Gamma_0} + \frac{p_1}{\Gamma_1}$, $\Gamma_0 = \Gamma$, $\Gamma_1 =  \gamma +  \frac{k}{q_{\rm off}} $ \\[-0.2cm] \\ 
Fast binding &  \raisebox{-0.3cm}{\includegraphics[width=2cm]{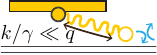}} & $\displaystyle \frac{1}{\Gamma_{\rm eff}}  = \frac{p_0}{\Gamma_0} + \frac{p_1}{\Gamma_1}$, $\Gamma_0 = \Gamma$, $\Gamma_1 =  \gamma +  k \left( \frac{\gamma}{k} \frac{\on}{\off} \right) $  \\[-0.2cm] \\ 
\hline
\multicolumn{3}{l}{\textit{Extended geometries}}   \\
\hline \\  [-0.3cm]
1-arm, 1-leg  &  \raisebox{-0.3cm}{\includegraphics[width=2cm]{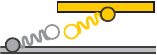}} &  $\displaystyle \frac{1}{\Gamma_{\rm eff}}  = \frac{p_0}{\Gamma_0} + \frac{p_1}{\Gamma_1}$, $\Gamma_0 = \Gamma$, $\Gamma_1 = \Gamma + \frac{1}{2}\gamma_{\rm eff}$ \\[-0.2cm] \\ 
%M-arms, 1-leg  &  \raisebox{-0.3cm}{\includegraphics[width=2cm]{Sketch/Sarmsleg.pdf}} &   \multirow{3}{*}{\shortstack[l]{$N$ legs, $M$ arms, }}    \\[-0.2cm] \\ 
M-arms, N-legs  &  \raisebox{-0.3cm}{\includegraphics[width=2cm]{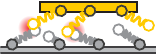}} & $\begin{cases}   & \displaystyle \frac{1}{\Gamma_{\rm eff}}  = \displaystyle \sum_{n=0}^N \frac{p_n}{\Gamma_n}, \Gamma_n = \Gamma + n \gamma_{\rm eff, n}(M,N) , \\ & \,\, (\gamma_{\rm eff, n}(M,N))^{-1} \simeq  (\gamma_{\rm eff, M,n})^{-1} + ( \gamma_{\rm eff, N,n})^{-1}, \gamma_{\rm eff, P,n}  = \gamma + k \left( \frac{1}{\off} + \frac{\gamma}{k} \frac{ (P-n)\on}{\off} \right) \end{cases}$ \\
\end{tabular*}
\end{table*}

\section*{Appendix}

\subsection*{Appendix A: Projection of the dynamics in the bound state}
%\label{app:projection}

To project the stochastic dynamics Eqns.~\eqref{eq:dlUnbound} and \eqref{eq:dxUnbound} in the bound case we use a formalism (and notations) similar to Ref.~\citenum{holmes2016stochastic}; see also \cite{ciccotti2008projection,HolmesCerfon:2013jw}. This projection consists in using stiff springs to impose each constraint, and considering the limit where the spring constants go to infinity. The resulting projected equations can be obtained by directly pursuing the steps below (without redoing the reasoning with stiff springs).

We start from stochastic equations in the $(x,l)$ space and seek to project them on the constraint manifold, defined by the constraint $q(x,l) = x + l - x_{\rm r} = 0$. The constraint matrix is therefore 
\begin{align}
C = (\nabla q)^T = \begin{pmatrix}
1 & 1
\end{pmatrix}.
\end{align}
We obtain the projector 
\begin{align}
P =  I - C^T(CC^T)^{-1}C = \frac{1}{2} \begin{pmatrix}
1 & -1 \\ -1 & 1 
\end{pmatrix}.
\end{align}
Initially the dynamics of $X = (x,l)^T$ may be written as 
\begin{align}
    \frac{dX}{dt} = - \tilde{\Gamma}^{-1} \nabla \mathcal{U}(X) + \sqrt{2k_BT \tilde{\Gamma}^{-1} } \eta_{xl}(t)
\end{align}
where the potential $\mathcal{U}(X) = k l^2/2$, the noise  $\eta_{xl} = (\eta_x, \eta_l)^T$ and the friction matrix is 
\begin{align}
\tilde{\Gamma} = \begin{pmatrix}
\Gamma & 0 \\ 0 & \gamma
\end{pmatrix}.
\end{align}
The projected friction and its Moore-Penrose pseudo-inverse are
\begin{eqnarray}
\Gamma_P = P \tilde{\Gamma} P &=  \frac{\Gamma+ \gamma}{4}\begin{pmatrix}
1 & -1 \\ -1 & 1 
\end{pmatrix}, \\
\Gamma_P^{\dagger} &= \frac{1}{\Gamma+ \gamma}\begin{pmatrix}
1 & -1 \\ -1 & 1 
\end{pmatrix}
\end{eqnarray}
with  a square root 
\begin{align}
\sigma_P = \sqrt{\Gamma_P^{\dagger}} = \frac{1}{\sqrt{\Gamma+ \gamma}} \begin{pmatrix}
1 & 0 \\ -1 & 0 
\end{pmatrix}.
\end{align}
We obtain the projected dynamics
\begin{eqnarray}
    \frac{dX}{dt} = & - \Gamma_P^{\dagger} \nabla\mathcal{U}(X)  + \sqrt{2k_BT \Gamma_P^{\dagger} } \eta_{xl}(t)
\end{eqnarray}
%\begin{eqnarray}
%    \frac{dX}{dt} = & - \Gamma_P^{\dagger} \nabla \left(\mathcal{U}(X) - k_BT \log |C| \right) +k_B T \nabla \Gamma_P^{\dagger} \nonumber \\
%    & + \sqrt{2k_BT \Gamma_P^{\dagger} } \eta_{xl}(t)
%\end{eqnarray}
%\sum_{i,j} P_{i,j} (\partial_j \Gamma_P^{\dagger})_{i,.} 
where additional terms are needed if $C$ is not constant over the constraint manifold \cite{ciccotti2008projection,HolmesCerfon:2013jw}. One can check that this exactly yields the bound dynamics Eq.~\eqref{eq:dxProjected}, with $\eta = \eta_x$ (this decomposition of the noise is not unique but this does not impact the dynamics in a weak sense).  %Note that usually one would have to add a term $-k_BT\Gamma_P^{\dagger} \nabla\log |C|$ to the equation above, where $|C|$ is the pseudodeterminant of $C$, to account for the fact that the constraints are imposed with stiff springs, but this term vanishes since $|C|$ is constant over the constraint manifold \cite{ciccotti2008projection,HolmesCerfon:2013jw}.

\subsection*{Appendix B: Numerical simulations}

Stochastic simulations of particle and leg dynamics are conducted using a custom made Fortran 90 routine. Fast random number generation is performed according to a Mersenne twister algorithm. Normally distributed random numbers are used for particle displacement while uniformly distributed random numbers are used to determine binding events. Equations are simulated in their non-dimensional form. The step $dt$ was chosen to be much small than all other time scales of the system. Typically $dt = \frac{1}{100} \min \left( \frac{\on \Gamma}{k}, \frac{\on \Gamma}{k}, \frac{\gamma}{\Gamma} \right)$. The system is simulated for $N_T = 10^8$ time steps, and the simulation is repeated over $N_{\rm runs} = 100$ independent runs (with renewed random number seed).

To simulate binding and unbinding events, for each leg, at each time step, we choose a random number $R$ uniformly distributed between 0 and 1 and then: 
\begin{itemize}
    \item if the leg is bound, and if $R > \off dt$ then the leg becomes unbound. Otherwise it remains bound.  
    \item if the leg is unbound, and if $R > \on dt$ then the leg becomes bound. Otherwise it remains unbound.
\end{itemize}
This simulation routine approximates well the exponential binding dynamics expected from the continuous equations since $dt \ll \off^{-1}, \on^{-1}$. To simulate all other stochastic equations we use a standard Euler-Maruyama discretization. 

The particle position $x$ is saved every $10^4$ time steps, and the mean squared displacement  $<(x(t+t_0) - x(t_0))^2 >_{t_0}$ (averaged over initial times $t_0$) is computed up to $N_T/100 = 10^6$ time steps. The effective diffusion coefficient for each run $D_{\rm eff,i}$ is obtained from the analytical least square regression of $< (x(t+t_0) - x(t_0))^2 >_{t_0}$ with time. The average value over the runs $D_{\rm eff} = \frac{1}{N_{\rm runs}} \sum_i D_{\rm eff,i}$ is retained as the effective long time diffusion coefficient. The standard deviation of $D_{\rm eff,i}$ allows to draw error bars in all simulation plots.

%%%END OF MAIN TEXT%%%


%\balance


%%%REFERENCES%%%
%\bibliography{Caterpillar} %You need to replace "rsc" on this line with the name of your .bib file


%\includepdf[pages=-]{Nanocaterpillar_SI.pdf}

\end{document}

\maketitle
%\subtitle{The nanocaterpillar's walk: motion with ligand-receptor contacts}

\tableofcontents

\clearpage

\section{N-legged, one dimensional, caterpillar model}

Note that unless specifically mentioned, in the entire supplementary information $l$ is used as a shortcut notation for spring length relative to its rest length $l-l_0$. 

\subsection{Agreement of simulation and analytical results regardless of the value of $\epsilon$ for the 1-legged caterpillar}

\begin{figure}[h!]
    \centering
    \includegraphics[width = 0.4\linewidth]{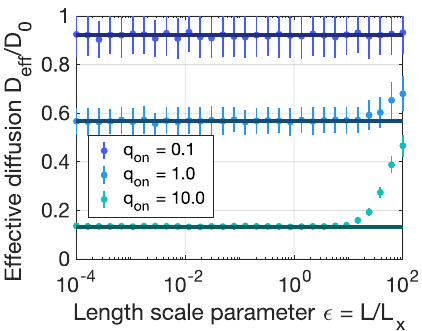}
    \caption{Simulation results for different values of the non-dimensionalizing parameter $\epsilon = L/L_x$ characterizing the difference between the length scale of oscillations of legs $L$ versus the length scale of particle displacement $L_x$, for the case of a 1-legged caterpillar. Various values of the attachment rate $q_{\rm on}$ are explored (given in non-dimensional units $k/\Gamma\epsilon^2$). The other numerical parameters are $\gamma/\Gamma = 0.1$ and $q_{\rm off} = 0.8 k/\Gamma\epsilon^2$. The lines correspond to the analytical formula Eq.~\gammaeff of the main paper. }
    \label{fig:epsilon}
\end{figure}

In Fig.~\ref{fig:epsilon} we present agreement between the effective diffusion evaluated using stochastic simulations and evaluated with the analytical formula Eq.~\gammaeff of the main paper. 

\subsection{N legs facing a uniformly sticky surface}

\subsubsection{Method on an example: 2 legs facing a sticky surface}

To investigate dynamics of caterpillars with multiple legs, we start by illustrating the framework on a 2 leg system. 
\paragraph{Projection of the dynamics in the bound state} 
The first step is to write the projected dynamics in the bound state. If there are 2 legs, when only 1 of them is bound, then the dynamics of the unbound leg are completely independent of the bound one and the projected bound equations are the same as those reported in the main paper. When 2 legs are bound however we must project again the dynamics. We therefore have 2 constraints $q_1(x,l_1,l_2) = x + l_1 + x_{r,1} = 0$ and  $q_2(x,l_1,l_2) = x + l_2 + x_{r,2} = 0$ where $x_r$ are reference positions when either of the legs first form their bond. The constraint matrix is therefore 
\begin{align}
C = (\nabla q)^T = \begin{pmatrix}
1 & 1 & 0 \\
1 & 0 & 1 
\end{pmatrix}.
\end{align}
We then get the projector 
\begin{align}
P =  I - C^T(CC^T)^{-1}C = \frac{1}{3} \begin{pmatrix}
1 & -1 & -1 \\ -1 & 1 & 1 \\ -1 & 1 & 1 
\end{pmatrix}
\end{align}
The friction matrix is in the unbound configuration 
\begin{align}
\tilde{\Gamma} = \begin{pmatrix}
\Gamma & 0 & 0  \\ 0 & \gamma & 0 \\ 0 & 0 & \gamma
\end{pmatrix}
\end{align}
giving a projected friction and its Moore-Penrose pseudo-inverse as
\begin{eqnarray}
\Gamma_P = P \tilde{\Gamma} P &=  \frac{\Gamma+ 2\gamma}{3} P, \\
\Gamma_P^{\dagger} &= \frac{1}{\Gamma+ 2\gamma} 3 P
\end{eqnarray}
with a square root
\begin{align}
\sigma_P = \sqrt{\Gamma_P^{\dagger}} = \frac{1}{\sqrt{\Gamma+ 2\gamma}} \begin{pmatrix}
1 & 0 & 0 \\ -1 & 0 & 0 \\ -1 & 0 & 0 
\end{pmatrix}.
\end{align}
We obtain the projected dynamics
\begin{eqnarray}
    \frac{dx}{dt} = - \frac{dl_1}{dt} = - \frac{dl_2}{dt} = \frac{k(l_1 + l_2)}{\Gamma+ 2\gamma} + \sqrt{\frac{2 k_B T}{\Gamma + 2\gamma}} \eta(t) 
\end{eqnarray}
where $\eta(t)$ is a white Gaussian noise. The friction in the bound state is therefore naturally the sum of the frictions $\Gamma + 2 \gamma$. 

\paragraph{Generator for the dynamics} 
For 2 legs we can write the full generator (in non-dimensional scales) $\mathcal{L}^{(2)} =  \frac{1}{\epsilon^2} \mathcal{L}^{(2)}_0 +  \frac{1}{\epsilon} \mathcal{L}^{(2)}_1 + \mathcal{L}^{(2)}_2$. The generator is now an operator acting on a space of 4 states ( $\# 1$ has no bond,  $\# 2-3$ have 1 bond, where the leg $1$ is bound in state $\# 2$ and reciprocally,  and $\# 4$ has 2 bonds). The lowest order generator is
\begin{equation}
    \mathcal{L}^{(2)}_0 = \mathcal{Q} + \mathcal{U}_0 = \begin{pmatrix}
    -2\on & \on & \on & 0 \\
    \off & -\off - \on & 0 & \on \\
    \off & 0 & -\off - \on & \on \\
    0 & \off & \off & 2 \off 
    \end{pmatrix} + \mathrm{diag}  \begin{pmatrix} \frac{\Gamma}{\gamma}  \left( D_{l_1} + D_{l_2} \right) \\
    \frac{\Gamma}{\Gamma + \gamma} D_{l_1} + \frac{\Gamma}{\gamma} D_{l_2} \\
    \frac{\Gamma}{\gamma} D_{l_1} + \frac{\Gamma}{\Gamma + \gamma} D_{l_2} \\
    \frac{\Gamma}{\Gamma + 2 \gamma} \left(- (l_1 + l_2) (\partial_{l_1} + \partial_{l_2}) + (\partial_{l_1} + \partial_{l_2})^2 \right) 
    \end{pmatrix} 
\end{equation}
where $D_{l_i} =  - l_i \partial_{l_i}  + \partial_{l_il_i}$ is an operator for the unbound tether $i$. The next orders are
\begin{equation}
\mathcal{L}^{(2)}_1 =  \mathrm{diag} \begin{pmatrix} 0 \\ \frac{\Gamma}{\Gamma + \gamma} (l_1 \partial_x  - 2 \partial_{xl_1}) \\ \frac{\Gamma}{\Gamma + \gamma} (l_2 \partial_x  - 2 \partial_{xl_2})  \\  \frac{\Gamma}{\Gamma + 2 \gamma} ((l_1 + l_2)\partial_x - 2\partial_{xl_1} - 2 \partial_{xl_2})
\end{pmatrix} \, \text{and} \,\, \mathcal{L}^{(2)}_2 = \mathrm{diag} \begin{pmatrix} \partial_{xx} \\ \frac{\Gamma}{\Gamma + \gamma} \partial_{xx} \\ \frac{\Gamma}{\Gamma + \gamma} \partial_{xx}  \\  \frac{\Gamma}{\Gamma + 2 \gamma} \partial_{xx}
\end{pmatrix}.
\end{equation}
The equilibrium distribution is simply
\begin{equation}
\pi \propto \begin{pmatrix}
(q_{\rm off}/q_{\rm on})^2 \\ q_{\rm off}/q_{\rm on} \\ q_{\rm off}/q_{\rm on} \\ 1
\end{pmatrix} e^{- l_1^2 /2 }e^{- l_2^2 /2 }.
\end{equation}

\paragraph{Long time solution} 
We now seek a solution as an expansion in $\epsilon$, $f = f_0 + \epsilon f_1 + ... $. In a very similar way as systematically observed in similar derivations we find $f_0 = a(x,t) \begin{pmatrix} 1 & 1 & 1 & 1 \end{pmatrix}^T$ at lowest order. The associated equilibrium distribution is $\pi_0 = \pi$.  At the following order we need to solve $\mathcal{L}^{(2)}_0 f_1 = - \mathcal{L}^{(2)}_1 f_0$ and we will seek a natural solution (that strongly reflects the symmetry of the problem) as
\begin{equation}
f_1 = \begin{pmatrix}
 u_0 l_1 + u_0 l_2 \\  b_1 l_1 + u_1 l_2 \\ u_1 l_1 + b_1 l_2  \\ b_2 l_1 + b_2 l_2
\end{pmatrix} \partial_x a
\end{equation}
where $u_n$ and $b_n$ are constants that solve a linear system of equations (with non zero determinant), and $u_n$ and $b_n$ refer respectively to unbound and bound contributions with $n$ bonds in the system. We do not report the equation system here but will come to it later on. 
At the next order, to find a solution for $f_2$ we require the Fredholm alternative, $\langle \partial_t f_0 - \mathcal{L}^{(2)}_2 f_0 - \mathcal{L}^{(2)}_1 f_1 , \pi_0 \rangle = 0$, which gives
\begin{equation}
\left( \frac{\off^2}{\on^2} + \frac{2\off}{\on} + 1 \right) \partial_t a =  \frac{\off^2}{\on^2} \partial_{xx} a  + 2 \frac{\off}{\on} \frac{\Gamma}{\Gamma + \gamma}  \left( 1 - b_1 \right) \partial_{xx} a + \left( 1 - 2 b_2 \right)\partial_{xx} a 
\end{equation}
which can be rewritten as a weighted sum (in dimensional scales) 
\begin{equation}
\partial_t a =  k_B T \left( \frac{p_0}{\Gamma} + \frac{p_1}{\frac{\Gamma + \gamma}{ \left( 1 - b_1 \right)}} + \frac{p_2}{\frac{\Gamma + 2\gamma}{ \left( 1 - 2 b_2 \right)}}  \right) \partial_{xx} a  
\end{equation}
where $p_{k}$ is the probability to have $k$ bonds ($p_0 = \off^2/Z$, $p_1 = 2\off \on/Z$ and $p_2 = \on^2/Z$ with $p_0 + p_1 + p_2 = 1$). The above expression clearly shows that the effective inverse friction is a weighted sum of inverse friction coefficients
\begin{equation}
    \frac{1}{\Gamma^{2\, \rm legs}_{\rm eff}} = \sum_{n=0}^2 \frac{p_n}{\Gamma_n} =  \sum_{n=0}^2 \frac{p_n}{\frac{\Gamma + n \gamma}{ (1 - n b_n)}}.
\end{equation}
We will show this expression for all $N$ below. The linear system of equations solved by the $u_k$ and $b_k$ can now be given 
\begin{equation}
\begin{split}
- 2 q_{\rm on} u_0  +  q_{\rm on} b_1  +  q_{\rm on} u_1 -  \frac{\Gamma}{\gamma} u_0 = 0 &  ,\,\,\text{(unbound  contributions in the 0 bond state)}\\
q_{\rm off} u_0 - q_{\rm off} u_1 - q_{\rm on} u_1 +  q_{\rm on} b_2 - \frac{\Gamma}{\gamma} u_1 =0 &  ,\,\,\text{(unbound  contributions in a 1 bond state)} \\
q_{\rm off} u_0 - q_{\rm off} b_1 - q_{\rm on} b_1 + q_{\rm on} b_2  - \frac{\Gamma}{\Gamma  + \gamma} b_1 = - \frac{\Gamma}{\Gamma + \gamma} &  ,\,\,\text{(bound contributions in a 1 bond state)}\\
q_{\rm off} b_1 + q_{\rm off} u_1 - 2 q_{\rm off} b_2 - 2 \frac{\Gamma}{\Gamma + 2 \gamma} b_2 = - \frac{\Gamma}{\Gamma + 2 \gamma}  &  ,\,\,\text{(bound contributions in the 2 bonds state)}.
\end{split}
\end{equation}
Solving the above linear system yields lengthy expressions for $b_k$ and $u_k$. One can show however that the effective contributions for the bound states can be expanded as 
\begin{equation}
\Gamma_1 = \Gamma + \gamma_{\rm eff} \left( 1 - O(\frac{\gamma_{\rm eff}}{\Gamma})\right) \,\, \Gamma_2 =\Gamma + 2\gamma_{\rm eff} \left( 1 + O(\frac{\gamma_{\rm eff}}{\Gamma})\right)
\end{equation}
such that we find already some a linear scaling as $\Gamma_n \sim \Gamma + n \gamma_{\rm eff}$.

\subsubsection{$N$ legs}

\paragraph{Projection of the dynamics with $N$ legs.} The projection formalism naturally extends to $N$ legs.  For $n$ bound legs we find that the friction is simply $\Gamma + n\gamma$, such that the projected dynamics are for the first $n$ bound legs
\begin{equation}
    \frac{dx}{dt} = - \frac{dl_1}{dt} = ... = - \frac{dl_n}{dt} = \frac{k \sum_{i=1}^n (l_i)}{\Gamma+ n \gamma} + \sqrt{2 \frac{k_B T}{\Gamma + n\gamma}} \eta(t). 
\end{equation}

\paragraph{System of equations for $N$ legs}
The generator is now an operator acting on $2^N$ states, and we order these states according to their number of bonds  (0 bonds,  all 1 bond states, all 2 bonds states, ...). For N legs we can write the full generator (in non-dimensional scales) $\mathcal{L}^{(N)} =  \frac{1}{\epsilon^2} \mathcal{L}^{(N)}_0 +  \frac{1}{\epsilon} \mathcal{L}^{(N)}_1 + \mathcal{L}^{(N)}_2$, where all $\mathcal{L}^{(N)}_i$ terms are very similar to the ones introduced for $N=2$ and can be naturally generalized. Similarly the equilibrium distribution is naturally extended as 
\begin{equation}
    \pi = \displaystyle e^{- \sum_{i=1}^N l_i^2 /2 }\begin{pmatrix}
    (\off/\on)^N &  (\off/\on)^{N-1} &  (\off/\on)^{N-1} & .... &  (\off/\on)^{N-2} & ... & 1
    \end{pmatrix}^T.
\end{equation}

\paragraph{Long time solution with $N$ legs}
We now seek a solution as an expansion in $\epsilon$, $f = f_0 + \epsilon f_1 + ... $. In a very similar way as systematically observed in similar derivations we find $f_0 = a(x,t) \begin{pmatrix} 1 & ... & 1 \end{pmatrix}^T$ at lowest order. The associated equilibrium distribution is $\pi_0 = \pi$.  At the following order we need to solve $\mathcal{L}^{(N)}_0 f_1 = - \mathcal{L}^{(N)}_1 f_0$ and we will seek a natural solution (that strongly reflects the symmetry of the problem) as
\begin{equation}
f_1 = \begin{pmatrix}
 u_0 l_1 + u_0 l_2 + ... + u_0 l_N \\  b_1 l_1 + u_1 l_2 + ... + u_1 l_N\\ u_1 l_1 + b_1 l_2  + ... + u_1 l_N \\ ... \\ b_2 l_1 + b_2 l_2 + u_2 l_3 ... + u_2 l_N \\ ... \\ b_N l_1 + b_N l_2 + ... + b_N l_N
\end{pmatrix} \partial_x a
\end{equation}
where $u_n$ and $b_n$ refer respectively to unbound and bound contributions with $n$ bonds in the system. We now seek the general system of equations satisfied by $u_n$ and $b_n$. 
For a number of bonds $n$, consider that a given focus tether is unbound, say $i$. This will therefore allow us to obtain an equation on the unbound contributions of that tether (that in $l_i$) so primarily on $u_n$. The tether is relaxing yielding a contribution $- \frac{\Gamma}{\gamma} u_n$. There are $n$ possible bonds to undo leading to a contribution ($ - n q_{\rm off} u_n$). In any $n-1$ bond configurations starting from our initial configuration, the focus tether will still be unbound ($u_{n-1}$), such that we get an ($+ n q_{\rm off} u_{n-1}$) contribution. There are $N-n$ bonds to form ($- (N-n) q_{\rm on} u_n$). In forming bonds, only 1 choice yields to bind the focus tether ($q_{\rm on} b_{n+1}$) while the other forming bonds will not be the focus tether ($(N-n-1)q_{\rm on} u_{n+1}$). The right hand side terms (from $\mathcal{L}^{(N)}_1 f_0$) corresponding to unbound tethers are $0$. This yields the first line of the system of equations Eq.~\eqref{eq:systemNlegs} below. If one considers a bound focus tether, similarly one can derive contributions due to binding and unbinding. The bound relaxation terms yield a contribution $(- \frac{n \Gamma}{\Gamma + n \gamma} b_n)$. Additionally, the right hand side terms (coming from $\mathcal{L}^{(N)}_1 f_0$) corresponding to the unbound tether is $- \frac{\Gamma}{\Gamma + n\gamma}$. We obtain
\begin{equation}
    \begin{cases}
        \displaystyle n q_{\rm off} u_{n-1} - n q_{\rm off} u_n  - (N-n) q_{\rm on} u_n + q_{\rm on} b_{n+1} + (N-n-1)q_{\rm on} u_{n+1} - \frac{\Gamma}{\gamma} u_n = 0& \\
        \displaystyle  q_{\rm off} u_{n-1} + (n-1) q_{\rm off} b_{n-1} - n q_{\rm off} b_n  - (N-n) q_{\rm on} b_n + (N-n) q_{\rm on} b_{n+1} - \frac{n \Gamma}{\Gamma + n \gamma} b_n = - \frac{\Gamma}{\Gamma + n\gamma}.
    \end{cases}
    \label{eq:systemNlegs}
\end{equation}
The system Eq.~\eqref{eq:systemNlegs} applies for all $n = 0 .. N$, taking as boundary equations $u_N = 0$ and $b_0 = 0$. 
Unfortunately the system does not simplify further but its determinant is non zero, showing that a non trivial solution  exists. We will study it further later but for now conclude on the long time solution. 
At the next order, to find a solution for $f_2$ we require the Fredholm alternative, $\langle \partial_t f_0 - \mathcal{L}^{(N)}_2 f_0 - \mathcal{L}^{(N)}_1 f_1 , \pi_0 \rangle = 0$, which yields after some algebraic manipulations (back in dimensional scales)
\begin{equation}
\partial_t a =  k_B T \left( \frac{p_0}{\Gamma} + \frac{p_1}{\frac{\Gamma + \gamma}{ \left( 1 - b_1 \right)}} + \frac{p_2}{\frac{\Gamma + 2\gamma}{ \left( 1 - 2 b_2 \right)}}  + ... + \frac{p_N}{\frac{\Gamma + N\gamma}{ \left( 1 - N b_N \right)}} \right) \partial_{xx} a  
\end{equation}
where $p_n = \frac{\binom{N}{n} x^n (1-x)^{N-n}}{Z}$ with $x = \on/\off$ is the probability to have $n$ bonds. Writing in full generality \begin{equation}
    \Gamma_n =  \frac{\Gamma + n\gamma}{ \left( 1 - n b_n \right)}
\end{equation}
we indeed recover Eq.~\gammaN of the main manuscript. We also see that the coefficients $\Gamma_n$ indeed correspond to friction contributions in a state with $n$ bonds as only $n$ and $b_n$, that corresponds to the bound contributions, intervene. 

\paragraph{Resolution when the system is dominated by the average number of bonds}

\begin{figure}[h!]
    \centering
    \includegraphics[width = 0.5\textwidth]{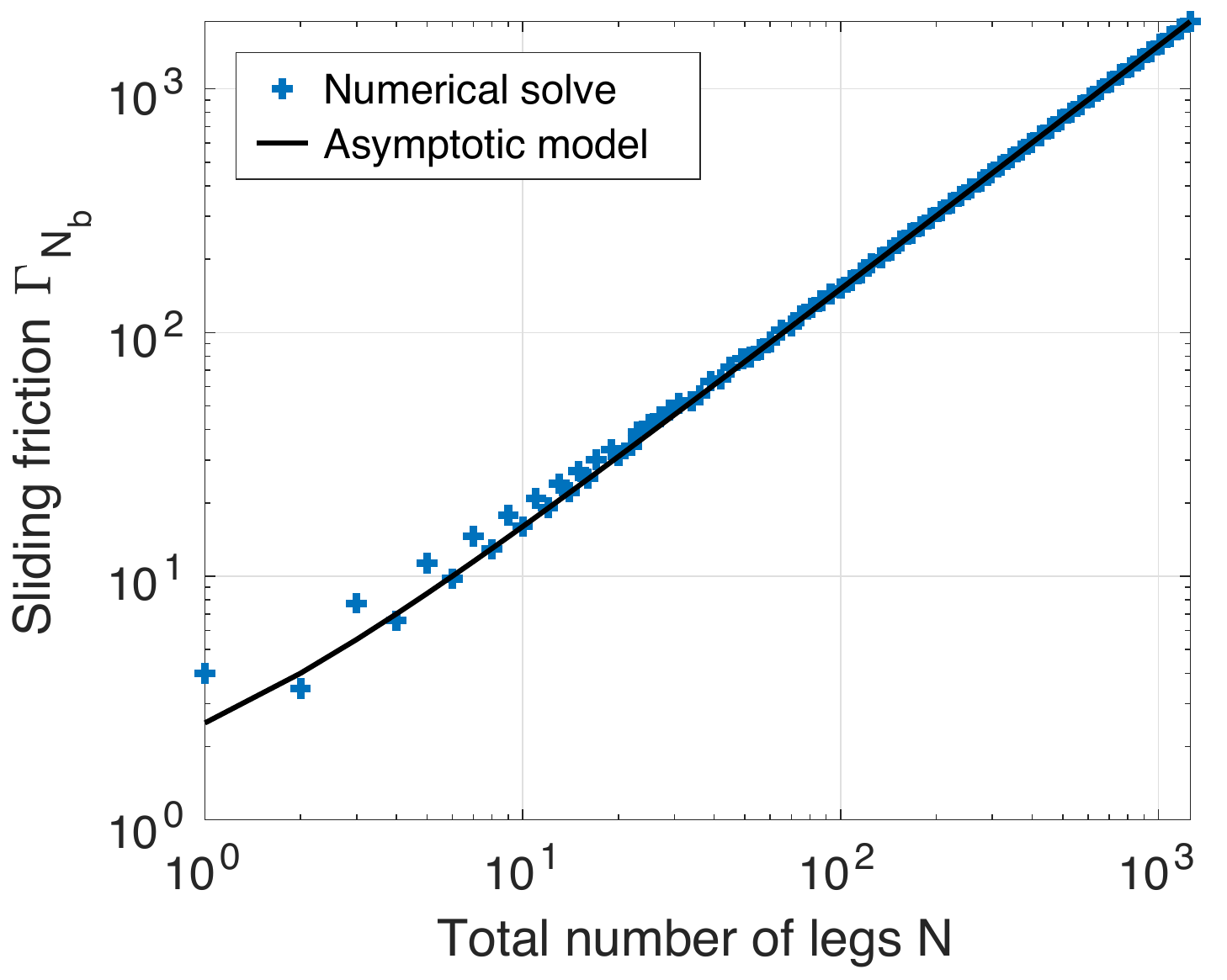}
    \caption{Value of friction coefficient for the average number of bonds as evaluated using Eq.~\eqref{eq:GNb} or equivalently Eq.~\gammaN of the main paper (``Asymptotic model'') and fully solving the system of equations Eq.~\eqref{eq:systemNlegs} and presenting the value $\Gamma_n$ for the index $n$ closest to $N_b$ (``Numerical solve''). Here the values of other parameters (in dimensional scales) are all set to $1 = \frac{q_{\rm on} \Gamma}{k} = \frac{q_{\rm off} \Gamma}{k} = \frac{\gamma}{\Gamma}$.}
    \label{fig:NlegsFig1}
\end{figure}

We can search for a closed (simpler) system for Eq.~\eqref{eq:systemNlegs} where the dominant terms will originate from the average number of bonds $N_b = \sum_{n=0}^N n p_n =  N \frac{q_{\rm on}}{q_{\rm on} + q_{\rm off}}$. We assume that, around this average number, terms do not change much (the derivatives are close to $0$), meaning we can approximate $u_{N_b} \simeq u_{N_b - 1} \simeq u_{N_b+1} \equiv \bar{u}$, and similarly for $b_{N_b} = \bar{b}$ leading to
\begin{equation}
    \begin{cases}
         - q_{\rm on} \bar{u} + q_{\rm on} \bar{b} - \frac{\Gamma}{\gamma}  \bar{u} = 0& \\
        q_{\rm off} \bar{u} - q_{\rm off} \bar{b}  - \frac{N_b \Gamma}{\Gamma + N_b \gamma} \bar{b} = - \frac{\Gamma}{\Gamma + N_b \gamma} & 
        \label{eq:systemNLegsshort}
    \end{cases}
\end{equation}
Solving the system for $\bar{b}$ and $\bar{u}$ yields then the value of the friction coefficient for the average number of bonds (back in dimensional scales)
\begin{equation}
    \Gamma_{N_b} = \frac{\Gamma + N_b \gamma}{1 - N_b \bar{b}} = \Gamma + N_b \left(\gamma + \frac{k}{q_{\rm off}}+ \gamma \frac{q_{\rm on}}{q_{\rm off}}  \right) 
    \label{eq:GNb}
\end{equation}
Eq.~\eqref{eq:GNb} is reported as Eq.~\gammaNb in the main text. It shows excellent agreement with the exact (numerical) solution to the full system of Eqs.~\ref{eq:systemNlegs} at large total number of legs $N$ (see Fig.~\ref{fig:NlegsFig1}). 

\paragraph{Empirical solution for an arbitrary number of bonds}

An interesting question is then to investigate $\Gamma_n$, the effective friction contributing to the state with $n$ bonds, in the large $N$ (total number of legs) limit. This requires solving the full system Eq.~\eqref{eq:systemNlegs}. This system is not easily amenable to analytical calculations, and instead we use it as a benchmark to explore a phenomenological law for  $\Gamma_n$.

\begin{figure}[h!]
    \centering
    \includegraphics[width = 0.9\textwidth]{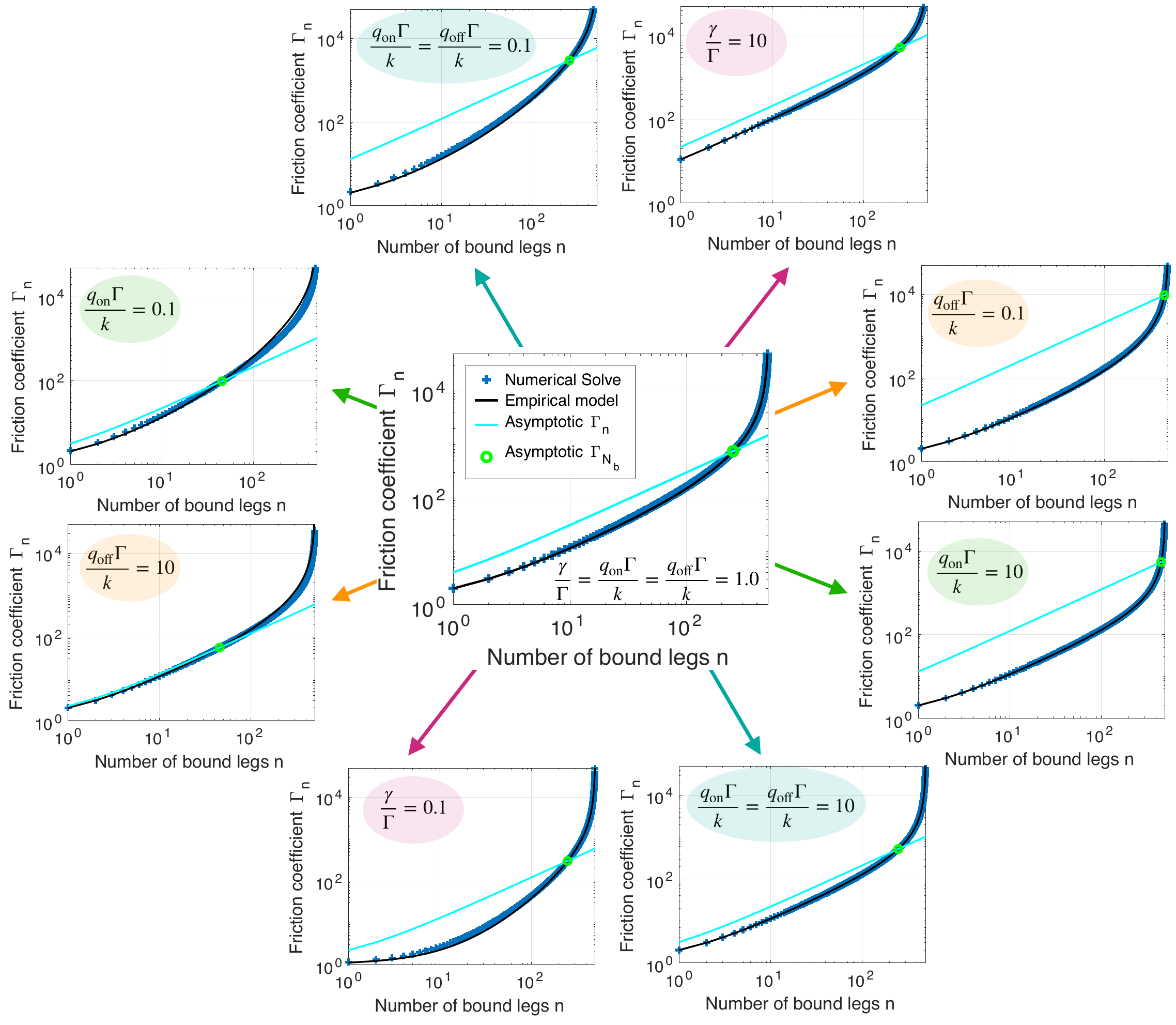}
    \caption{Value of friction coefficients $\Gamma_n$ for all possible number of bonds (for a maximum of $N = 500$) as evaluated using Eq.~\eqref{eq:GammanEmpirical} (``Empirical model'') and fully solving for the system of equations Eq.~\eqref{eq:systemNlegs} and presenting the values $\Gamma_n$ (``Numerical Solve''). Index $N_b$ is highlighted in green in each plot, and calculated from Eq.~\gammaNb. The ``Asymptotic $\Gamma_n$'' result corresponds to Eq.~\eqref{eq:asymptotic}. Here the values of other parameters (in dimensional scales) are all set to $1 = \frac{q_{\rm on} \Gamma}{k} = \frac{q_{\rm off} \Gamma}{k} = \frac{\gamma}{\Gamma}$ unless another indication is given.}
    \label{fig:NlegsFig2}
\end{figure}

First, it is natural to assume that the correction $\Gamma_n - \Gamma$ typically contains a term $n\gamma$ coming from added friction of the $n$ bonds (as is noted already in the projected dynamics). Then, recall forces are also exerting friction. Typically $n$ bonds are exerting friction due to recall forces. Yet for this final contribution to $\Gamma_n$, the situation is not the same for $n$ bonds as for $N_b$ bonds. Around $n=N_b$, the probabilities to be in a state with one more bond or one bond less are more or less the same, $p(N_b-1) \simeq p(N_b) \simeq p(N_b+1)$. For $n$ bonds, we have in general (for example investigating the probability to undo a bond)
\begin{equation}
\begin{split}
\frac{p(n)}{p(n-1)} &= \frac{\binom{N}{n} p_0^n (1-p_0)^{N-n} }{\binom{N}{n-1} p_0^{n-1} (1-p_0)^{N-n+1} } \\
& = \frac{\binom{N}{n} p_0^n (1-p_0)^{N-n}  p_0}{ \binom{N}{n}  \frac{n}{N-n+1} p_0^{n} (1-p_0)^{N-n} (1-p_0) } \\
& = \frac{p_0}{(1-p_0)} \frac{N-n+1}{n} = \frac{N_b}{N - N_b} \frac{N-n+1}{n}
\end{split}
\end{equation}
We expect that the typical time over which the spring resistance acts $\tau_{\rm eff}$ has to be modified by the propensity to unbind (coming from the state with $n$ bonds) as $\tau_{\rm eff} \rightarrow \tau_{\rm eff} \frac{p(n-1)}{p(n)}$. We obtain, wrapping up all contributions 
\begin{equation}
     \Gamma_n = \Gamma + n \left(\gamma + k \tau_{\rm eff} \frac{p(n-1)}{p(n)} \right) = \Gamma + n \left(\gamma + k   \frac{p(n-1)}{p(n)}\left[ \frac{1}{q_{\rm off}}+ \frac{\gamma}{k} \frac{q_{\rm on}}{q_{\rm off}} \right]  \right).
\end{equation}
which explicitly writes as
\begin{equation}
    \Gamma_n = \Gamma + n \left(\gamma + k \frac{N-N_b}{N-(n-1)} \frac{n}{N_b} \left[ \frac{1}{q_{\rm off}}+ \frac{\gamma}{k}  \frac{q_{\rm on}}{q_{\rm off}} \right]  \right) 
    \label{eq:GammanEmpirical}
\end{equation}
Eq.~\eqref{eq:GammanEmpirical} is compared to the full solution of the linear system in Fig.~\ref{fig:NlegsFig2}. We find excellent agreement over a broad range of parameters. Notice that also around $n \simeq N_b$ and for $N_b \ll N$ we find the limit behavior 
\begin{equation}
     \Gamma_n = \Gamma + n \left(\gamma + \left[ \frac{k}{q_{\rm off}}+ \gamma \frac{q_{\rm on}}{q_{\rm off}} \right]  \right) 
     \label{eq:asymptotic}
\end{equation}
which allows us to recover, as anticipated, the result for $n = N_b$ of Eq.~\gammaNb.

\section{Comparison to experimental data for diffusion of DNA-coated colloids}

\subsection{Experimental data for the diffusion of DNA-coated colloids: I. Additional data}

\subsubsection{Preparation of material}

\paragraph{DNA coated polystyrene colloids}
We synthesize DNA-coated polystyrene (PS) spheres using the
swelling/deswelling method reported in Ref.~\cite{oh2015high}. Polystyrene-b-poly(ethylene oxide) copolymer PS(3800 g/mol)-b-PEO(6500 g/mol) is purchased from Polymer Source Inc, and is first functionalized with azide at the end of the PEO chain~\cite{hiki2007facile}. PS-b-PEO-N3 are then attached to the PS particles using the swelling/deswelling method. In the synthesis, 15~$\mathrm{\mu L}$ of $1 \mu m$ particles (10 w/v, purchased from Thermo Scientific), 125~$\mathrm{\mu L}$ Deionized (DI) water, 160~$\mathrm{\mu L}$ tetrahydrofuran (THF) and 100~$\mathrm{\mu L}$ of PS-b-PEO-N3 are mixed at room temperature. The mixture is placed on a horizontal shaker (1000 rpm) for 1.5 hours to fully swell the PS particles and absorb the PS block of the PS-b-PEO-N3 molecules. Then THF is slowly removed from the solution via evaporation by adding DI water, leaving the hydrophobic PS blocks physically inserted into the particles and the hydrophilic PEO chains extending out into the solution. The particles are washed with DI water three times to remove excess polymers.

Single stranded DNA (ssDNA, 20 bases, purchased from Integrated DNA Technologies) with 5’ dibenzocyclooctyne (DBCO) end modification, is clicked to the N3 (at the end of PS-b-PEO-N3) through strain promoted alkyne-azide cycloaddition~\cite{oh2015high}. PS particles previously coated with the PS-b-PEO-N3 polymer brush are dispersed in 200~$\mathrm{\mu L}$ of 500~mM PBS buffer, at pH 7.4. Then 10~$\mathrm{\mu L}$ of DBCO-DNA (0.1~mM) are added to the suspension. The mixture is left to react for 48 hours on a horizontal shaker (1000 rpm). The final product is washed in DI water three times and stored in 140~mM PBS buffer. The DNA coverage density is measured using flow cytometry and we obtain $\sigma = 1/(3.27~\mathrm{nm}^2)$. The DNA sequence used on the colloids is    5'-/DBCO/-T$_{14}$-ACCGCA-3'.

\paragraph{DNA coated glass substrate}
DNA coated substrates are prepared using the same swelling/deswelling method. First, an ultra thin PS layer is spin-coated to a cleaned 22~mm x 22~mm glass coverslip (purchased from Bioscience Tools). The substrate is then swelled in the same PS-b-PEO-N3 solution in THF for 4 hours. Then THF is slowly removed from the solution via evaporation. DNA clicking is performed in a home made PDMS reaction chamber for 48 hours on a shaking stage, then washed 10 times in DI water to remove extra DNA. The entire sample is sealed in the 140~mM PBS buffer (ph 7.4) with  0.3$\%$ w/v pluronic F127 surfactants, using UV glue to avoid any external flow or evaporation of the buffer. The DNA sequence used on the glass substrate is complementary to that on the particles, 5'-/DBCO/-T$_{14}$-TGCGGT-3'.

\subsubsection{Tracking DNA coated colloids}

\begin{figure}[h!]
\centering
\includegraphics[width = 0.7\linewidth]{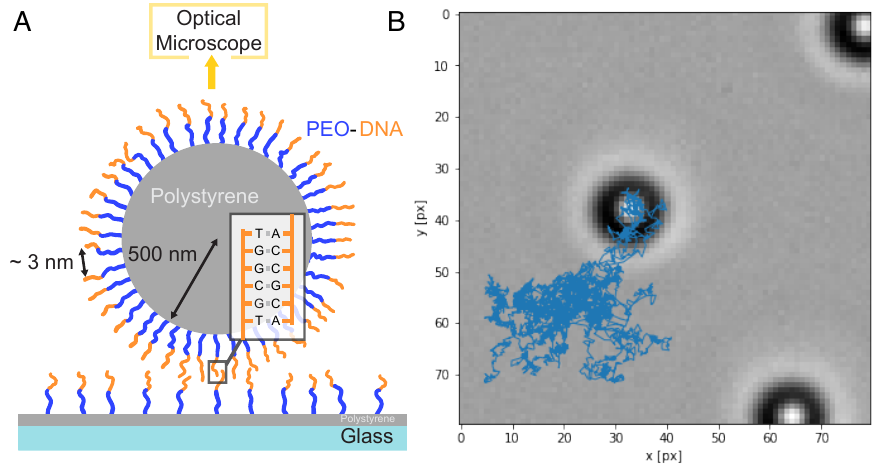}% Here is how to import EPS art
\caption{\label{fig:figExpSketch} \textbf{Experimental setup to measure diffusion of DNA-coated colloids on DNA-coated surfaces.} (A) Schematic of a DNA-coated colloid attaching to a DNA-coated substrate, with the specific DNA sequence used in this study. Diffusion of the colloids is tracked from on-top. (B) Example of a colloid trajectory over an $ 18~\mathrm{min}$ time frame (in blue) overlaid on the bright-field microscope image corresponding to the colloid's initial position. Here 1 px corresponds to $0.108~\mathrm{\mu m}$. }
\end{figure}

\paragraph{Particle positions measurements}
To study the diffusion of DNA coated colloids, we track the motion of about 500 particles as they bind and diffuse on the DNA coated substrate -- see Fig.~\ref{fig:figExpSketch}-A. The sample is mounted on a homemade lab microscope (Nikon  Eclipse  Ti  60X,  72nm  pixel  size, depth of focus $560~\mathrm{nm}$)  thermal stage with a temperature controller. Tracer particles fixed on the substrate are used to substract camera drift during the tracking. Displacement measurements are performed by tracking particles over the  temperature range 28-62 $^\circ$C -- see Fig.~\ref{fig:figExpSketch}-B. At each temperature, particles are tracked over a time range of 20 min at a frame rate of 5 images per second. For the highest temperature reported here, $T = 59.1~^\circ$C, particles diffuse faster and we only track them over 5 min, with 10 images per second. Images are then analyzed using the TrackPy software to obtain individual particle positions with time. Particles that do not move at all even at high temperatures are removed from the analysis. These particles are likely in a low density area where steric repulsion is not sufficient to screen van der Waals attraction, and therefore are ``crashed'' on the surface.

%obtain the fraction of bound and unbound particles we use a similar thermodynamic approach as in Ref.~\cite{fan2021microscopic}. At the lowest recorded temperature we count the average total number of particles in focus per frame say $N_{\rm tot}$. For each higher temperature we count the average number of particles in focus per frame say $N_T$. We then obtain the percentage of bound particles as $N_T/N_{\rm tot}$ and of unbound particles as $1 - N_T/N_{\rm tot}$. This approach effectively counts the number of particles within typically $280~\mathrm{nm}$ (half the depth of focus) from the surface. At high temperatures some particles remain close to the surface due to gravity.  

\paragraph{Mean square displacement analysis}

We fit the ensemble mean-squared displacement to a power law as $<\bm{x}^2(t)>  = 4Dt^{\alpha}$, where $\bm{x}$ is the position of each particle on the surface plane, using a linear regression in log space to get the diffusion coefficient $D$ and the power on time $\alpha$. Typically, $\alpha$ decreases from $\sim 1$ at high temperatures to values $<1$ at lower temperatures. Around the melting transition however, there exists a window of a few degrees where the motion is diffusive and we obtained $1.02 > \alpha > 0.94$. On this temperature window we then fix $\alpha = 1$ and the effective diffusivity $D_{\rm eff}$ is obtained by fitting the ensemble-averaged mean-square displacement to the power law $\langle \bm{x}^2(t) \rangle = 4 D_{\rm eff} t$.

\paragraph{Melting curve}
To compare different measurements with one another we define a kinetic melting temperature. This temperature $T_m$ corresponds to the temperature for which the measured diffusion coefficient is half that of the high temperature diffusion coefficient (the latter corresponding to the hydrodynamic diffusion coefficient). 

\paragraph{Reproducibility and error bars}

The entire experimental process (synthesis and mean square displacement measurements) are reproduced 3 times and the results are reported with different symbols in Fig.~7A of the main manuscript. Note that the synthesis is performed with slight variations of the coating process (shaking time), yet very similar behavior is obtained -- for example the melting temperatures for each of the samples are within $3^{\circ}$C of each other. Such disparity in melting temperature can occur due to density differences originating while performing the same synthesis and hence we do not report further details here.  

As the average for $D_{\rm eff}$ is calculated over a great number of particles, the typical error on $D_{\rm eff}$, for example due to the fitting procedure, is very small -- typically smaller than the size of the points used to represent data and also much smaller than intrinsic disparities from sample to sample due to density fluctuations on the surface coverage during sample preparation. Therefore we do not report any vertical error bars. 
The exact temperature measured can slightly fluctuate, due to potential drift of the temperature controller, thus it is reasonable to assume a $0.2^{\circ}$C error bar on each experimental data point.

\subsection{Experimental data for the diffusion of DNA-coated colloids: II. Existing data}

Diffusion of DNA-coated colloids from existing data was obtained from 2 published references~\cite{xu2011subdiffusion,wang2015crystallization}.

\paragraph{High coating density diffusion coefficients from Ref.~\cite{wang2015crystallization}}

Diffusion coefficients from Ref.~\cite{wang2015crystallization} were obtained 
by fitting a linear law through extracted mean square displacement data using WebPlotDigitizer~\cite{rohatgi2017webplotdigitizer} -- when the diffusion exponent $\alpha$ is greater than $\sim 0.8$. Mean square displacement data in Ref.~\cite{wang2015crystallization} represents the projected mean square displacement $\langle \bm{r}^2 \rangle$ covered on a half-sphere, when the displacement on the half-sphere is observed from on-top. The actual surface covered $\langle \bm{x}^2 \rangle$ is therefore larger than that measured on the projected area, and we can write $\langle \bm{x}^2 \rangle = \mathcal{A} \langle \bm{r}^2 \rangle$ where $\mathcal{A}$ is an area correction number. If the particle covers the entire area typically $\mathcal{A} \simeq 2$ since $2\pi R^2$ is the actual area of the half sphere of radius $R$, and $\pi R^2$ is the projected area. This typically accounts for the fact that the particles does not spend the same amount of time on the sides of the half-sphere and on the top, and that on the sides displacements can be fully orthogonal to the observation projection plane. Additionally, since motion is constrained to the half-sphere, in practice the random walk is constrained and folded back onto the sphere. If it were unconstrained on the sphere we would typically have $4\pi R^2$ of area covered projected on $\pi R^2$ so we take $\mathcal{A} = 4$ as an upper bound on $\mathcal{A}$. $\mathcal{A} = 2$ is our lower bound. These bounds allow us to define error bars for the diffusion data of Ref.~\cite{wang2015crystallization}. Again, considering potential density fluctuations on either surfaces and other experimental uncertainties due to calibration of the temperature controller, it is reasonable to assume a $0.2^{\circ}$C error bar on each experimental data point.  %Typically, if $\theta$ is the azimuthal angle measuring the location of the particle on the half sphere (where $\theta = 0$ corresponds to the summit) the actual displacement $d\bm{x}$ is $d\bm{x} \cos \theta = d\bm{r}$ where $d\bm{r}$ is the projected displacement. Since the particle has equal probability  % that reasoning reaches the same conclusion

The detailed parameters of the DNA-coated colloids used in Ref.~\cite{wang2015crystallization} are provided in that reference and we use their specific values to perform analytical predictions for $D_{\rm eff}$, see details below. The melting temperature in Ref.~\cite{wang2015crystallization} is defined as the temperature for which the fraction of single particles is 50\%, since the particles can self-assemble in arrays. This is a typical thermodynamic quantity hence we use a the thermodynamic definition of melting~\cite{fan2021microscopic}
\begin{equation}
    p^{Ref.~\cite{wang2015crystallization}}_{\rm unbound} = 1 - \frac{1}{Z} \int_0^{h_c} e^{-\phi(h)/k_BT}dh
\end{equation}
where $\phi(h)$ is a particle-particle interaction potential, $h_c \simeq 20~\mathrm{nm}$ is a typical interaction range and $Z$ a normalization constant. We find without any fitting that $T^{\rm theo}_m = 25.3^{\circ}$C which is not too far from the experimental melting temperature $T_m = 28.9^{\circ}$C. The difference is likely due to the slightly different method used to quantify $T_m$. We align experimental data relative to $T_m$ and theoretical data relative to $T^{\rm theo}_m$. 

\paragraph{Low coating density diffusion coefficients from Ref.~\cite{xu2011subdiffusion}}

Ref.~\cite{xu2011subdiffusion} provides the diffusion coefficients for their DNA-coated particles above the melting temperature $T_m = 44.7^{\circ}$C as $D_{\rm eff}(47^{\circ}~\mathrm{C}) = 0.38~\mathrm{\mu m^2/s}$, and at the melting temperature $D_{\rm eff}(44.7^{\circ}~\mathrm{C}) = 1.4\times 10^{-3}~\mathrm{\mu m^2/s}$. Additionally, for the data provided $0.27^{\circ}~\mathrm{C}$ below $T_m$, the exponent for diffusion is $\alpha \sim 0.8$ and we can estimate the diffusion coefficient from a linear fit to the data. We obtain $D_{\rm eff}(44.5^{\circ}~\mathrm{C}) = \frac{30~\mathrm{\mu m^2}}{30 000 \mathrm{s}\times 4} = 0.25 \times 10^{-3}~\mathrm{\mu m^2/s}$.

The detailed parameters of the DNA-coated colloids used in Ref.~\cite{xu2011subdiffusion} are provided in that reference and we use their respective values to perform analytical predictions for $D_{\rm eff}$, see details below. The melting temperature in Ref.~\cite{xu2011subdiffusion} is defined as the temperature for which the fraction of moving particles is 50\%, where `` Moving is defined as a displacement larger than
50 nm (1 pixel) between frames (frame rate = 1 Hz)''. We can relate to this moving quantity by defining the unbound probability as 
\begin{equation}
    p^{Ref.~\cite{xu2011subdiffusion}}_{\rm unbound} = 1 - \mathrm{erf} \left(\frac{\Delta x_{\rm max}}{\sqrt{4 D_{\rm eff} \Delta t_{\rm max}}}\right)
\end{equation}
measuring the probability that a step is larger than $\Delta x_{\rm max} = 50~\mathrm{nm}$ during a time interval $ \Delta t_{\rm max} = 1~\mathrm{s}$ where the diffusion coefficient of the particle is $D_{\rm eff}$. Here we use $D_{\rm eff}$ as predicted by hopping motion only since only hopping is occurring in this sample due to geometrical constraints. We find $T^{\rm theo}_m = 44.2^{\circ}$C close to the experimental measurement of $T_m = 44.7^{\circ}$C. We align experimental data relative to $T_m$ and theoretical data relative to $T^{\rm theo}_m$. In line with previous analysis we also add $0.2^{\circ}$C error bar on each experimental data point.

\subsection{Modeling tools for DNA-coated colloids}

\subsubsection{Number of legs and average number of bonds}
To evaluate $D_{\rm eff}$ from Eq.~\gammaN, we must evaluate the parameters of the 1D nanocaterpillar model. As mentioned in the main manuscript, some parameters, such as $N$ and $N_b$ (or equivalently $N$ and the ratio $\on/\off$) require careful modeling of the detailed leg-arm interactions~\cite{fan2021microscopic} to be estimated. 

We thus calculate the detailed DNA-DNA brush interactions, accounting for leg density, leg length and DNA sequence, by evaluating the interaction energy $\phi(h)$ of the DNA-coated colloid with another coated surface at separation distance $h$. Following Ref.~\cite{fan2021microscopic}, $\phi(h)$ includes repulsive steric interactions~\cite{milner1988theory} and attractive binding interactions, with entropic terms due to loss of degrees of freedom upon binding and competition for binding partners~\cite{varilly2012general}.

\begin{figure}[h!]
\centering
\includegraphics[width = 0.6\linewidth]{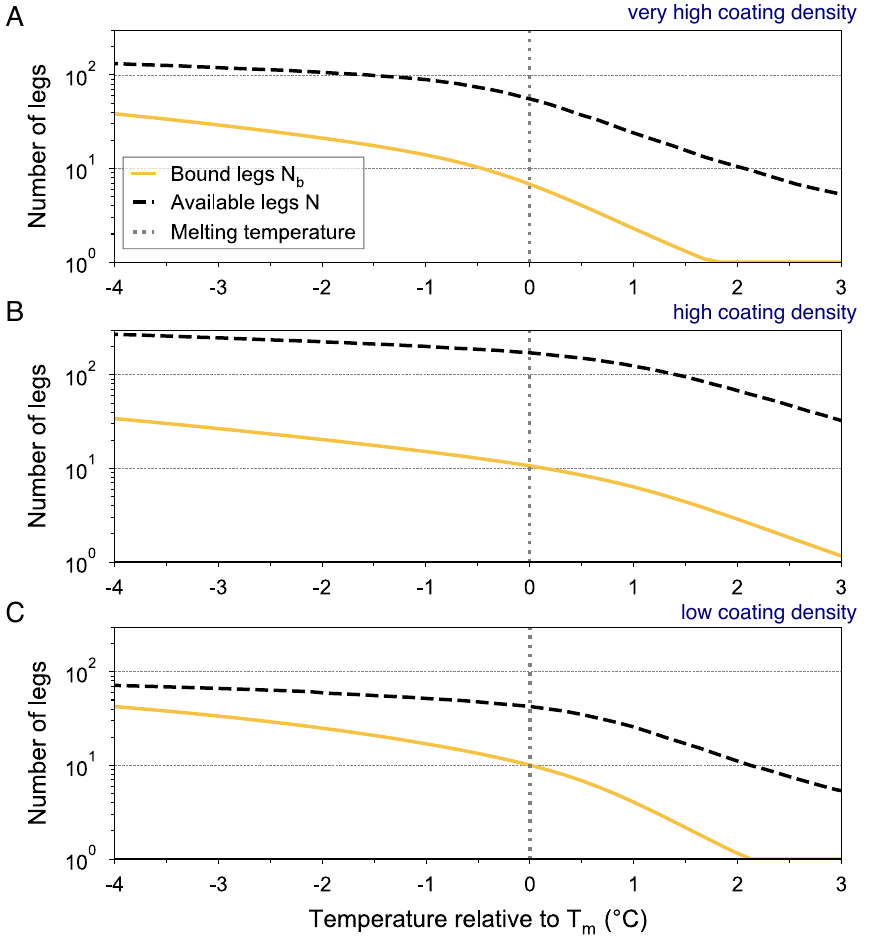}% Here is how to import EPS art
\caption{\label{fig:Nbonds} \textbf{Number of legs $N$ and average number of bonds $N_b$ involved in the binding process predicted from theory.} $N$ and $N_b$ are evaluated from detailed microscopic interactions for each system corresponding to detailed design parameters of DNA-coated colloids used (A) in this work (B) in Ref.~\cite{wang2015crystallization} and (C) in Ref.~\cite{xu2011subdiffusion}. }
\end{figure}

All parameters are precisely known in the experimental system, but one: the coating density. We can estimate it by looking at the thermodynamic unbound probability obtained from the Boltzmann distribution as $p_u \sim \int_0^{h_c} e^{-\beta \phi(h)}dh$, and comparing it to the similar quantity calculated experimentally. The experimental data shows that the unbound probability transitions around $T_m = 55-58^{\circ}~\mathrm{C}$ which corresponds to theoretical curves obtained with a surface density ranging from 1 DNA per $(9~\mathrm{nm})^2$ to $(12~\mathrm{nm})^2$. We therefore use 1 DNA per $(10.5~\mathrm{nm})^2$ as a center value and the extremal values to calculate a fidelity interval for $D_{\rm eff}$ (gray area in Fig.~6A). These obtained values are within the range of expected values~\cite{fan2021microscopic}. 

The average number of bonds $N_b$ (gold in Fig.~\ref{fig:Nbonds}) and the number of legs within reach $N$ (dashed black) with respect to temperature are then readily evaluated from the model leading to $\phi(h)$. The number of bonds at the melting temperature is only $N_b(T_m) \simeq 10$ while the number of available legs can be quite high $N(T_m) \simeq 100$. The number of bonds $N_b$ increases strongly with decreasing temperature, from $0$ to $40$ over the 4$^\circ$C window, thus potentially accounting for most of the decrease in diffusion.

\subsubsection{3D geometry}

In general the particle can explore positions not just in one dimension but in 3D. Here we discuss how to take into account this full geometry.

\paragraph{Second lateral dimension} The second lateral dimension is -- to some extent -- a trivial extension of the 1-lateral dimension model derived in the main manuscript. 
We consider the constitutive equations for the leg and the particle in 2 lateral dimensions ($x,y$), simplifying here to $\ell_0 = 0$, namely considering that the rest state of the tether lies right above the surface and that deformations are still quadratic in the leg extension. We have
\begin{eqnarray*}
    &\frac{d\bm{l}}{dt} = - \frac{k}{\gamma} |\bm{l}| \frac{\bm{l}}{|\bm{l}| } + \sqrt{2 \frac{k_B T}{\gamma}} \bm{\eta_l} \\ 
    &\frac{d\bm{x}}{dt} = + \sqrt{2\frac{k_B T}{\gamma}} \bm{\eta_x} .
\end{eqnarray*}
The leg extension $\bm{l}$ is readily projected on both coordinates:
$$
|\bm{l}| \frac{\bm{l}}{|\bm{l}| } = \bm{l} = l_x \bm{u}_x +   l_y \bm{u}_y 
$$
and similarly for the noise operators. We obtain
\begin{eqnarray*}
    &\frac{dl_i}{dt} = - \frac{k}{\gamma} l_i + \sqrt{2\frac{k_B T}{\gamma}} \eta_{l,i} \\ 
    &\frac{dx_i}{dt} = + \sqrt{2\frac{k_B T}{\gamma}} {\eta_{x,i}} .
\end{eqnarray*}
where $i = (x,y)$ refers to both lateral dimensions. The equations are fully uncoupled and hence it is not necessary to conduct further calculations to conclude that the effective long time motion should write as 
\begin{equation*}
    \frac{d\bm{x}}{dt} = + \sqrt{2D_{\rm eff}} \bm{\eta_x} .
\end{equation*}
where $D_{\rm eff}$ has the same expression as in the main manuscript. 

\paragraph{Vertical dimension}
The particle may also venture far from the surface, where binding is not possible. To account for this 3D geometry, we use an extension of our main model. 

One option to account for such a 2D dependence is to add a vertical degree of freedom say $z$ for the particle, together with spatially dependent rates $q_{\rm on}(z),q_{\rm off}(z)$. This is not a trivial modification, especially as there are different ways to set the spatial dependence $z$ of $q_{\rm on}(z),q_{\rm off}(z)$ (see for example the variability between Refs.~\cite{goodrich2018enhanced,noe2018,fogelson2019transport}). 

Instead we rely on a simplified geometrical approach, that has been shown to accurately reproduce a 2D geometry in another context~\cite{marbach2021intrinsic}, where we describe the system with 2 $\times$ 1D lines. For a 1-legged particle, we consider that the particle can switch between two regions where its dynamics are constrained to 1D: surface and bulk regions. The particle enters the surface region with rate $Q_{\rm on}$, and then can bind to the surface with rate $\on$. If the particle is \textit{unbound} in the surface region, it may lift off from the surface region to the bulk region with rate $Q_{\rm off}$. 
%Detailed balance requires $\frac{Q_{\rm on}}{Q_{\rm off}} = \frac{\pi_u}{\pi_V}$ where  $\pi_V$ is the equilibrium probability to be far from the surface. This ratio 
$\frac{Q_{\rm on}}{Q_{\rm off}}$ corresponds to the ratio of positions where the receptors are within and beyond reach. In DNA-coated colloid explorations, where particles are considered on top of a sticky surface, the ratio can be small or large depending on the density mismatch between the particle material and the surrounding fluid, that is otherwise described by the particle's gravitational height~\cite{xu2011subdiffusion}. For other systems, such as white blood cells that are confined within blood vessels, particles are always close to the wall~\cite{sun2003red} and hence the ratio is quite large.  Approach and lift-off from the surface are slow processes that scale like the diffusive dynamics of the particle and thus we may assume $Q_{i} \tau \sim O_{\epsilon}(1)$. 

Performing similar coarse-graining steps (see following paragraph), we obtain an effective friction 
\begin{equation}
    \frac{1}{\Gamma_{\rm eff}^{2\times 1D}} = \frac{p^{2\times 1D}_0}{\Gamma_0} + \frac{p^{2\times 1D}_1}{\Gamma_1} 
    \label{eq:gammaeff2d}
\end{equation}
where the probability to be in either states takes into account the added degree of freedom, $p^{2\times 1D}_1 = Q_{\rm on}q_{\rm on}/Z$ and $p^{2\times 1D}_0 = q_{\rm off}(Q_{\rm on} + Q_{\rm off})/Z$ with $Z$ a normalization constant such that $p^{2\times 1D}_0 + p^{2\times 1D}_1 = 1$. The added degree of freedom does not change the result of Eq.~\gammaeff, simply the mathematical interpretation of the probability factors. Note that this framework has been verified against numerical simulations. 

The values of $Q_{\rm on}$ and $Q_{\rm off}$ can be evaluated from the detailed interaction potential $\phi(h)$ of a DNA-coated colloid and the surface. In fact, the probability to be near the surface, in the absence of binding, is measured by
\begin{equation}
    \frac{Q_{\rm on}}{Q_{\rm on} + Q_{\rm off}} = \int_0^{h_p} e^{-\beta (\phi(h)-\phi_{\rm bind}(h))} dh/Z
\end{equation}
where $\phi_{\rm bind}(h)$ measures the contributions to the interaction potential due to binding, $h_p \simeq 20~\mathrm{nm}$ measures the typical width of attractive interactions (region of space where binding could happen) and $Z$ is a normalizing factor. For our DNA-coated colloids we find $\frac{Q_{\rm on}}{Q_{\rm on} + Q_{\rm off}}  \simeq 0.0015$ and that the ratio does not depend much on temperature. It also does not depend significantly on the exact value of $h_p$ for $h_p = 2 - 40~\mathrm{nm}$. 

For an $N$-legged caterpillar, the result generalizes to a change in the probability factors $p_n$ in Eq.~\gammaN for $D_{\rm eff}$. We have $p^{2\times 1D}_0 = q_{\rm off}^N(Q_{\rm on} + Q_{\rm off})/Z$ and $p^{2\times 1D}_n = \binom{N}{n} q_{\rm off}^{N-n} q_{\rm on}^n Q_{\rm on}/Z$ such that $Z = Q_{\rm on}(\on +\off)^N + Q_{\rm off}\off^N$.

\subsubsection{2$\times$1D, 1 legged nanocaterpillar model}

In this section we derive the effective 1-legged long term caterpillar dynamics in an effective ``2D'' geometry by using the 2$\times$1D mapping. The steps are carefully detailed so as to serve as an additional pedagogical explanation of the coarse-graining procedure introduced in the main text.

\paragraph{Constitutive equations of the 2$\times$1D, 1-legged caterpillar model}

Let $p(x,l,t) = (p_V, p_u, p_b)^T$ be the probability distribution function of finding the system at time $t$ in state $x,l$ far from the surface (V), or close to the surface with a bound (b) or an unbound (u) leg. It obeys the Schmoluchowski equation
\begin{equation}
    \partial_t p = \mathcal{L}^{\star} p \,\, \mathrm{with} \,\,\,  \mathcal{L}^{\star} = \mathcal{Q}^{\star} + \mathcal{U}^{\star}
    \label{eq:FPE}
\end{equation}
where $\mathcal{Q}^{\star}$ is the matrix of rates to going from one state to another
\begin{equation}
\mathcal{Q}^{\star} = \begin{pmatrix} - Q_{\rm on} & Q_{\rm off} & 0 \\ Q_{\rm on} & -Q_{\rm off} - q_{\rm on} & q_{\rm off} \\ 0  & q_{\rm on} &  - q_{\rm off} \end{pmatrix}
\end{equation}
and $\mathcal{U}^{\star}$ contains the dynamics in each state
\begin{equation}
    \mathcal{U}^{\star} = \mathrm{diag} \begin{pmatrix} \partial_l \left( \frac{k}{\gamma} (l-l_0) + \frac{k_B T}{\gamma} \partial_l\right) + \frac{k_B T}{\Gamma} \partial_{xx} \\
    \partial_l \left( \frac{k}{\gamma} (l-l_0) + \frac{k_B T}{\gamma} \partial_l\right) + \frac{k_B T}{\Gamma} \partial_{xx} \\
    (\partial_l -  \partial_x) \left( \frac{k}{\Gamma + \gamma} (l-l_0) + \frac{k_B T}{\Gamma + \gamma} (\partial_l -  \partial_x)  \right)
    \end{pmatrix}.
\end{equation}
Consistently, the equilibrium distribution $\pi = \frac{e^{-\beta k (l-l_0)^2/2}}{Z} \begin{pmatrix} \frac{Q^{\rm off}}{Q^{\rm on}}, & 1, & \frac{q_{\rm on}}{q_{\rm off}} \end{pmatrix}^T$ is indeed a stationary solution of Eq.~\eqref{eq:FPE}. Note that here $Q_{\rm on}$ and $Q_{\rm off}$ represent respectively the rates at which the particle approaches and leaves the vicinity of the surface, namely the region of space where binding is possible.

\paragraph{Non-dimensionalization}

Using the non-dimensional notation introduced in the main text allows to separate the Schmoluchowski operator $\mathcal{L}^{\star}$ in fast and slow operators. In the following, it will be somewhat easier to conduct the reasoning not on $\mathcal{L}^{\star}$ but on its adjoint $\mathcal{L}$, the generator of the system, defined such that for appropriate functions $f$, we have $\langle f , \mathcal{L}^{\star} p \rangle = \langle \mathcal{L} f , p \rangle$, where $\langle f , g \rangle = \iint (f_Vg_V + f_u g_u + f_b g_b) dl dx$ is the inner product. We therefore seek a solution $f$ of the dynamics
\begin{equation}
    \partial_t f = \mathcal{L} f = \left( \frac{1}{\epsilon^2} \mathcal{L}_0 + \frac{1}{\epsilon} \mathcal{L}_1 + \mathcal{L}_2 \right)
    \label{eq:Generator1}
\end{equation}
where
\begin{equation}
    \mathcal{L}_0 = \begin{pmatrix} \frac{\Gamma}{\gamma} ( - l \partial_ l + \partial_{ll} ) & 0 & 0 \\
    0 & - q_{\rm on} + \frac{\Gamma}{\gamma} ( - l \partial_ l + \partial_{ll} ) & q_{\rm on} \\
    0 & q_{\rm off} & - q_{\rm off} + \frac{\Gamma}{\Gamma + \gamma} ( - l \partial_ l + \partial_{ll} ) 
    \end{pmatrix},
\end{equation}
\begin{equation}
    \mathcal{L}_1= \mathrm{diag} \left(0 , 0,  \frac{\Gamma}{\Gamma + \gamma} \left( l \partial_x - 2 \partial_{lx} \right) \right)
\end{equation}
and 
\begin{equation}
    \mathcal{L}_2 = \begin{pmatrix} -Q_{\rm on} & Q_{\rm on} & 0 \\
     Q_{\rm off} & - Q_{\rm off} & 0 \\
     0 & 0 & 0 \end{pmatrix}+ \mathrm{diag}\left( 1,1,\frac{\Gamma}{\Gamma + \gamma}\right) \partial_{xx} .
\end{equation}

Additionally, $f$ has to satisfy boundary conditions of no flux at infinity
\begin{equation}
    \partial_l f(x,l,t)|_{l= \pm \infty} = 0
\end{equation}
which correspond to the usual no flux in probability space (where the flux in probability space satisfies $\left( lp + \partial_l p\right) |_{l= \pm \infty} = 0$). This condition is physical as it imposes conservation of probability. Note that we expect these boundary conditions to be satisfied only at lowest order in $\epsilon$. 

\paragraph{Homogenization}

We seek a solution to Eq.~\eqref{eq:Generator1} as an expansion in the small parameter $\epsilon$, as $f = f_0 + \epsilon f_1 + \epsilon f_2 + ...$. At lowest order we need to satisfy $\mathcal{L}_0 f_0 = 0$ which yields the general solution
\begin{equation}
    f_0 = \begin{pmatrix} a_V(x,t) \\ a_S(x,t) \\ a_S(x,t) \end{pmatrix} +  \int_0^l e^{y^2/2} dy \begin{pmatrix} b_V(x,t) \\ b_S(x,t) \\ b_S(x,t) \end{pmatrix}
\end{equation}
where $a_S, a_V, b_S$ and $b_V$ are all integration ``constants'' and $S$ and $V$ denote surface and volume terms.
With the boundary conditions on $f$ we get $b_S(x,t) = b_V(x,t) = 0$. Note that such boundary conditions also allow the cross product $\langle f_0 , \pi \rangle$ to remain finite which will be expected later to use the Fredholm alternative. 

The associated equilibrium distribution at lowest order $\pi_0$ spans a two dimensional space described by $(\pi_{0,V}(x,t),\pi_{0,S}(x,t))$ such that
\begin{equation}
    \pi_0 = \frac{1}{Z} \begin{pmatrix} \pi_{0,V}(x,t) \\ \pi_{0,S}(x,t)) \\ \pi_{0,S}(x,t)) \frac{q_{\rm on}}{q_{\rm off}} \end{pmatrix} e^{-l^2/2}.
\end{equation}
We therefore expect that our long time dynamics will consist in a $2\times2$ matrix describing the joint evolution of surface variables ($a_S(x,t)$) and volume variables ($a_V(x,t)$). 

At the next order we need to satisfy $\mathcal{L}_0 f_1 = -\mathcal{L}_1 f_0$. $f_1$ is the sum of a particular integral and a complementary function (\textit{i.e.} a function in the nullspace of $\mathcal{L}_0$). The complementary function can be taken to be 0 otherwise $f_1$ would contain terms that are redundant with $f_0$. One can check that the particular integral to this equation is simply 
\begin{equation}
    f_1 =\begin{pmatrix} 0 \\ \gamma q_{\rm on} \\ \Gamma + \gamma q_{\rm on}  \end{pmatrix} \frac{l \partial_x a}{\Gamma (1 + q_{\rm off}) + \gamma (q_{\rm on} + q_{\rm off}) }.
\end{equation}

At the following order we need to find a solution to $\mathcal{L}_0 f_2 = \partial_t f_0 -\mathcal{L}_2 f_0 - \mathcal{L}_1 f_1$. This equation has a solution if the right hand side terms of the equal sign satisfy the Fredholm alternative~\cite{pavliotis2008multiscale}, namely
\begin{equation}
    \langle \partial_t f_0 -\mathcal{L}_2 f_0 - \mathcal{L}_1 f_1 , \pi_0 \rangle = 0.
\end{equation}
As $\pi_0$ spans a 2D space described by $(\pi_{0,V}(x,t),\pi_{0,S}(x,t))$ we can evaluate the Fredholm alternative on an orthogonal basis of this space; specifically here we will investigate the Fredholm alternative on $(1,0)$ then $(0,1)$. On the volume space we have 
\begin{equation}
    \partial_t a_V = \partial_{xx} a_V - Q_{\rm on} a_V + Q_{\rm on} a_S.
\end{equation}
On the surface space the terms are more lengthy and we split them for readability
\begin{equation}
    \langle \partial_t f_0  , \pi_0 \rangle = \left( 1 + \frac{q_{\rm on}}{q_{\rm off}} \right) \partial_t a_S,
\end{equation}
\begin{equation}
    \langle \mathcal{L}_2 f_0  , \pi_0 \rangle = - Q_{\rm on} a_S + Q_{\rm on} a_V + \partial_{xx} a_S + \frac{q_{\rm on}}{q_{\rm off}} \partial_{xx} a
\end{equation}
and finally 
\begin{equation}
    \langle \mathcal{L}_1 f_1  , \pi_0 \rangle = - \frac{q_{\rm on}}{q_{\rm off}}
\frac{\Gamma + \gamma q_{\rm on}}{\Gamma (1 + q_{\rm off}) + \gamma (q_{\rm on} + q_{\rm off}) } \partial_{xx} a_S.
\end{equation}
Compiling all contributions on the surface we find
\begin{equation}
\begin{split}
     \partial_t a_S &= - Q_{\rm off} a_S + Q_{\rm off} a_V \\
    &+  \frac{q_{\rm off}}{q_{\rm off}+ q_{\rm on}}  \partial_{xx} a_S  + \frac{q_{\rm off}}{q_{\rm off}+ q_{\rm on}}  \frac{\Gamma}{\Gamma + \gamma  \frac{1 + q_{\rm off} + q_{\rm on} }{q_{\rm off}}} \partial_{xx} a_S.
\end{split}
\end{equation}

Overall we have found effective long time dynamics described by the generator (in dimensional scales)
\begin{equation}
    \mathcal{L}_{\rm eff} = \begin{pmatrix} - Q_{\rm on} + \frac{k_B T}{\Gamma} \partial_{xx} & Q_{\rm on} \\ Q_{\rm off}  \frac{q_{\rm off}}{q_{\rm on} + q_{\rm off}}& - Q_{\rm off} \frac{q_{\rm off}}{q_{\rm on} + q_{\rm off}} + \frac{k_B T}{\Gamma_{\rm eff}} \partial_{xx}  \end{pmatrix}, \,\, \text{such that} \,\, \partial_t a = \mathcal{L}_{\rm eff} a .
\end{equation}
Here $\Gamma_{\rm eff}^{-1} = p_0 \Gamma_0^{-1} + p_1 \Gamma_1^{-1}$ with $\Gamma_0 = \Gamma$, $\Gamma_1 = \Gamma + \gamma + k \left( \frac{1}{q_{\rm off}} + \frac{k}{\gamma}\frac{q_{\rm on}}{q_{\rm off}}\right)$, $p_0 = q_{\rm off}/(q_{\rm off} + q_{\rm on})$  is the probability to have no bond near the surface and $p_1 = 1 - p_0$ is the probability to have 1 bond near the surface. Note that the expression of $\Gamma_{\rm eff}$ is exactly that when focusing only on surface dynamics and discarding effective 2D dynamics (see Eq.~\gammaeff of the main paper). This shows that homogenization steps in the embedded 2$\times$1D geometry do not entangle with surface dynamics. To understand the meaning of this effective generator, we go one step further. 

\paragraph{Long (long) times}

We now wish to understand the long time dynamics of the generator $\mathcal{L}_{\rm eff}$. We search for long (long) time dynamics by using a non-dimensionalization that seeks even longer times as
\begin{equation}
    t \rightarrow \tilde{t} \frac{\tau}{\epsilon}, \,\, x \rightarrow \tilde{x} L_x
\end{equation}
where $\tau = L_y^2/D_0 = L_y^2 \Gamma/k_B T$ and $L_y/L_x = 1/\sqrt{\epsilon}$ corresponds to the far horizontal scales $x$ is going to explore (compared to the shorter vertical scales). $Q_{\rm on}$ and $Q_{\rm off}$ are typically associated with the time scale that the particle takes to diffuse vertically and therefore $Q_{\rm on} \tau \sim 1$ and likewise $Q_{\rm off} \tau \sim 1$. 
We obtain the non-dimensional generator
\begin{equation}
    \mathcal{L}_{\rm eff} = \frac{1}{\epsilon} \begin{pmatrix} - Q_{\rm on}  & Q_{\rm on} \\ Q_{\rm off}  \frac{q_{\rm off}}{q_{\rm on} + q_{\rm off}}& - Q_{\rm off} \frac{q_{\rm off}}{q_{\rm on} + q_{\rm off}}   \end{pmatrix} + \begin{pmatrix} \partial_{xx} & 0 \\ 0 &  \frac{\Gamma}{\Gamma_{\rm eff}} \partial_{xx} \end{pmatrix} = \frac{1}{\epsilon} \mathcal{L}_0 + \mathcal{L}_1
\end{equation}
and we search for a solution $f$ of the equation $\partial_t f =  \mathcal{L}_{\rm eff}  f$ expanded in $\epsilon$ as $f = f_0 + \epsilon f_1 + ...$. 

At lowest order we obtain from $\mathcal{L}_0 f_0 = 0$, $f_0 = a(x,t) \begin{pmatrix} 1 \\ 1 \end{pmatrix}$, with the associated equilibrium distribution \begin{equation}
    \pi_0 = \frac{1}{Z} \begin{pmatrix} Q_{\rm off}  \frac{q_{\rm off}}{q_{\rm on} + q_{\rm off}} \\ Q_{\rm on} \end{pmatrix}.
\end{equation}
At the next order we need to satisfy the Fredholm alternative, namely $\langle \partial_t f_0 - \mathcal{L}_1 f_0 , \pi_0 \rangle = 0$ leading to (back in dimensional scales)
\begin{equation}
    \partial_t a = \frac{k_BT}{\Gamma_{\rm eff}^{\rm 2\times 1D}} \partial_{xx} a
\end{equation}
where
\begin{equation}
   \frac{1}{  \Gamma_{\rm eff}^{\rm 2\times 1D} }=  \frac{Q_{\rm off} q_{\rm off}}{Q_{\rm off} q_{\rm off} + Q_{\rm on} \left(q_{\rm on} + q_{\rm off}\right)} \frac{1}{\Gamma} + \frac{Q_{\rm on} \left(q_{\rm on} + q_{\rm off}\right)}{Q_{\rm off} q_{\rm off} + Q_{\rm on} \left(q_{\rm on} + q_{\rm off}\right)} \frac{1}{\Gamma_{\rm eff}}.
\end{equation}
Expanding terms with the expression of $\Gamma_{\rm eff}$ and rearranging we can summarize the result in the explicit form, similarly as in Eq.~\gammaeff of the main paper, 
\begin{equation}
   \boxed{\frac{1}{  \Gamma_{\rm eff}^{\rm 2\times 1D} }=  \frac{p^{2\times 1D}_0}{\Gamma_0} + \frac{p^{2\times 1D}_1}{\Gamma_1}}
\end{equation}
where
\begin{equation}
    \boxed{p^{2\times 1D}_0 = \frac{(Q_{\rm off} + Q_{\rm on}) q_{\rm off}}{Z} \,\, \mathrm{and} \,\, p^{2\times 1D}_1 = \frac{Q_{\rm on}q_{\rm on}}{Z}}
\end{equation}
 are the probabilities to have respectively 0 and 1 bond, $Z = (Q_{\rm off} + Q_{\rm on}) q_{\rm off} + Q_{\rm on} q_{\rm on}$ and $\Gamma_0 = \Gamma$ is the friction in the unbound state and $\Gamma_1 = \Gamma + \gamma + k \left( \frac{1}{q_{\rm off}} + \frac{k}{\gamma}\frac{q_{\rm on}}{q_{\rm off}}\right)$ is that contributing to the bound state. A similar result for an $N$ legged caterpillar, simply adapting the probabilities, is thus used to quantify diffusion of DNA-coated colloids on surfaces.

To recover surface only dynamics, one simply has to take $Q_{\rm on}/Q_{\rm off} \rightarrow \infty$ in the above expression. In that case one can easily obtain the surface only effective friction Eq.~\gammaeff of the main text. 

\clearpage

\section{List of parameters for typical biological and artificial systems}

%Note that for DNA coated colloids, we expect the number of available legs to be smaller around the melting temperature, due to increased separation around the melting temperature~\cite{fan2021microscopic}. For the investigation of Fig.~5 of the main paper we therefore attribute only $50\%$ of the values for $N$ reported in Tables S1 and S2. Actually numbers are already right

%DNA kinetics may have binding rates that can be as fast as $10^5/s$, very strongly dependent on temperature~\cite{bonnet1998kinetics} and also can be non arrhenius like~\cite{wallace2001non}. There's also another paper by Ben rogers on that. 
% \cite{zoli2018end} shorter lengths expected for dsDNA, especially because of double helix. 

% I may have to rearrange all of this into multiple tables... 

\begin{table}[h!]
\footnotesize
%\begin{tabularx}{\textwidth}{lcc}
\begin{tabular}{p{0.15\textwidth}|>{\centering}p{0.15\textwidth}|>{\centering\arraybackslash}p{0.65\textwidth}}
\textbf{Quantity} & \textbf{Range of values} & \textbf{Details}  \\
\hline 
\hline 
\multicolumn{3}{l}{\textit{DNA coated colloids - low coverage large colloids}} \\
\hline
 $R$ & $525~\mathrm{nm}$ & Low coverage streptavidin beads of~\cite{xu2011subdiffusion}\\
 $\eta$ & $ 0.0006  \mathrm{Pa.s}$ & Viscosity of water at the melting temperature  44$~^{\circ}C$ \\
 $a$ & $10-15~\mathrm{nm}$ & estimated with for double stranded DNA, 60 nucleotides~\cite{zoli2018end} \\
 $k$ & $1 \times 10^{-4}~\mathrm{N/m}$ & $k \simeq 3\frac{k_B T}{2L l_p}$ where $l_p \simeq 3.4~\mathrm{nm}$ is the typical persistence length corresponding to 10 base pairs in a helix and $L \simeq 60 (l_p/10) \simeq 20~\mathrm{nm}$~\cite{xu2011subdiffusion} \\
 $\on$ & $2500~\mathrm{s^{-1}}$ & from $q_{\rm on} \sim k_{\rm on} \sigma /a \mathcal{N_A}$  with $k_{\rm on} = 2.2 \times 10^{6} M^{-1} s^{-1}$ (44.7$~^{\circ}C$) (value predicted from Ref.~\cite{zhang2018predicting} for the CCAAGTTATGA sequence used in \cite{xu2011subdiffusion}, measurements on sequences with a similar number of bases show slightly lower hybridization rates~\cite{xu2017real}) \\
 $\off$ & $25000~\mathrm{s^{-1}}$ & estimated unbinding rate around the melting temperature with a bound probability of $~10 \%$ (evaluated using full potential profile estimates following~\cite{fan2021microscopic}). Expected $\off\simeq \on$ at lower temperatures \\
 $\sigma$ & $1/(12 ~\mathrm{nm} - 18~\mathrm{nm})^2$& coating density~\cite{xu2011subdiffusion} \\
 $D$ & $1.4\times 10^{-3}~\mathrm{\mu m^2/s}$& at the melting temperature~\cite{xu2011subdiffusion} \\
 $D_0$ & $0.37~\mathrm{\mu m^2/s}$ & calculated as $D_0 = k_B T / 12 \pi \eta R $ with $T = 44\,^{\circ}$C (since close to the surface, longitudinal friction is doubled due to hydrodynamic interactions) \\
 $D_0/D$ & $270$ & diffusion decrease factor, calculated with range of above values \\
 $N$ & $40 - 70$ & estimated from $N \simeq 2 \pi R h \sigma$ taking $h \simeq a/3$ the typical penetration length of the layers, and density limited by the surface density, probably in the lower range $\sigma = 1/(16~\mathrm{nm})^2$ \\
 $N(T_m)$ & $40$ &  number of involved legs at the melting point, using the methodology described in Ref.~\cite{fan2021microscopic}. Taking symmetric (particle and flat surface) density with $\sigma = 1/(17~\mathrm{nm})^2$ and the exact DNA sequences and polymer types detailed in Ref.~\cite{xu2011subdiffusion} \\
 $D_0/D_{\rm hop}$ & $45-800$ & calculated with $D_{\rm hop} = D_0 \off^N/(\off+\on)^N$ with values above. \\
\hline 
\multicolumn{3}{l}{\textit{DNA coated colloids - high coverage large colloids}} \\
\hline 
$R$ & $500~\mathrm{nm}$ & High coverage beads of~\cite{fan2021microscopic}\\
 $\eta$ & $ 0.0005  \mathrm{Pa.s}$ & Viscosity of water at melting temperatures 50-60$~^{\circ}C$ \\
 $a$ & $13-18~\mathrm{nm}$ & measured lengths of $6.5$ k PEO strands tethered with 20 nucleotides of single stranded DNA~\cite{fan2021microscopic} \\
 $k$ & $2 \times 10^{-4}~\mathrm{N/m}$ & $k \simeq 3\frac{k_B T}{2L \ell_p}$ expected spring constant with $\ell_p \simeq 0.5~\mathrm{nm}$ and $L = 80~\mathrm{nm}$ (mixed brush with PEO and DNA)\\
 $\on$ & $1.3-1.9 \times 10^4~\mathrm{s^{-1}}$ & from $q_{\rm on} \sim k_{\rm on} \sigma /a \mathcal{N_A}$  with $k_{\rm on} = 1.6 \times 10^{6} M^{-1} s^{-1}$ (55$~^{\circ}C$) (value predicted from Ref.~\cite{zhang2018predicting} for the ACCGCA sequence used in \cite{fan2021microscopic}) \\
 $\off$ & $1.3-1.9 \times 10^5~\mathrm{s^{-1}}$ & estimated unbinding rate at the melting temperature with a bound probability of $~10 \%$ (evaluated using full potential profile estimates following~\cite{fan2021microscopic}). Expected $\off \simeq \on$ at lower temperatures \\
 $\sigma$ & $1/(3.27 ~\mathrm{nm})^2$& measured coating density~\cite{fan2021microscopic} \\
  $N$ & $140 - 190$ & estimated from $N \simeq 2 \pi R h \sigma$ taking $h \simeq a/3$ the typical penetration length of the layers, and density limited by the surface density, probably in the lower range $\sigma = 1/(10~\mathrm{nm})^2$ 
  \end{tabular}
%\end{tabularx}
\caption{\label{tab:parameters} Parameter values for DNA coated colloids (large colloids)}
\end{table}

\begin{table}[h!]
\footnotesize
%\begin{tabularx}{\textwidth}{lcc}
\begin{tabular}{p{0.15\textwidth}|>{\centering}p{0.15\textwidth}|>{\centering\arraybackslash}p{0.65\textwidth}}
\textbf{Quantity} & \textbf{Range of values} & \textbf{Details}  \\
\hline 
\hline 
\multicolumn{3}{l}{\textit{DNA coated nanoparticles}} \\
\hline
 $R$ & $7.5~\mathrm{nm}$ & Gold nanoparticles of Ref.~\cite{park2008dna}\\
 $\eta$ & $ 0.0006  \mathrm{Pa.s}$ & Viscosity of water at the melting temperature  44$~^{\circ}C$ \\
 $a$ & $10~\mathrm{nm}$ & Estimated in Ref.~\cite{park2008dna} \\
 $k$ & $2 \times 10^{-4}~\mathrm{N/m}$ & $k \simeq 3\frac{k_B T}{2L l_p}$ where $l_p \simeq 3.4~\mathrm{nm}$ is the typical persistence length corresponding to 10 base pairs in a helix and $L \simeq 30 (l_p/10) \simeq 10~\mathrm{nm}$~\cite{park2008dna} \\
 $\on$ & $2\times10^4~\mathrm{s^{-1}}$ & from $q_{\rm on} \sim k_{\rm on} \sigma /a \mathcal{N_A}$  with $k_{\rm on} = 1.0 \times 10^{6} M^{-1} s^{-1}$ (44$~^{\circ}C$) (value predicted from Ref.~\cite{zhang2018predicting} for the CGCG sequence used in \cite{park2008dna}) \\
 $\off$ & $2\times10^5~\mathrm{s^{-1}}$ & estimated unbinding rate around the melting temperature with a bound probability of $~10 \%$ (evaluated using full potential profile estimates following~\cite{fan2021microscopic}). Expected $\off\simeq \on$ at lower temperatures \\
 $\sigma$ & $1/(3~\mathrm{nm})^2$& from 80 strands/ 15 nm particle~\cite{hurst2006maximizing} \\
 $N$ & $15-20$ & estimated from $N \simeq 2 \pi R_{\rm eff} h \sigma_{\rm eff}$ taking $h \simeq a/3$ the typical penetration length of the layers, effective radius of coated nanocolloid $R_{\rm eff} = R + h$ and density at the outer layer $\sigma_{\rm eff} = \sigma R^2/R_{\rm eff}^2$ 
\end{tabular}
%\end{tabularx}
\caption{\label{tab:parameters2} Parameter values for DNA coated nanoparticles}
\end{table}

\begin{table}[h!]
\footnotesize
%\begin{tabularx}{\textwidth}{lcc}
\begin{tabular}{p{0.15\textwidth}|>{\centering}p{0.15\textwidth}|>{\centering\arraybackslash}p{0.65\textwidth}}
\textbf{Quantity} & \textbf{Range of values} & \textbf{Details}  \\
\hline 
\hline 
\multicolumn{3}{l}{\textit{Leukocyte adhesion mediated by P-selectin or L-selectin}} \\
\hline
 $R$ & $4.15~\mathrm{\mu m}$ & typical cell size~\cite{ting1993volume} \\
 $\eta$ & $ 0.001 \mathrm{Pa.s}$ & typical physiological conditions~\cite{chen2001selectin} \\
 $a$ & $300~\mathrm{nm}$ & typical microvillus length~\cite{shao1998static} \\
 $k$ & $4 \times 10^{-5} - 5 \times 10^{-3}~\mathrm{N/m}$ & typical spring constant  of microvilli~\cite{shao1998static} up to spring constant of the bond itself (excluding microvilli)~\cite{fritz1998force} \\
 $N_b$ & 1-2 & typical number of contacts~\cite{schwarz2004selectin,alon1995lifetime} \\
  $\sigma$ & $15-30~\mathrm{\mu m}^{-2}$& typical density of ligands~\cite{alon1997kinetics,alon1995lifetime} \\
  $N$ & $40 - 80$ & calculated from $N \simeq 2 \pi R h \sigma$ taking $h \simeq a/3$ the typical penetration length of the layers. Coherent with Ref.~\cite{korn2008dynamic}. \\ 
 \hline 
\multicolumn{3}{l}{\textit{P-selectin}} \\
\hline
 $\on$ & $4-300~\mathrm{s^{-1}}$ & from $q_{\rm on} \sim k_{\rm on} \sigma /a \mathcal{N_A}$ with measured binding rates $k_{\rm on} = (4 - 100) \times 10^5~\mathrm{M^{-1}s^{-1}}$~\cite{fritz1998force,mehta1998affinity,alon1995lifetime} \\
 $\off$ & $0.02-1.6~\mathrm{s^{-1}}$ & measured unbinding rate~\cite{fritz1998force,mehta1998affinity,chen2001selectin} \\
 \hline 
\multicolumn{3}{l}{\textit{L-selectin}} \\
\hline
$\on$ & $(0.4 - 4)\times 10^4~\mathrm{s}^{-1}$ &  measured binding rates~\cite{schwarz2004selectin} \\
 $\off$ & $7 - 250~\mathrm{s^{-1}}$ & measured unbinding rate~\cite{dwir2003avidity,alon1997kinetics}, shorter lifetime than P-selectin
\end{tabular}
%\end{tabularx}
\caption{\label{tab:parameters3} Parameter values for white blood cells}
\end{table}

\begin{table}[h!]
\footnotesize
%\begin{tabularx}{\textwidth}{lcc}
\begin{tabular}{p{0.15\textwidth}|>{\centering}p{0.15\textwidth}|>{\centering\arraybackslash}p{0.65\textwidth}}
\textbf{Quantity} & \textbf{Range of values} & \textbf{Details}  \\
\hline 
\hline 
\multicolumn{3}{l}{\textit{E. Coli motility mediated by adhesion between FimH adhesin at the tip of pili and glycoproteins on the surface}} \\
\hline
 $R$ & $350~\mathrm{nm}$ and $L = 3~\mathrm{\mu m}$ & cylindrical features~\cite{bakshi2012superresolution} \\
 $\eta$ & $ 0.001 \mathrm{Pa.s}$ & typical physiological conditions~\cite{chen2001selectin} \\
 $a$ & $100~\mathrm{nm}$ & pili characteristic size to calculate hydrodynamic resistance, from $\sqrt{2 L_c \ell}$, see below~\cite{miller2006mechanical} \\
 $k$ & $0.6~\mathrm{\mu N/m}$ & typical spring constant as $k_B T /L_c \ell$ where $L_c = 2~\mathrm{\mu m}$ is the contour length and $\ell = 3.3~\mathrm{nm}$ is the persistence length \cite{miller2006mechanical} \\
 $\on$ & $5-125~\mathrm{s^{-1}}$ & from $q_{\rm on} \sim k_{\rm on} \sigma /a \mathcal{N_A}$ with measured binding rates $k_{\rm on} = 2-50\times 10^5~\mathrm{M^{-1}s^{-1}}$~\cite{yakovenko2015inactive,sauer2018binding,sauer2016catch} \\
 $\off$ & $1-100~\mathrm{s^{-1}}$ & measured unbinding rate~\cite{yakovenko2015inactive,sauer2018binding,sauer2016catch} \\
 $\sigma$ & $15-45~\mathrm{\mu m}^{-2}$ & Pili density, calculated from 100-300 pili over the cylinder surface~\cite{neidhardt1990physiology} \\
  $N$ & $2 - 7$ & calculated from $N \simeq 2 r L \sigma$ taking $r \simeq \sqrt{2aR/3}$ the typical half width of the cylinder in contact. 
\end{tabular}
%\end{tabularx}
\caption{\label{tab:parameters4} Parameter values for Escherichia Coli}
\end{table}

\begin{table}[h!]
\footnotesize
%\begin{tabularx}{\textwidth}{lcc}
\begin{tabular}{p{0.15\textwidth}|>{\centering}p{0.15\textwidth}|>{\centering\arraybackslash}p{0.65\textwidth}}
\textbf{Quantity} & \textbf{Range of values} & \textbf{Details}  \\
\hline 
\hline 
\multicolumn{3}{l}{\textit{Cargo transport by molecular motors}} \\
\hline
 $R$ & $1~\mathrm{\mu m}$ & typical cargo size~\cite{korn2009stochastic} \\
 $\eta$ & $ 0.001 \mathrm{Pa.s}$ & typical physiological conditions~\cite{chen2001selectin} \\
 $a$ & $25~\mathrm{nm}$ & microtubule diameter~\cite{gibbons2001dynamical} \\
 $k$ & $0.2-0.5~\mathrm{mN/m}$ & typical spring constant~\cite{gibbons2001dynamical,duke1996motor} \\
 $\on$ & $0.4~\mathrm{s^{-1}}$ & measured individual binding rate~\cite{vale1996direct} \\
 $\off$ & $4~\mathrm{s^{-1}}$ & measured unbinding rate~\cite{leduc2004cooperative}
\end{tabular}
%\end{tabularx}
\caption{\label{tab:parameters5} Parameter values for molecular motors}
\end{table}

%\cite{ziebert2021influenza} typical parameters for influenza are here. I would need to get more. typically this one can't slide for sure because it's super stuck because it's damn slow. like the motors. they have more references for the rates etc.  -numbers kept are coherent with whatever is in this work

\begin{table}[h!]
\footnotesize
%\begin{tabularx}{\textwidth}{lcc}
\begin{tabular}{p{0.15\textwidth}|>{\centering}p{0.15\textwidth}|>{\centering\arraybackslash}p{0.65\textwidth}}
\textbf{Quantity} & \textbf{Range of values} & \textbf{Details}  \\
\hline 
\hline 
\multicolumn{3}{l}{\textit{Protein cargos in the nuclear pore complex}} \\
\hline
 $R$ & $50~\mathrm{nm}$ & estimated from $D = 4~\mathrm{\mu m^2/s}$ for karyopherin-$\beta$ which is a major transporter in the NPC~\cite{wu2009intracellular,aramburu2017floppy}, $R = k_B T/6\pi \eta D$, and using $\eta$ as below \\
 $\eta$ & $ 0.001 \mathrm{Pa.s}$ & typical physiological conditions~\cite{chen2001selectin} \\
 $a$ & $8-10~\mathrm{nm}$ & estimated with worm like chain model in~\cite{lim2006flexible} \\
 $k$ & 0.06-0.1$~\mathrm{mN/m}$ & typical spring constant of the Nucleoporin Nup 153~\cite{lim2006flexible} \\
 $\on$ & $3\times 10^5 - 3 \times 10^8~\mathrm{s^{-1}}$ & from $q_{\rm on} \sim k_{\rm on} \rho$ with $k_{\rm on} =  10^7-10^9~\mathrm{M^{-1}s^{-1}}$~\cite{aramburu2017floppy} \\
 $\off$ & $10^4 - 10^7~\mathrm{s^{-1}}$ & estimated unbinding rate~\cite{aramburu2017floppy} \\
 $\rho$ & $30 - 250 \mathrm{mM}$ & concentration of Nup in pore~\cite{aramburu2017floppy,tu2013large}
\end{tabular}
%\end{tabularx}
\caption{\label{tab:parameters6} Parameter values for nuclear pore transport}
\end{table}

\begin{table}[h!]
\footnotesize
%\begin{tabularx}{\textwidth}{lcc}
\begin{tabular}{p{0.15\textwidth}|>{\centering}p{0.15\textwidth}|>{\centering\arraybackslash}p{0.65\textwidth}}
\textbf{Quantity} & \textbf{Range of values} & \textbf{Details}  \\
\hline 
\hline 
\multicolumn{3}{l}{\textit{Influenza A}} \\
\hline
 $R$ & $60~\mathrm{nm}$ & average measured diameter~\cite{harris2006influenza} \\
 $\eta$ & $ 0.001 \mathrm{Pa.s}$ & typical physiological conditions~\cite{chen2001selectin} \\
 $a$ & $12-15~\mathrm{nm}$ & typical height of Hemagglutinin~\cite{reiter2019force,harris2006influenza} \\
 $k$ & $1-2~\mathrm{mN/m}$ & typical spring constant~\cite{reiter2019force,sieben2012influenza} \\ %40pN/20nm %20pN/20nm\\ 
 $\on$ & $0.07-130$~$\mathrm{s^{-1}}$ & calculated from  $q_{\rm on} \sim q_{\rm off} / K_D \sigma /a \mathcal{N_A}$ with $K_D = 2-950~\mathrm{mM}$~\cite{sauter1989hemagglutinins,reiter2019force} \\
 $\off$ & $0.1-30~\mathrm{s^{-1}}$ & measured unbinding rate (focused on Hemagglutinin to Sialic Acid)~\cite{reiter2019force,muller2019mobility,sieben2012influenza,sakai2017influenza} \\
 $x = \frac{\on}{\on + \off}$ & 0.2 & Bound fraction at equilibrium~\cite{reiter2019force} \\
 $\sigma$ & $6800~\mathrm{\mu m^{-2}}$ & average measured coverage of HA proteins on virus ($~85\%$ of total (HA + NA))~\cite{harris2006influenza} \\
 $D$ & $0.05-0.5~\mathrm{\mu m^2/s}$ & measured on typical lipid bilayers with sialic acid receptors~\cite{muller2019mobility}, strongly dependent on $\off$ in the same qualitative way as predicted in out theory \\
 $D$ &$0.01~\mathrm{\mu m^2/s}$ & also measured on typical surfaces~\cite{sakai2017influenza}\\
 $D_0$ & $1.9~\mathrm{\mu m^2/s}$ & calculated as $D_0 = k_B T / 12 \pi \eta R $ with $T = 37\,^{\circ}$C (since close to the surface, longitudinal friction is doubled due to hydrodynamic interactions) \\
 $D_0/D$ & $ 4 - 190 $ & diffusion decrease factor, calculated with range of above values \\
 $N$ & $10 - 13$ & calculated from $N \simeq 2 \pi R h \sigma$ taking $h \simeq a/3$ the typical penetration length of the layers. \\
 $D_0/D_{\rm hop}$ & $9 - 20$ & calculated with $D_{\rm hop} = D_0 \off^N/(\off+\on)^N$ with values above. \\
 \hline 
\multicolumn{3}{l}{\textit{SARS CoV 2}} \\
\hline
 $R$ & $50~\mathrm{nm}$ & typical virus size~\cite{laue2021morphometry} \\
 $\eta$ & $ 0.001 \mathrm{Pa.s}$ & typical physiological conditions~\cite{chen2001selectin} \\
 $a$ & $6-23~\mathrm{nm}$ & ligand protein characteristic size~\cite{yang2020molecular,laue2021morphometry} \\
 $k$ & $0.1 - 0.4~\mathrm{N/m}$ & typical spring constant from simulation results~\cite{cao2021biomechanical} and \cite{ponga2020quantifying} \\ %400pN/1nm and 100pN/1nm \\
 $\on$ & $770 - 1500 ~\mathrm{s^{-1}}$ & from $q_{\rm on} \sim k_{\rm on} \sigma /a \mathcal{N_A}$ with $k_{\rm on} = 0.7 - 1.4\times 10^5~\mathrm{M^{-1}s^{-1}}$~\cite{yang2020molecular,walls2020structure} \\ % not retained from walls are off rates for the full SARS virus which do not correspond to off rates of the individual
 $\off$ & $0.05~\mathrm{s^{-1}} - 0.01~\mathrm{s^{-1}}$ & measured unbinding rate of individual ligand-receptor pairs~\cite{cao2021biomechanical,yang2020molecular} \\  
  $N$ & $1 - 2$ & calculated from $N \simeq 2 \pi R h \sigma$ taking $h \simeq a/3$ the typical penetration length of the layers. \\
 $ x$ & 0.1-0.2 & Bound fraction at equilibrium~\cite{yang2020molecular} in AFM experiments\\
 $\sigma$ & $1000~\mathrm{\mu m^{-2}}$ & density of spike proteins on the surface~\cite{laue2021morphometry} \\
 \hline 
\multicolumn{3}{l}{\textit{SARS CoV 1}} \\
\hline
 $R$ & $50~\mathrm{nm}$ & typical virus size~\cite{laue2021morphometry} \\
 $\eta$ & $ 0.001 \mathrm{Pa.s}$ & typical physiological conditions~\cite{chen2001selectin} \\
 $a$ & $6-23~\mathrm{nm}$ & ligand protein characteristic size~\cite{yang2020molecular,laue2021morphometry} \\
 $k$ & $0.6~\mathrm{N/m}$ & typical spring constant from simulation results~\cite{cao2021biomechanical} \\ %400pN/1nm \\
 $\on$ & $1200 - 2500 ~\mathrm{s^{-1}}$ & from $q_{\rm on} \sim k_{\rm on} \sigma /a \mathcal{N_A}$ with $k_{\rm on} = 0.7-1.4 \times 10^5~\mathrm{M^{-1}s^{-1}}$~\cite{yang2020molecular,walls2020structure} (similar to Sars CoV 2~\cite{walls2020structure}) \\ %~\cite{wu2011virus}  \\
 $\off$ & $0.6~\mathrm{s^{-1}}$ & measured unbinding rate of individual ligand-receptor pairs~\cite{cao2021biomechanical} \\
 $N$ & $1 - 4$ & calculated from $N \simeq 2 \pi R h \sigma$ taking $h \simeq a/3$ the typical penetration length of the layers. \\
 $ x$ & 0.1-0.2 & Bound fraction at equilibrium is similar to Sars CoV 2~\cite{yang2020molecular,walls2020structure} \\
 $\sigma$ & $1600~\mathrm{\mu m^{-2}}$ & density of spike proteins on the surface~\cite{laue2021morphometry}

\end{tabular}
%\end{tabularx}
\caption{\label{tab:parameters7} Parameter values for viruses}
\end{table}

\clearpage

\section{Coarse-graining under different models and assumptions}

\subsection{Coarse-graining with particle inertia}
\label{sec:inertia}

While more details on inertial effects with multivalent receptor contacts will be written in a separate paper~\cite{marbach2021inertial}, here we briefly recapitulate some results of this work to support claims made in the main manuscript. 

\subsubsection{Equations set up with particle inertia}

We consider now that the particle has inertia, described by a mass $m$. To simplify derivations we can neglect inertial effects from the legs, as in general the legs are much smaller than the particle itself, and hence have much lower mass. Alternatively, in Ref.~\cite{marbach2021inertial} we will show that one can start with inertia on all components, and take the limit of small mass of the legs relative to particle mass on the final result, and obtain the same result as if the limits were inverted. 

We thus write the unbound equations for a particle with a single leg as
\begin{equation}
    \begin{cases}  \frac{dl}{dt} &=  - \frac{k}{\gamma} l + \sqrt{2 \frac{k_B T}{\gamma}} \eta_l  \\   \frac{dx}{dt} &= v \\     \frac{dv}{dt} &= \frac{1}{m} \left( - \Gamma v + \sqrt{2 k_B T \Gamma } \eta_x \right)  \end{cases}
\end{equation}
where $m$ is the mass of the particle and $v$ the velocity of the particle. 

When the leg is bound to the surface, it is not necessary to project the dynamics. Writing Newton's second law on the system of the (particle+leg bound to surface) one finds the bound equations
\begin{equation}
    \begin{cases}  \frac{dl}{dt} &= v \\  \frac{dx}{dt} &= v \\    \frac{dv}{dt} &=  \frac{1}{m} \bigg[ - \Gamma v + \sqrt{ 2 k_B T \Gamma } \eta_x  + \left( - \gamma v - k l + \sqrt{2 k_B T \gamma} \eta_l \right) \bigg]  \end{cases}.
\end{equation}

\subsubsection{Possible resolution with particle inertia following Ref.~\cite{lee2018modeling}}

The generator for the system is 
\begin{equation}
\mathcal{L} = \begin{pmatrix}
- q_{\rm on}  - \frac{k}{\gamma} l \partial_l + \frac{k_B T}{\gamma} \partial_{ll} +  v \partial_x - \frac{\Gamma}{m} v \partial_v + \frac{k_B T \Gamma}{m^2} \partial_{vv}    & + q_{\rm on} \\
+q_{\rm off} & -q_{\rm off}  +  v \partial_x + v \partial_l - \frac{k}{m}l \partial_v - \frac{\Gamma+\gamma}{m} v \partial_v + \frac{k_B T (\Gamma+\gamma)}{m^2} \partial_{vv} 
\end{pmatrix}
\label{eq:GeneratorInertia}
\end{equation}
with a natural stationary distribution (now including a Boltzmann factor corresponding to the kinetic energy of the particle)
\begin{equation}
\pi = \frac{1}{Z} \begin{pmatrix}
q_{\rm off}/q_{\rm on} \\ 1 
\end{pmatrix} e^{-kl^2/2k_B T} e^{-m v^2/2k_B T}.
\end{equation}
In addition to non-dimensionalizing space and time we need to non-dimensionalize the velocity. We then take (on top of usual non-dimensional quantities in the paper, reported here for completeness)
\begin{equation}
   x \rightarrow L_x \tilde{x}, \,\, l \rightarrow L \tilde{l}, \,\,  t \rightarrow \tau \tilde{t},  v \rightarrow \tilde{v} L_x/\tau  = \tilde{v} \frac{L}{\epsilon} \frac{\epsilon^2 k}{\Gamma}.
\end{equation}
Mass also needs to be non-dimensionalized. Here we write, following Ref.~\cite{lee2018modeling}, $m = \tilde{m} L k \tau^2 / L_x$. The dimensionless number $m L_x/(L k \tau) 1/\tau = \tau_v /\tau$ can be interpreted as the ratio of the correlation time of the velocity $\tau_v$ to the time scale of observation $\tau$. We require $\tau/\tau_v = \frac{1}{\tilde{m} \epsilon}$ such that we may observe coarse grained dynamics. Dropping the $\tilde{\cdot}$ notation we find the non-dimensional generator
\begin{equation}
\begin{split} \mathcal{L} = \frac{1}{\epsilon^2} \mathcal{L}_0 + \frac{1}{\epsilon} \mathcal{L}_1 +  \mathcal{L}_2 =  &  \frac{1}{\epsilon^2}  \begin{pmatrix}
- q_{\rm on}  - \frac{\Gamma}{\gamma} l \partial_l + \frac{\Gamma}{\gamma} \partial_{ll}   & + q_{\rm on} \\
+q_{\rm off} & -q_{\rm off}  
\end{pmatrix} + \frac{1}{\epsilon}  \begin{pmatrix}
0   & 0 \\
0 &  v \partial_l - \frac{1}{m} l \partial_v 
\end{pmatrix} \\
&+  \begin{pmatrix}
 v \partial_x - \frac{1}{m} v \partial_v + \frac{1}{m^2} \partial_{vv}    &  0\\
0 &   v \partial_x - \frac{\Gamma+\gamma}{\Gamma}\frac{1}{m} v \partial_v + \frac{\Gamma+\gamma}{\Gamma}\frac{1}{m^2} \partial_{vv} 
\end{pmatrix} 
\end{split}
\end{equation}
We can then set up a similar step by step search of a solution at multiple orders seeking a solution $f = f_0 + \epsilon f_1 + \epsilon f_2 + ...$

At lowest order solving $\mathcal{L}_0 f_0 = 0$ yields simply $f_0 = a(x,t) \begin{pmatrix} 1 \\1 \end{pmatrix}$ and the associated equilibrium distribution $\pi_0 = \frac{1}{Z'} \begin{pmatrix}
q_{\rm off}/q_{\rm on} \\ 1 
\end{pmatrix} e^{-l^2/2}$. 

At the next order we need to solve $\mathcal{L}_0 f_1 = \frac{l}{m} \partial_v a \begin{pmatrix} 0 \\1 \end{pmatrix}$ that is easily shown to yield
\begin{equation}
f_1 = - \frac{1}{1 + \frac{(\gamma + \Gamma)q_{\rm off}}{\gamma q_{\rm on}+ \Gamma}} \begin{pmatrix} \frac{q_{\rm off}}{q_{\rm on} + \Gamma/\gamma}\end{pmatrix} \frac{l \partial_v a}{m}.   
\end{equation}

To find a solution at the following order, we need to satisfy the Fredholm alternative $\langle \partial_t f_0 - \mathcal{L}_2 f_0 - \mathcal{L}_1 f_1 , \pi_0 \rangle = 0$. Standard algebra yields an equation for the function $a(x,v,t)$ as (back in dimensional scales)
\begin{equation}
\partial_t a =  v \partial_x a - \frac{\Gamma_{\rm eff}^{m}}{m} v \partial_v a + \frac{k_B T \Gamma_{\rm eff}^{m}}{m^2} \partial_{vv} a
\label{eq:LangevinNotOverdamped}
\end{equation}
which corresponds to an inertial motion with friction
\begin{equation}
\boxed{\Gamma_{\rm eff}^{m}= \frac{q_{\rm off}}{q_{\rm on} + q_{\rm off}} \Gamma  + \frac{q_{\rm on}}{q_{\rm on} + q_{\rm off}} \left( \Gamma + \gamma + k \left( \frac{1}{q_{\rm off}}  + \frac{\gamma}{k} \frac{q_{\rm on}}{q_{\rm off}} \right) \right) }
\label{eq:inertiaEffWrongLimit}
\end{equation}
which writes with the notations of the main paper
\begin{equation}
    \boxed{\Gamma_{\rm eff}^{m}= p_0 \Gamma_0 + p_1 \Gamma_1}
\end{equation}
which is exactly Eq.~\gammainf of the main paper. 

%\subsubsection{Critical point: scalings for homogenization to go from inertial to overdamped dynamics}

As highlighted in the main paper, there is a notable difference between $\Gamma_{\rm eff}^{m} = p_0 \Gamma_0 + p_1 \Gamma_1$, with inertia, and $\Gamma_{\rm eff}^{-1} = p_0\Gamma_0^{-1} + p_1 \Gamma_1^{-1}$ when inertia is neglected. In particular, the results are not equivalent when  unbinding rates are slow such as $q_{\rm off} \Gamma/k \ll 1$. We will reconcile these results in a separate paper~\cite{marbach2021inertial}.

\subsection{Choice of time-scale hierarchy}

\subsubsection{Averaging with a different choice of scaling $\varepsilon = \gamma/\Gamma$}

It is common to assume a different choice of scalings assuming fast unbound tether dynamics. This choice of assumptions can be formulated mathematically as $\gamma/\Gamma = \gamma_r \epsilon^2$, where $\gamma_r$ is a non-dimensional number of order 1. This typically corresponds to short legs on a large particle, as $\gamma$ and $\Gamma$ are expected to scale with leg size and particle size via Stokes law. When doing such a reasoning, it is also common to lighten the assumption on scale separation for $x$ and $l$ and take $L = L_x$~\cite{fogelson2018enhanced}. We keep other non-dimensional scalings. For simplicity we will write $\varepsilon = \epsilon^2$ as no terms in $\epsilon$ appear now. We then obtain the non-dimensional generator as an expansion $\mathcal{L} =\frac{1}{\varepsilon} \mathcal{L}_0 + \mathcal{L}_1 + \varepsilon  \mathcal{L}_2 + \varepsilon^2  \mathcal{L}_3 + ...$, with the first terms as
\begin{equation}
\mathcal{L} =\frac{1}{\varepsilon} \begin{pmatrix}
-q_{\rm on} - \frac{1}{\gamma_r} l  \partial_l + \frac{1}{\gamma_r}  \partial_{ll}   & q_{\rm on} \\
q_{\rm off} & -q_{\rm off} \end{pmatrix} +  \begin{pmatrix}  \partial_{xx} & 0 \\ 0 &  l ( \partial_x - \partial_l) + (\partial_{x} - \partial_l)^2
\end{pmatrix} + \varepsilon \begin{pmatrix}  0 & 0 \\ 0 &  - \gamma_r l ( \partial_x - \partial_l) - \gamma_r  (\partial_{x} - \partial_l)^2
\end{pmatrix} + ... 
\end{equation}
Notice above that the non-dimensionalization for the binding rates $q_{\rm on}, q_{\rm off}$ is such that it assumes binding and unbinding to be much faster than the long time dynamics searched for. 

We look for a solution as $f = f_0 + \varepsilon f_1 + \varepsilon^2 f_2 + ...$ At first order we find easily $f_0 = a(x,t) \begin{pmatrix} 1 \\ 1 \end{pmatrix}$ associated with the equilibrium distribution $\pi_0 \propto \begin{pmatrix} q_{\rm off}/q_{\rm on} \\ 1 \end{pmatrix}e^{-l^2/2}$.

At the following order, to find a solution $f_1$, we require the Fredholm alternative, namely $ \langle \partial_t f_0 - \mathcal{L}_1 f_0 , \pi_0 \rangle = 0$, yielding
\begin{equation}
    \partial_t a - \partial_{xx} a = 0.
\end{equation}
We can now solve for $\mathcal{L}_0 f_1 = -  \mathcal{L}_1 f_0 + \partial_t f_0$ making use of this first order equation on $a$. The equation to be solved simplifies to $\mathcal{L}_0 f_1 = - \begin{pmatrix} 1 \\ 0 \end{pmatrix} l \partial_x a$. This gives
\begin{equation}
    f_1 = \frac{l \partial_x a}{q_{\rm off}} \begin{pmatrix} \gamma_r q_{\rm on } \\ 1 + \gamma_r q_{\rm on} \end{pmatrix}.    
\end{equation}

To solve for $f_2$ we require the Fredholm alternative at the following order, namely  $ \langle \partial_t f_0 + \varepsilon \partial_t f_1 - \mathcal{L}_1 f_0 - \varepsilon \mathcal{L}_1 f_1 - \varepsilon \mathcal{L}_2 f_0 , \pi_0 \rangle = 0$. We focus on specific terms
\begin{equation}
    \langle \partial_t f_1 , \pi_0 \rangle = 0,
\end{equation}
then 
\begin{equation}
    \langle - \mathcal{L}_1 f_1 , \pi_0 \rangle = + \partial_{xx}a \frac{\gamma_r q_{\rm on} + 1}{q_{\rm off}},
\end{equation}
and
\begin{equation}
    \langle - \mathcal{L}_2 f_0 , \pi_0 \rangle = + \gamma_r \partial_{xx}a
\end{equation}
such that summing up all contributions and reverting to original dimensions we obtain
\begin{equation}
    \partial_t a = \frac{k_B T}{\Gamma_{\rm eff}^{\gamma/\Gamma = \varepsilon}} \partial_{xx} a
\end{equation}
with
\begin{equation}
    \boxed{\frac{1}{\Gamma_{\rm eff}^{\gamma/\Gamma = \varepsilon} } = \left( \frac{q_{\rm off}}{q_{\rm on} + q_{\rm off}} \right) \frac{1}{\Gamma}  + \left( \frac{q_{\rm on}}{q_{\rm on} + q_{\rm off}} \right) \frac{1}{\Gamma} \left( 1 - \frac{\gamma}{\Gamma} \left[ 1 + \frac{k}{\gamma} \left( \frac{1}{q_{\rm off}} + \frac{\gamma}{k} \frac{q_{\rm on}}{q_{\rm off}} \right) \right] \right)}
\end{equation}
with is exactly Eq.~\gammaeps of the main paper. We note that this is exactly the $\gamma/\Gamma$ first order Taylor expansion of the equation obtained without assuming $\gamma/\Gamma \ll 1$, namely of Eq.~\gammaeff of the main paper. 

\subsubsection{Averaging with pre-averaging of tether dynamics (fast tether relaxation dynamics compared to all other dynamics)}

\paragraph{Equation set up with pre-averaging and resolution}

A commonly used framework is to assume that unbound leg dynamics are so fast that essentially when a new bond is created, the leg length may be sampled from its (bare) equilibrium distribution. This may be formally obtained using homogenization as well by assuming unbound relaxation is very fast compared to binding dynamics,  $\gamma/k \ll 1/q_{\rm on/ off}$, though we do not report the details here. It is a commonly used framework~\cite{fogelson2018enhanced,jana2019translational}.

The unbound state is described by the variables $(x,t)$ while the bound state is described with $(x,l,t)$. We write the equations for the probability distributions in each state
\begin{align}
\partial_t p_u &= - p_u \int e^{-kl^2/2k_B T} q_{\rm on} dl + \int q_{\rm off}   p_b(x,l,t) dl + \frac{k_B T}{\Gamma} \partial_{xx} p_u  \\
\partial_t p_b &=  + p_u e^{-kl^2/2k_B T} q_{\rm on} +  q_{\rm off}   p_b(x,l,t)  + \frac{k_B T}{\Gamma + \gamma} \partial_{xx} p_b + (...)
\label{eq:preaveraged}
\end{align}
where the $(...)$ denote the rest of the bound projected dynamics and $Z$ is some normalization constant that does not depend on $l$. Notice that here we kept the $\Gamma + \gamma$ in the bound state to highlight that such a term would have to be kept in the case of a great number of springs $N$, as this would become $\Gamma + N\gamma$ and would therefore have to remain. The equilibrium distribution associated with these dynamics is simply $\pi = \frac{1}{Z}\begin{pmatrix} \frac{\off}{\on} \\ e^{-\beta k l^2} \end{pmatrix}$ and satisfies detailed balance: 
\begin{equation}
    \pi_u \times q_{\rm on} e^{-\beta kl^2/2} = \frac{1}{Z} \frac{\off}{\on}  \times q_{\rm on} e^{-\beta kl^2/2} = \frac{e^{-\beta kl^2/2}}{Z} \times \off = \pi_b \times \off.
\end{equation}
% \begin{center}
% \begin{tikzcd}
% \pi_u \sim \frac{q_{\rm off}}{q_{\rm on}} \arrow[r, bend left = 50, "q_{\rm on} e^{-\beta kl^2/2}"] & \arrow[l, bend left = 50, "q_{\rm off}"]  \pi_b \sim e^{-\beta k l^2 } .
% \end{tikzcd}
% \end{center}

With this approach (compared to the main paper derivation), the only part of the generator that changes is the lowest order term $\mathcal{L}_0$. In particular one should determine the non-dimensionalization. Importantly here one should notice that $\on$ and $\off$ do not have the same units. The ratio $\off/\on = O(L)$ has units of a lengthscale. We can therefore keep the usual non-dimensionalization for $\off \frac{\Gamma}{k \epsilon^2} = \frac{\tilde{\off}}{\epsilon^2}$ but not for the binding rate, which we take as $\on \frac{\Gamma}{k  L \epsilon^2} = \frac{\tilde{\on}}{\epsilon^2}$. We find (dropping the $\tilde{\cdot}$)
\begin{equation}
 \mathcal{L}_0 = \begin{pmatrix}
- \int q_{\rm on} e^{-l^2/2} dl   & + \int q_{\rm on} e^{- l^2/2} dl  \\
+q_{\rm off} & -q_{\rm off} - l \partial_l +  \partial_{ll} .
\end{pmatrix}
\end{equation}

Resolution for $f_0$ does not change and we get $f_0 = a(x,t) \begin{pmatrix} 1 \\ 1 \end{pmatrix} $, with associated equilibrium distribution $\pi_0 = \pi$. 

At the next order we get the solution $f_1 = l \partial_x a \frac{1}{1 + q_{\rm off}} \begin{pmatrix} 0 \\ 1 \end{pmatrix}$. 

Finally at second order we require the Fredholm alternative $\langle \partial_t f_0 -\mathcal{L}_2 f_0 - \mathcal{L}_1 f_1 , \pi_0 \rangle = 0$ yielding
\begin{equation}
    \partial_t a = \frac{k_B T}{\Gamma_{\rm eff}^{\rm k/\gamma \gg q}} \partial_{xx} a
\end{equation}
with
\begin{equation}
    \boxed{\frac{1}{\Gamma_{\rm eff}^{\rm k/\gamma \gg q}} = \left( \frac{q_{\rm off}}{q_{\rm on} + q_{\rm off}} \right) \frac{1}{\Gamma}  + \left( \frac{q_{\rm on}}{q_{\rm on} + q_{\rm off}} \right) \frac{1}{\Gamma + \gamma + \frac{k}{q_{\rm off}}}.}
\end{equation}
This is nearly exactly the result obtained without pre-averaging but for the $ k \tau^{\rm relax}_{\rm u} =  k \left(\frac{\gamma}{k}\frac{\on}{\off}\right)$ contribution corresponding to the time the tether is allowed to relax between 2 binding periods. It is exactly the result reported in Eq.~\gammapreavera of the main manuscript. 

\paragraph{Relation to Ref.~\cite{fogelson2019transport}}

In this paragraph we relate our results to the results obtained in Ref.~\cite{fogelson2019transport}. Eq. (2.48) of Ref.~\cite{fogelson2019transport} finds an effective long time diffusion, starting from similar equations as Eq.~\eqref{eq:preaveraged}, 
\begin{equation}
    D_{\rm eff}^{\rm Ref.~\cite{fogelson2019transport}} = D_0 \left( 1 + \varepsilon \frac{\nu - 2}{\beta_0 (1+ \beta_0)\lambda} \right)
\end{equation}
where we will give the meaning of the new notations ($\nu$, $\varepsilon$, $\beta_0$, $\lambda$) by expressing them with respect to our notations. Here, $\nu = k/k_{\rm tether} = 1$ in our case because there is no change in recall spring force between the bound ($k_{\rm tether}$) and unbound states ($k$). We also have $\beta_0 = \frac{q_{\rm off}}{q_{\rm on}}$, here $\varepsilon = \frac{D_0}{q_{\rm on} L^2}$ and $\lambda = k_B T/kL^2$ such that the effective diffusion writes with our notations
\begin{equation}
    D_{\rm eff}^{\rm Ref.~\cite{fogelson2019transport}} = D_0 \left( 1 - \frac{k/q_{\rm off}}{\Gamma} \frac{q_{\rm on}}{q_{\rm on} + q_{\rm off}} \right).
\end{equation}
Compared to the previous derivation, this result corresponds to an effective friction with pre-averaging of tether dynamics (which is indeed what is done in Ref.~\cite{fogelson2019transport}) \textit{and} scales with $k/\off$ similarly as the derivation assuming $\gamma/\Gamma = \varepsilon$. In Ref.~\cite{fogelson2019transport}, as the dynamics are already pre-averaged, they are expressed at $0^{\rm th}$ order in $\gamma/\Gamma$. Therefore the key common point of these derivations (Ref.~\cite{fogelson2019transport} and Sec.~2.2 here) is to assume similar spatial scales, namely $L_x = L$. This highlights that the assumption $L/L_x = \epsilon$ allows one to "safely" average dynamics without specific assumptions on other physical parameters. %( which is similar to taking $\varepsilon = \frac{D_0}{q_{\rm on} L^2}$ as $\gamma/k$ and $q_{\rm on}^{-1}$ are considered to be time scales of similar order in Ref.~\cite{fogelson2019transport}. 

\paragraph{Relation to N legs facing a uniformly sticky membrane}

\begin{figure}[h!]
    \centering
    \includegraphics[width = 0.8\textwidth]{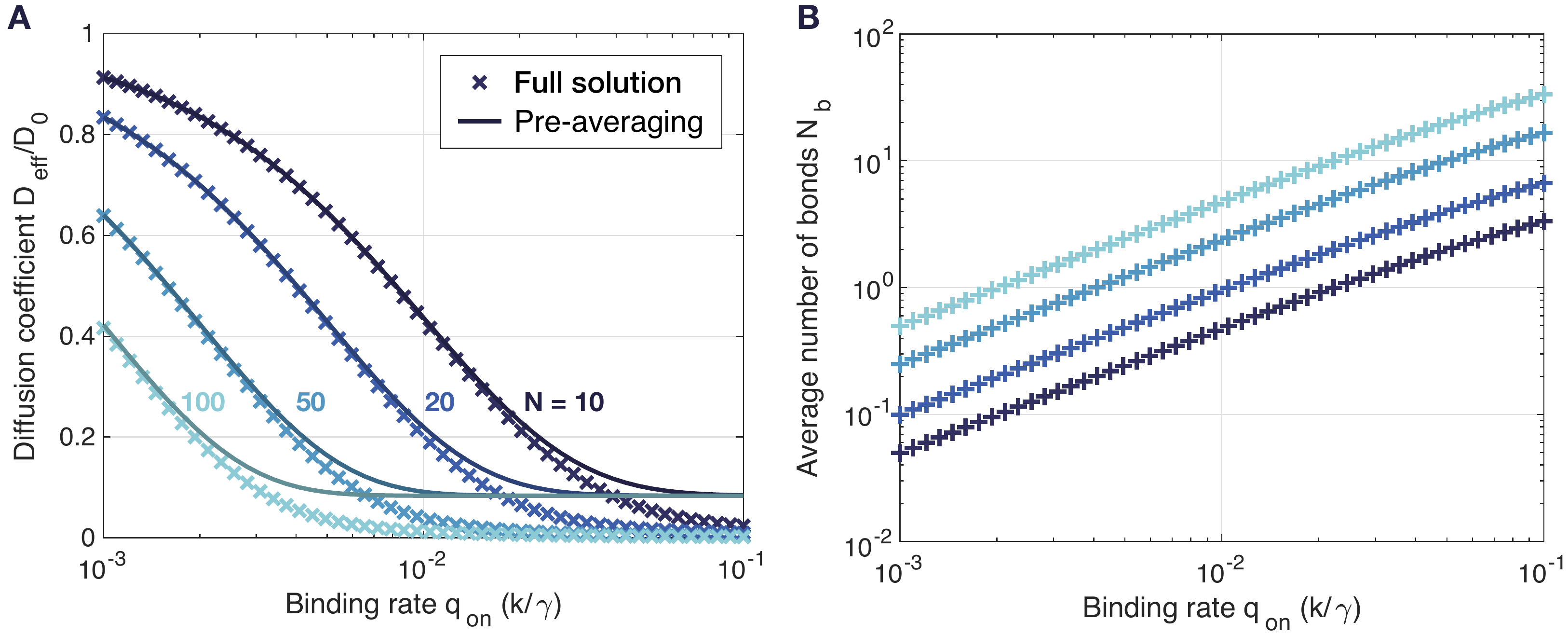}
    \caption{Pre-averaged results correspond to N legs facing a uniformly sticky surface when the average number of bonds legs $N_b \lesssim 1$. (A) $D_{\rm eff}$ as calculated with a numerical resolution of Eq.~\gammaN of the main paper (``Full solution'') or with the pre-averaged result of Eq.~\gammapreavera of the main paper (``Pre-averaging'') with respect to the binding rate $\on$. Other parameters are $\frac{\gamma}{\Gamma} = 1$ and $\frac{\off \Gamma}{k} = 0.1$; (B) Corresponding average number of bonds $N_b$.}
    \label{fig:preaveraging}
\end{figure}

In Fig.~\ref{fig:preaveraging} we show that the pre-averaged result corresponds to the predictions for N legs facing a uniformly sticky surface when the average number of bonds legs $N_b \lesssim 1$.

\subsubsection{Averaging with fast binding dynamics compared to relaxation dynamics}

To understand how fast binding dynamics affect this system, we use the same non-dimensionalization as in the main text but for $L = L_x$ (thereby allowing relaxation dynamics to be of similar order as the long time mobility of the particle). 
We obtain the non-dimensional generator
\begin{equation}
    \mathcal{L} = \frac{1}{\epsilon^2} \begin{pmatrix} - \on & \on \\ \off & -\off \end{pmatrix} + \begin{pmatrix}
    \frac{\Gamma}{\gamma} \left( - l \partial_l + \partial_{ll} \right) + \partial_{xx} & 0 \\ 0 & \frac{\Gamma}{\Gamma + \gamma} \left( l (\partial_x - \partial_l)  + (\partial_x - \partial_l)^2 \right) \end{pmatrix}
    = \frac{1}{\varepsilon} \mathcal{L}_0 + \mathcal{L}_1
\end{equation}
where we used $\varepsilon= \epsilon^2$. 

We then seek a solution as an expansion $f = f_0 + \varepsilon f_1 + \varepsilon^2 f_2 + ... $. At lowest order we need to solve
\begin{equation}
    \begin{pmatrix} - \on & \on \\ \off & -\off \end{pmatrix} f_0 = 0
\end{equation}
which simply yields $f_0 = a(x,l,t)\begin{pmatrix}
1 \\ 1
\end{pmatrix}$ and the associated equilibrium distribution $\pi_0 = \begin{pmatrix}
\frac{\off}{\on} \\ 1 
\end{pmatrix}$. Notice how here the equilibrium distribution at lowest order does not correspond to the full equilibrium distribution $\pi_0 \neq \pi$. At the following order to find a solution to the problem we require the Fredholm alternative $\langle \partial_t f_0 - \mathcal{L}_1 f_0 , \pi_0 \rangle = 0$, which gives the equation 
\begin{equation}
\begin{split}
    \partial_t a =  &p_0\frac{\Gamma}{\gamma} \left( - l \partial_l + \partial_{ll} \right) + p_0\partial_{xx} \\
&+ p_1 \frac{\Gamma}{\Gamma + \gamma} \left( l (\partial_x - \partial_l)  + (\partial_x - \partial_l)^2 \right)
\end{split}
\end{equation}
where $p_0 = \frac{\off}{\off + \on} = 1 - p_1$ is the probability to be unbound and $p_1$ to be bound. This new effective equation can now be explored by means of time and length scale separation by setting a new small scale parameter $\epsilon_2 = \frac{L}{L_x}$. This small scale operator allows us to write
\begin{equation}
    \partial_t a = \frac{1}{\epsilon_2^2} \mathcal{L}_{0,2} +  \frac{1}{\epsilon_2}\mathcal{L}_{1,2} + \mathcal{L}_{2,2}
\end{equation}
where $\mathcal{L}_{0,2} = \left(p_0 \frac{\Gamma}{\gamma} + p_1 \frac{\Gamma}{\Gamma + \gamma}\right)\left( - l \partial_l + \partial_{ll} \right) $, $\mathcal{L}_{1,2} = p_1 \frac{\Gamma}{\Gamma + \gamma} \left( l \partial_x - 2 \partial_{xl} \right)$ and $\mathcal{L}_{2,2} = \left(p_0  + p_1 \frac{\Gamma}{\Gamma + \gamma}\right) \partial_{xx}$. 
We seek a solution $a = a_0 + \epsilon_2 a_1 + \epsilon^2_2 a_2 + ...$. 

At lowest order we need to solve $\mathcal{L}_{0,2}a_0 = 0$ which implies $a_0 = a_0(x,t)$ making use of vanishing flux boundary conditions at infinity. The associated equilibrium distribution is now $\pi_{0,2} = e^{-l^2/2}/Z$. 

At the next order we need to solve $\mathcal{L}_{0,2} a_1 = - \mathcal{L}_{1,2} f_0 = - p_1 \frac{\Gamma}{\Gamma + \gamma} l \partial_x a $ such that $a_1 = - \frac{p_1 \frac{\Gamma}{\Gamma + \gamma}}{p_0 \frac{\Gamma}{\gamma} + p_1 \frac{\Gamma}{\Gamma + \gamma}} l \partial_x a = - \frac{l \partial_x a}{1 + \frac{p_0}{p_1} \frac{\gamma}{\Gamma + \gamma}}$. 
At the following order, to find a solution we require the Fredholm alternative, namely $\langle \partial_t a_0 - \mathcal{L}_{2,2} a_0 - \mathcal{L}_{1,2} a_1 , \pi_{0,2} \rangle = 0$. After some standard algebra one finds
\begin{equation}
    \partial_t a_0 = \frac{1}{\Gamma^{q\, \rm fast}} \partial_{xx} a 
\end{equation}
where
\begin{equation}
    \frac{1}{\Gamma^{q\, \rm fast}}  = \frac{p_0}{\Gamma} + \frac{p_1}{\Gamma + \gamma/p_0}
\end{equation}
which, reverting to dimensional scales, is exactly Eq.~\gammaqfast of the main paper.

\subsection{Arm and/or legs}

\subsubsection{Arm or leg}

In this part we show precisely how having an arm (spring always attached to the surface) or a leg (spring always attached to the particle) affects the dynamics. 
There are several ways one can consider to obtain the dynamics of the leg or the arm (referred to henceforth as spring). Either assume (1) that the center of mass of the spring is attached to the particle (or the surface), (2) either that the center of mass is located at the free end of the spring (when it is attached to the particle or the surface). Both assumptions do not yield exactly the same dynamics but the differences in the long time effective dynamics are minor. 

\paragraph{(1) Arm or leg (spring) attached by their center of mass}

Consider in general a free spring, where motion is confined to a line but none of the spring ends are attached. The length  of the spring $l$ (more accurately here $l$ represents the length imbalance compared to the rest length of the spring $l - l_0$ but we take $l_0 = 0$ for simplicity) obeys the overdamped Langevin equation
\begin{equation}
    \frac{dl}{dt} = - \frac{k l}{\gamma} + \sqrt{ \frac{2 k_BT}{\gamma}}\eta_l.
\end{equation}
The center of mass $c$ of the spring similarly obeys an overdamped Langevin equation, with diffusion only
\begin{equation}
    \frac{dc}{dt} = \sqrt{ \frac{2 k_BT}{\gamma}}\eta_c
\end{equation}
and we considered that the diffusion coefficient of the center of mass is similar to that of the spring length. For simplicity we consider here that the center of mass of the spring is located at one of its ends, namely the end that will be permanently attached to a surface in this paragraph.

In addition the particle also diffuses as
\begin{equation}
    \frac{dx}{dt}  = \sqrt{\frac{2 k_BT}{\Gamma}}\eta_x.
\end{equation}

\underline{Arm configuration.} If we consider the arm configuration, then the center of mass $c$ is attached to the surface and satisfies the constraint $q(x,l,c) = c - x_{\rm surface} = 0$. The projected dynamics in that case are trivial and sum up to the ones detailed in the main text and recalled here for consistency:
\begin{equation}
    \begin{cases}
        \frac{dx}{dt}  = \sqrt{\frac{2 k_BT}{\Gamma}}\eta_x, \\
        \frac{dl}{dt} = - \frac{k l}{\gamma} + \sqrt{\frac{2 k_BT}{\gamma}}\eta_l, \\
        \frac{dc}{dt} = 0 .
    \end{cases}
\end{equation}

\underline{Leg configuration.} If we consider the leg configuration, then the center of mass $c$ is attached to the particle and satisfies the constraint $q(x,l,c) = x - c = 0$. This constraint is similar to the one for the bound spring for which the projection formalism is described in Appendix A of the main text. The projected dynamics are therefore
\begin{equation}
    \begin{cases}
        \frac{dx}{dt}  = \sqrt{\frac{2 k_BT}{\Gamma + \gamma}}\eta_x, \\
        \frac{dl}{dt} = - \frac{k l}{\gamma} + \sqrt{\frac{2 k_BT}{\gamma}}\eta_l, \\
        \frac{dc}{dt} =  \frac{dx}{dt}.
    \end{cases}
\end{equation}
These dynamics are exactly equivalent to the arm configuration but for the change $\Gamma \rightarrow \Gamma + \gamma$, and therefore yield the same resulting effective long time dynamics with a similar change $\Gamma \rightarrow \Gamma + \gamma$. One can thus simply consider that $\Gamma$ is indeed the friction coefficient of the unbound particle, which potentially includes corrections to friction due to legs being attached to the surface.

\paragraph{(2) Arm or leg (spring) with center of mass at the free end}

We now consider the situation where the center of mass of the spring is located at its free end, and the other end is attached to the particle or the surface.

\underline{Leg configuration.} 
Consider in general a spring, attached to one end to the particle (in $x$) and to the other end to the spring's mass (in $x + l$). Newton's second law on each mass, and taking loosely overdamped dynamics with masses going to 0, yields the system of equations
\begin{equation}
\begin{cases}
    0  & = - \Gamma \frac{dx}{dt} + kl + \sqrt{2 k_BT \Gamma}\eta_x, \\
    0  & = - \gamma \frac{d(x + l)}{dt} - kl + \sqrt{2 k_BT \gamma}\eta_l.
\end{cases}   
\end{equation}
The system simplifies for each variable into
\begin{equation}
    \begin{cases}
        \frac{dx}{dt}  =  + \frac{k l}{\Gamma} + \sqrt{\frac{2 k_BT}{\Gamma}}\eta_x, \\
        \frac{dl}{dt} = - kl \left(\frac{1}{\Gamma} + \frac{1}{\gamma}\right) - \sqrt{\frac{2 k_BT}{\Gamma}}\eta_x + \sqrt{\frac{2 k_BT}{\gamma}}\eta_l.
    \end{cases}
\end{equation}
This system corresponds to a friction matrix and force field
\begin{equation}
    \tilde{\Gamma}^{-1} = \begin{pmatrix}
    \frac{1}{\Gamma}  & - \frac{1}{\Gamma} \\
    -\frac{1}{\Gamma}  & \frac{1}{\gamma} + \frac{1}{\Gamma}
    \end{pmatrix} 
    \,\, \mathrm{and} \,\, 
    \nabla \mathcal{U} = \begin{pmatrix}
    0 \\ k l 
    \end{pmatrix}.
\end{equation}
When the spring (here the leg) becomes temporarily bound to the surface, the bound equations then simply read (provided appropriate projection, following Appendix A, is made)
\begin{equation}
        \begin{cases}
        \frac{dx}{dt}  =  + \frac{k l}{\Gamma} + \sqrt{\frac{2 k_BT}{\Gamma}}\eta_x, \\
        \frac{dl}{dt} = - \frac{dx}{dt} = -  \frac{k l}{\Gamma} - \sqrt{\frac{2 k_BT}{\Gamma}}\eta_x.
    \end{cases}
\end{equation}

In non-dimensional scales, the hierarchy of generators now reads
\begin{equation}
    \mathcal{L}_0 = \begin{pmatrix}
    - \on + \frac{\Gamma + \gamma}{\Gamma} \left( -l \partial_l + \partial_{ll} \right) & \on \\
    \off & \off + \left( -l \partial_l + \partial_{ll} \right) 
    \end{pmatrix},
\end{equation}
\begin{equation}
    \mathcal{L}_1 = \begin{pmatrix}
    \left( l \partial_x - 2 \partial_{lx} \right) & 0 \\
    0 &  \left(l \partial_x - 2 \partial_{lx} \right) 
    \end{pmatrix}, \,\,
\mathrm{and} \,\,  
    \mathcal{L}_2 = \begin{pmatrix}
    \partial_{xx} & 0 \\
    0 &  \partial_{xx} 
    \end{pmatrix}.
\end{equation}

Using a coarse-graining method as usual, we obtain simply the usual harmonic sum $\Gamma_{\rm eff}^{-1} = p_0 \Gamma_0^{-1} + p_1 \Gamma_1^{-1}$ where the friction coefficients write
\begin{equation}
    \begin{cases}
         \Gamma_0 =& \Gamma + \gamma + k \left( \frac{\gamma}{k} \frac{\on}{\off + k/\Gamma}\right) \\
         \Gamma_1 =& \Gamma + \gamma + k \left( \frac{1}{\off} + \frac{\gamma}{k}  \frac{\on + k/\Gamma}{\off}  \right) \,\, \text{(model (2))}.
    \end{cases}
\end{equation}
Compared to the coefficients obtained if the spring is attached to the particle by its center of mass (model (1)), namely
\begin{equation}
    \begin{cases}
         \Gamma_0 =& \Gamma + \gamma  \\
         \Gamma_1 =& \Gamma + \gamma + k \left( \frac{1}{\off} + \frac{\gamma}{k}  \frac{\on + \off}{\off}  \right), \,\, \text{(model (1))}, 
    \end{cases}
\end{equation}
the results are quite similar independent of the attaching model. Qualitatively, in the non-center of mass case (model (2)), we obtain additional feedback friction terms due to the spring, as $k \tau_{\rm eff}$ where $\tau_{\rm eff}$ is a typical time over which the spring relaxes, scaling naturally as $\gamma/k$ multiplied by a ratio of characteristic times $\frac{\tau_{...}}{\tau_{...}}$. This ratio corresponds to the fact that the spring may only relax in the other state, and hence different ratios appear according to the different modeling options and bound states. Be that as it may, such contributions are generally minor. In fact, one can verify (not shown here) that the numerical values of $\Gamma_{\rm eff}/\Gamma$ according to model (1) or (2) show very little difference over full $\mathbb{R}^3$ space described by the parameters. 

We now explore potential differences when the number of legs is increased. For $2$ legs, the dynamics in the unbound state are
\begin{equation}
    \begin{cases}
        \frac{dx}{dt}  =  + \frac{k (l_1 + l_2)}{\Gamma} + \sqrt{\frac{2 k_BT}{\Gamma}}\eta_x, \\
        \frac{dl_1}{dt} = - \frac{dx}{dt} -  \frac{k l_1}{\gamma}  + \sqrt{\frac{2 k_BT}{\gamma}}\eta_1 \\ 
        \frac{dl_2}{dt} = - \frac{dx}{dt} -  \frac{k l_2}{\gamma}  + \sqrt{\frac{2 k_BT}{\gamma}}\eta_2 \\ 
    \end{cases}
\end{equation}
and when leg $1$ is bound to the surface simply
\begin{equation}
    \begin{cases}
        \frac{dx}{dt}  = - \frac{dl_1}{dt} =   + \frac{k (l_1 + l_2)}{\Gamma} + \sqrt{\frac{2 k_BT}{\Gamma}}\eta_x, \\
        \frac{dl_2}{dt} = - \frac{dx}{dt} -  \frac{k l_2}{\gamma}  + \sqrt{\frac{2 k_BT}{\gamma}}\eta_2 \\ 
    \end{cases}
\end{equation}
and when both legs are bound
\begin{equation}
    \begin{cases}
        \frac{dx}{dt}  = - \frac{dl_1}{dt}  =  -\frac{dl_2}{dt} =   + \frac{k (l_1 + l_2)}{\Gamma} + \sqrt{\frac{2 k_BT}{\Gamma}}\eta_x.    \end{cases}
\end{equation}
Dynamics are then easily extended to $N$ legs using the free spring end model applied to each leg. Coarse-graining and asymptotics (around the average number of bonds) then easily lead to
\begin{equation}
    \Gamma_{\rm eff} \simeq \Gamma_{N_b} = \Gamma + N \gamma + N_b \left[ \gamma +  k \left( \frac{1}{\off} + \frac{\gamma}{k} \frac{\on }{\off}\right) \right] \,\, \text{(model (2))}.
    \label{model2eff}
\end{equation}
Eq.~\eqref{model2eff} is exactly the result obtained by attaching legs by their center of mass (model (1)), provided the suitable change $\Gamma \rightarrow \Gamma + N \gamma$ is done for $N$ legs. There is thus no difference between the different models in the leg configuration when a large number of legs are involved. 

\underline{Arm configuration.} The arm configuration with mass at the free end (model (2)) is trivially equivalent to that attached by the center of mass as
\begin{equation}
    \begin{cases}
        \frac{dx}{dt}  = \sqrt{\frac{2 k_BT}{\Gamma}}\eta_x, \\
        \frac{dl}{dt} = - \frac{k l}{\gamma} + \sqrt{\frac{2 k_BT}{\gamma}}\eta_l.
    \end{cases}
\end{equation}
There is thus no difference between the different models in the arm configuration. 

\subsubsection{Arm and leg}

\paragraph{Equations set up}

\begin{figure}
    \centering
    \includegraphics[width = 0.6\textwidth]{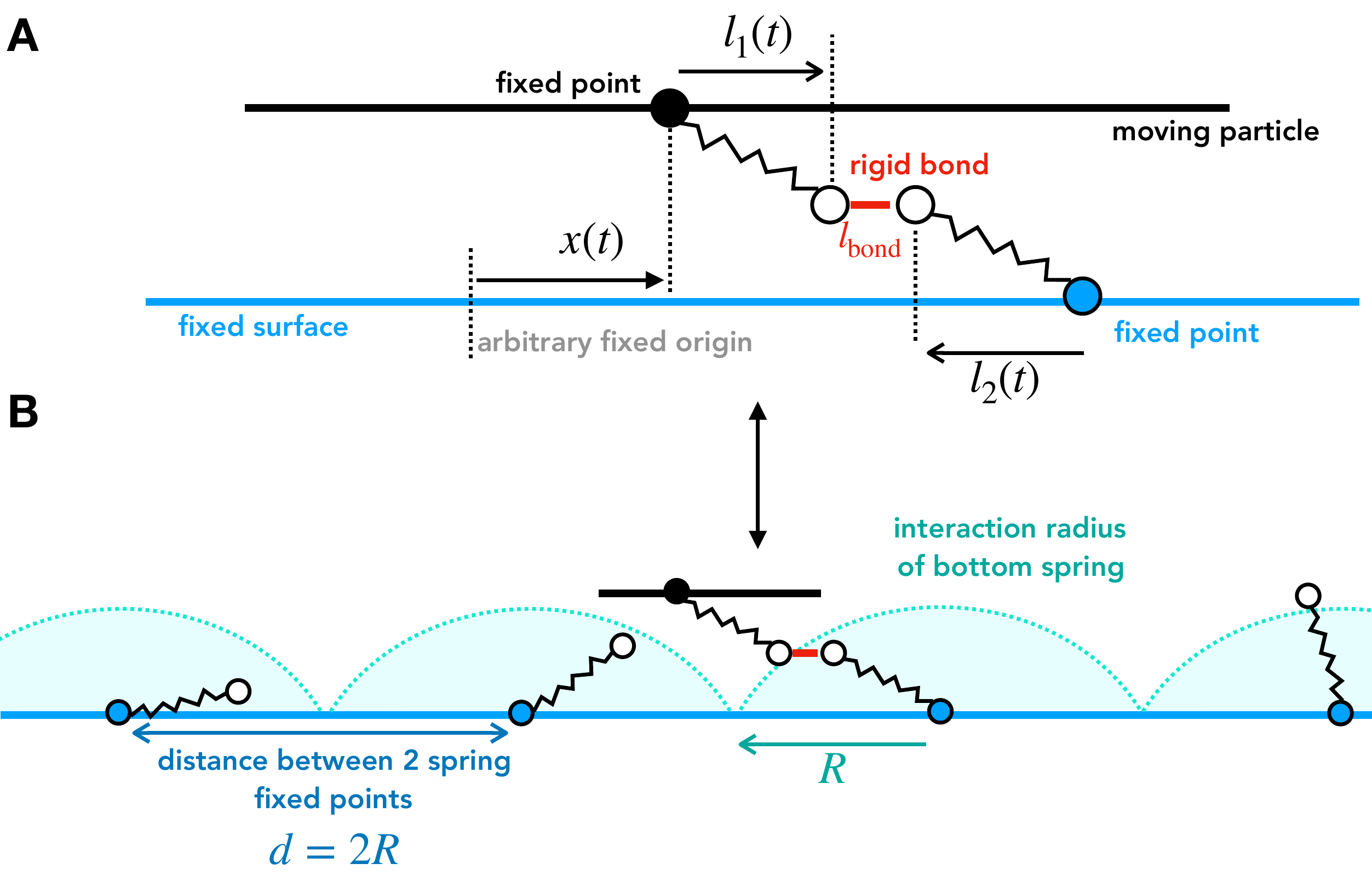}
    \caption{Geometry of binding with a particle having an arm and a leg. (A) The spring attached to the moving particle may bind to the bottom spring by forming a rigid bond that ``fills in the distance" between the separated springs. Such a model is equivalent to (B) where the bond is formed with the ``closest" available bond. Here if springs on the surface are evenly spaced with a typical spacing $d = 2R$ we consider that the top spring's sticky end binds to a surface spring whose fixed point is closest, and always closer than $R$. Switching events between one spring and then another are long if the distance between two surface springs is large and are ignored. The equivalence between A and B could be shown more systematically, but is beyond the scope of this manuscript.}
    \label{fig:armleg}
\end{figure}

We consider random attachement and detachement of two springs to one another, in the leg and arm geometry -- see Fig.~\ref{fig:armleg}-A.

When the springs are unbound the dynamic equations are 
\begin{equation}
\begin{cases}
\frac{dl_1}{dt} &= - \frac{k}{\gamma} l_1 + \sqrt{\frac{2 k_B T}{\gamma}} \eta_1(t) \\
\frac{dl_2}{dt} &= - \frac{k}{\gamma} l_2 + \sqrt{\frac{2 k_B T}{\gamma}} \eta_2(t) \\
\frac{dx}{dt} &= \sqrt{\frac{2 k_B T }{\Gamma}} \eta_x(t)
\end{cases}
\end{equation}
where $l_1$ is the length of the top spring, $l_2$ the length of the bottom spring, and for simplicity here we took $l_0 = 0$.

In the bound state we need to project the dynamics. When the springs bind, we consider that a rigid bond is formed between the springs' sticky ends that keeps the distance constant -- see Fig.~\ref{fig:armleg}. The dynamic constraint is then 
\begin{equation}
    q(x,l_1,l_2) = x + l_1 - l_2 + l_{\rm bond} = 0
\end{equation}
where $ l_{\rm bond}$ is the length of the bond and remains constant until the springs detach and reattach to form another bond length. If we imagine that the bottom spring is part of a periodic array of springs, such that at any time, only one bottom spring is accessible to the top spring, $l_{\rm bond}$ is typically of order $L$ -- see Fig.~\ref{fig:armleg} -- and thus a reasonable physical assumption.

The constraint matrix is then $C = (1,1,-1)$ and the projection matrix
\begin{equation}
    P = 1 - \frac{1}{3} \begin{pmatrix}
    1 & 1 & -1 \\
    1 & 1 & -1 \\
    -1 & -1 & 1 
    \end{pmatrix}  = \frac{1}{3} \begin{pmatrix}
    2 & -1 & 1 \\
    -1 & 2 & 1 \\
    1 & 1 & 2 
    \end{pmatrix}
\end{equation}
such that the Moore-Penrose pseudo inverse of the projected friction is 
\begin{equation}
    \Gamma_P^{\dagger} = \frac{1}{\gamma + 2\Gamma} \begin{pmatrix}
    2 &  -1 & 1 \\
    -1 & \frac{\gamma + \Gamma}{\gamma} & \frac{\Gamma}{\gamma} \\
    1 & \frac{\Gamma}{\gamma} & \frac{\gamma + \Gamma}{\gamma}
    \end{pmatrix}
\end{equation}
with a square root
\begin{equation}
    \sigma_P = \frac{1}{\gamma + 2 \Gamma} \begin{pmatrix}
   2 \sqrt{\Gamma} & \sqrt{\gamma} & - \sqrt{\gamma} \\
    \sqrt{\Gamma} & \sqrt{\gamma}/2 + \frac{\gamma + 2 \Gamma}{\sqrt{4\gamma}} & -\sqrt{\gamma}/2 + \frac{\gamma + 2 \Gamma}{\sqrt{4\gamma}} \\
    -\sqrt{\Gamma} & -\sqrt{\gamma}/2 + \frac{\gamma + 2 \Gamma}{\sqrt{4\gamma}} & \sqrt{\gamma}/2 + \frac{\gamma + 2 \Gamma}{\sqrt{4\gamma}}
    \end{pmatrix}.
\end{equation}
%Note that this Cholesky decomposition is actually unique because it is the only one that satisfies proper symmetry requirements. 

The dynamics in the bound state are therefore
\begin{equation}
\begin{cases}
\frac{dx}{dt} &= - \frac{k}{2\Gamma+\gamma}(l_1-l_2)+  \sqrt{\frac{8k_B T \Gamma}{(2\Gamma+\gamma)^2}} \eta_x + \sqrt{\frac{2 k_B T \gamma}{(2\Gamma+\gamma)^2}} (\eta_1 - \eta_2) \\
\frac{dl_1}{dt} &=  \frac{1}{2}\frac{dx}{dt} - \frac{k}{2 \gamma} (l_1+l_2) + \sqrt{\frac{2 k_B T}{4 \gamma}} (\eta_1 + \eta_2) \\
\frac{dl_2}{dt} &= - \frac{1}{2}\frac{dx}{dt}  - \frac{k}{2 \gamma}  (l_1+l_2) + \sqrt{\frac{2 k_B T}{4 \gamma}} (\eta_1 + \eta_2) 
\end{cases}
\end{equation}

\paragraph{Generator}

The generator is then
%from the projection operators through the formula~\cite{ciccotti2008projection}
% \begin{equation}
% \mathcal{L} f = -  \Gamma_P^{\dagger} \nabla \mathcal{U}_{\rm eff}(x) . \nabla f + k_B T  \mathrm{Tr} \left[  \nabla . ( \Gamma_P^{\dagger} . \nabla f) \right]
% \end{equation}
% where $\mathcal{U}_{\rm eff}(x) = \mathcal{U}(x) - k_B T \log |C|$ and $|C|  = |C C^T|^{1/2}$ is the pseudo-determinant of $C$. 

% Here the projection operators are spatially independent and therefore the generator is
\begin{equation}
\begin{split}
\mathcal{L}= &\begin{pmatrix}
-q_{\rm on} & q_{\rm on} \\
q_{\rm off} & -q_{\rm off}\end{pmatrix} +  \begin{pmatrix}
 -\frac{k}{\gamma}l_1  \partial_{l_1} + \frac{k_B T}{ \gamma} \partial_{l_1l_1}  -\frac{k}{\gamma}l_2  \partial_{l_2} + \frac{k_B T}{ \gamma} \partial_{l_2l_2} + \frac{k_B T }{\Gamma}\partial_{xx}  & 0 \\
0 & 0 \end{pmatrix}   \\
&+ \begin{pmatrix}
0  & 0 \\
0 & - \frac{k}{2\Gamma+\gamma}(l_1-l_2) (\partial_x - \frac{1}{2}\partial_{l_2} + \frac{1}{2}\partial_{l_1}) - \frac{k}{2\gamma}(l_1+l_2) (\partial_{l_2} + \partial_{l_1}) \end{pmatrix}  \\
&+ \begin{pmatrix}
0  & 0 \\
0 &  + \frac{2 k_B T}{2\Gamma+\gamma}(\partial_x - \frac{1}{2}\partial_{l_2} + \frac{1}{2}\partial_{l_1})^2 + \frac{k_B T}{2\gamma} (\partial_{l_2} + \partial_{l_1})^2 \end{pmatrix}  
\end{split}
\end{equation}
With this generator one can check that $\mathcal{L}^{\star}\pi = 0$ with the natural equilibrium distribution 
\begin{equation}
\pi = \frac{1}{Z} \begin{pmatrix}
q_{\rm off}/q_{\rm on} \\ 1
\end{pmatrix}e^{-kl_1^2/2k_B T }e^{-kl_2^2/2k_B T }
\end{equation}
where $Z$ is a normalization constant.

\paragraph{Homogenization}

Taking the usual scalings we get the following non-dimensional, expanded generator
\begin{equation}
\begin{split}
\mathcal{L} =&\frac{1}{\epsilon^2} \bigg[ \begin{pmatrix}
-q_{\rm on} & q_{\rm on} \\
q_{\rm off} & -q_{\rm off} 
\end{pmatrix} +  \frac{\Gamma}{\gamma}\begin{pmatrix}
 - l_1  \partial_{l_1} + \partial_{l_1l_1}  - l_2  \partial_{l_2} + \partial_{l_2l_2} & 0 \\
0 & 0
\end{pmatrix} \\
& \,\,\,\, + \begin{pmatrix} 0 & 0 \\
0&  - \frac{\Gamma}{2(2\Gamma+\gamma)}(l_1-l_2) (\partial_{l_1}-\partial_{l_2} ) - \frac{\Gamma}{2\gamma}(l_1+l_2) (\partial_{l_2} + \partial_{l_1}) + \frac{ \Gamma}{2(2\Gamma+\gamma)}(\partial_{l_2} - \partial_{l_1})^2 + \frac{\Gamma}{2\gamma} (\partial_{l_2} + \partial_{l_1})^2 \end{pmatrix}  \bigg]  \\
 &+ \frac{1}{\epsilon}  \begin{pmatrix}
 0  & 0 \\
0 &  - \frac{\Gamma}{2\Gamma+\gamma}(l_1-l_2) \partial_x +  \partial_x(\partial_{l_1} -\partial_{l_2}) \frac{2 \Gamma}{\gamma + 2\Gamma}
\end{pmatrix}  \\
&+ 1 \begin{pmatrix}
   \partial_{xx}  & 0 \\
0 &   \frac{2 \Gamma}{\gamma+2\Gamma} \partial_{xx}
\end{pmatrix} = \frac{1}{\epsilon^2} \mathcal{L}_0 + \frac{1}{\epsilon} \mathcal{L}_1 +\mathcal{L}_2
\end{split}
\end{equation}
We now seek an expanded solution $f$ of $\partial_t f = \mathcal{L} f$ as $f = f_0 + \epsilon f_1 + \epsilon^2 f_2 + ...$

At lowest order, $\mathcal{L}_0 f_0 = 0$ gives $f_0 = a(x,t) \begin{pmatrix} 1 \\ 1 \end{pmatrix}$, and the associated equilibrium distribution $\pi_0 = \pi$. At first order we need to solve
\begin{equation}
\mathcal{L}_0 f_1 = - \mathcal{L}_1 f_0 =  + \begin{pmatrix}
0 \\ 1
\end{pmatrix} \frac{\Gamma}{\gamma + 2\Gamma} (l_1 - l_2) \partial_x a
\end{equation}
which has a unique solution
\begin{equation}
f_1 = - \begin{pmatrix}
\frac{\gamma q_{\rm on}}{\Gamma+ \gamma q_{\rm on}} \\ 1 
\end{pmatrix}\frac{(l_1 - l_2) \partial_x a}{1 + \frac{\Gamma q_{\rm off}}{\Gamma + \gamma q_{\rm on}}\frac{\gamma+2\Gamma}{\Gamma}}.
\end{equation}

At 2nd order we need to satisfy the Fredholm alternative $\langle \partial_t f_0 - \mathcal{L}_2 f_0 - \mathcal{L}_1 f_1 , \pi_0 \rangle = 0$. We split up the terms to highlight calculation steps (discarding $Z$ terms to simplify notations, as they would cancel out eventually)
\begin{equation}
\begin{split}
\langle \mathcal{L}_1 f_1  , \pi_0 \rangle = &- \langle \left( - (l_1-l_2)^2 \frac{\Gamma}{\gamma + 2\Gamma}  + 2 \frac{2\Gamma}{\gamma+2\Gamma}\right) e^{-l_1^2/2} e^{-l_2^2/2} \rangle \frac{ \partial_{xx} a}{1 + \frac{\Gamma q_{\rm off}}{\Gamma + \gamma q_{\rm on}}\frac{\gamma+2\Gamma}{\Gamma}} \\
= & - \left( - 2 \frac{\Gamma}{\gamma + 2\Gamma}  + 2 \frac{2\Gamma}{\gamma+2\Gamma}\right) \frac{ \partial_{xx} a}{1 + \frac{\Gamma q_{\rm off}}{\Gamma + \gamma q_{\rm on}}\frac{\gamma+2\Gamma}{\Gamma}} \\
= & - \frac{2 \Gamma}{\gamma+2\Gamma} \frac{ \partial_{xx} a}{1 + \frac{\Gamma q_{\rm off}}{\Gamma + \gamma q_{\rm on}}\frac{\gamma+2\Gamma}{\Gamma}}
\end{split}
\end{equation}
and
\begin{equation}
\langle \mathcal{L}_2 f_0 , \pi_0 \rangle = \left( \frac{q_{\rm off}}{q_{\rm on}}  + \frac{2 \Gamma}{\gamma+2\Gamma} \right) \partial_{xx} a .
\end{equation}
Gathering terms as $\langle \partial_t f_0  ,  \pi_0 \rangle = \langle \mathcal{L}_2 f_0  ,  \pi \rangle  + \langle \mathcal{L}_1 f_1  , \pi_0 \rangle $  we get
\begin{equation}
\left(1 + \frac{q_{\rm off}}{q_{\rm on}}\right)\partial_t a  =  \left( \frac{q_{\rm off}}{q_{\rm on}} + \frac{2\Gamma}{\gamma+2\Gamma} \frac{\frac{\Gamma q_{\rm off}}{\Gamma + \gamma q_{\rm on}}\frac{\gamma+2\Gamma}{\Gamma}}{1 + \frac{\Gamma q_{\rm off}}{\Gamma + \gamma q_{\rm on}}\frac{\gamma+2\Gamma}{\Gamma}}\right) \partial_{xx} a .
\end{equation}
Now shifting back to dimensional scales and reorganizing terms slightly we obtain
\begin{equation}
\partial_t a = \left(\frac{q_{\rm off}}{q_{\rm on} + q_{\rm off}}\frac{k_B T}{\Gamma} + \frac{q_{\rm on}}{q_{\rm on} + q_{\rm off}}\frac{k_B T}{\Gamma + \frac{1}{2}\left(\frac{k}{q_{\rm off}} + \frac{q_{\rm on}}{q_{\rm off}}\gamma + \gamma\right)} \right) \partial_{xx} a.
\end{equation}
Using the notations of the main papaer $p_0 = \off/(\on + \off)$ the probability to be unbound, $p_1 = 1 - p_0$ the probability to have 1 bond and $\gamma_{\rm eff} = \gamma  + \frac{k}{q_{\rm off}} + \gamma\frac{q_{\rm on}}{q_{\rm off}}$ we obtain
 \begin{equation}
 \boxed{\frac{1}{\Gamma_{\rm eff}^{\rm leg+arm}} = \frac{p_0}{\Gamma} + \frac{p_1}{\Gamma + \frac{1}{2}\gamma_{\rm eff}}}
 \end{equation}
that is exactly Eq.~\gammaarmleg in the main paper. 

\subsubsection{Several arms for 1 leg}

\paragraph{Equations set up}

We consider random attachment and detachment of two springs to one another, in the leg and arm geometry, but now when there are possibly $M$ arms to attach to. 

When the springs are unbound the dynamic equations are 
\begin{equation}
\begin{cases}
\frac{dl}{dt} &= - \frac{k}{\gamma} l + \sqrt{\frac{2 k_B T}{\gamma}} \eta(t) \\
\frac{dl_i}{dt} &= - \frac{k}{\gamma} l_i + \sqrt{\frac{2 k_B T}{\gamma}} \eta_i(t) \,\, \text{for} \,\, i = 1\ldots M \\
\frac{dx}{dt} &= \sqrt{\frac{2 k_B T }{\Gamma}} \eta_x(t)
\end{cases}
\end{equation}
where $l$ is the length of the top spring, $l_i$ are the lengths of all the bottom springs, and for simplicity here we took $l_0 = 0$.

In the bound state we need to project the dynamics. The dynamic constraint with the bound bottom spring indexed by b is then 
\begin{equation}
    q(x,l,l_b) = x + l - l_b + l_{\rm bond} = 0
\end{equation}
where $ l_{\rm bond}$ is the length of the bond, similarly as in the previous section. The constraint process leaves the unbound spring equations completely unaffected and we find after the projection step
\begin{equation}
\begin{cases}
\frac{dx}{dt} &= - \frac{k}{2\Gamma+\gamma}(l_1-l_2)+  \sqrt{\frac{8k_B T \Gamma}{(2\Gamma+\gamma)^2}} \eta_x + \sqrt{\frac{2 k_B T \gamma}{(2\Gamma+\gamma)^2}} (\eta - \eta_b) \\
\frac{dl}{dt} &=  \frac{1}{2}\frac{dx}{dt} - \frac{k}{2 \gamma} (l+l_b) + \sqrt{\frac{2 k_B T}{4 \gamma}} (\eta + \eta_b) \\
\frac{dl_b}{dt} &= - \frac{1}{2}\frac{dx}{dt}  - \frac{k}{2 \gamma}  (l+l_b) + \sqrt{\frac{2 k_B T}{4 \gamma}} (\eta + \eta_b) \\
\frac{dl_i}{dt} &= - \frac{k}{\gamma} l_i + \sqrt{\frac{2 k_B T}{\gamma}} \eta_i(t) \,\, \text{for} \,\, i = 1 .. M \,\, \text{and} \,\, i \neq b 
\end{cases}
\end{equation}

\paragraph{Generator}

The generator is similarly
\begin{equation}
\mathcal{L}= \mathcal{Q} +  \mathcal{U}  = \begin{pmatrix}
-Mq_{\rm on} & q_{\rm on} & \on & .. & \on \\
q_{\rm off} & -q_{\rm off} & 0 & ... & 0 \\
q_{\rm off} & 0 & -q_{\rm off} & ... & 0 \\
... \\
q_{\rm off} & 0 & 0 & ... & -q_{\rm off} 
\end{pmatrix} + \mathcal{U} 
\end{equation}
and $\mathcal{U}$ is a diagonal matrix. The first term of $\mathcal{U}$ corresponds to fully unbound dynamics
\begin{equation}
   \mathcal{U}_{00} =  -\frac{k}{\gamma}l \partial_{l} + \frac{k_B T}{ \gamma} \partial_{ll} + \sum_i \left( -\frac{k}{\gamma}l_i  \partial_{l_i} + \frac{k_B T}{ \gamma} \partial_{l_il_i} \right) + \frac{k_B T }{\Gamma}\partial_{xx} 
\end{equation}
and further terms correspond each to a bond with the $b^{th}$ arm
\begin{equation}
\begin{split}
   \mathcal{U}_{bb} = & - \frac{k}{2\Gamma+\gamma}(l-l_b) (\partial_x - \frac{1}{2}\partial_{l_b} + \frac{1}{2}\partial_{l}) - \frac{k}{2\gamma}(l+l_b) (\partial_{l_b} + \partial_{l})   \\
&+ \frac{2 k_B T}{2\Gamma+\gamma}(\partial_x - \frac{1}{2}\partial_{l_b} + \frac{1}{2}\partial_{l})^2 + \frac{k_B T}{2\gamma} (\partial_{l_b} + \partial_{l})^2 + \sum_{i\neq b} \left( -\frac{k}{\gamma}l_i  \partial_{l_i} + \frac{k_B T}{ \gamma} \partial_{l_il_i} \right) 
\end{split}
\end{equation}
With this generator one can check that $\mathcal{L}^{\star}\pi = 0$ with the natural equilibrium distribution 
\begin{equation}
\pi = \frac{1}{Z} \begin{pmatrix}
1 \\ \on/\off \\ \on/\off \\ ...
\end{pmatrix}e^{-kl^2/2k_B T }e^{-\sum_i kl_i^2/2k_B T }
\end{equation}
where $Z$ is a normalization constant.

\paragraph{Homogenization}

Taking the usual scalings we get the non-dimensional generator $\mathcal{L} = \frac{1}{\epsilon^2} \mathcal{L}_0 + \frac{1}{\epsilon} \mathcal{L}_1 +\mathcal{L}_2$, where $\mathcal{L}_0 = \mathcal{Q} + \mathcal{U}_0$ where $\mathcal{U}_0$ is diagonal with 
\begin{equation}
(\mathcal{U}_0)_{00} = \frac{\Gamma}{\gamma} \left(- l  \partial_{l} + \partial_{ll}  + \sum_i \left( - l_i  \partial_{l_i} + \partial_{l_il_i} \right) \right) 
\end{equation}
\begin{equation}
\begin{split}
(\mathcal{U}_0)_{bb} &=- \frac{\Gamma}{2(2\Gamma+\gamma)}(l-l_b) (\partial_{l}-\partial_{l_b} ) - \frac{\Gamma}{2\gamma}(l+l_b) (\partial_{l_b} + \partial_{l}) + \frac{ \Gamma}{2(2\Gamma+\gamma)}(\partial_{l_b} - \partial_{l})^2 + \frac{\Gamma}{2\gamma} (\partial_{l_b} + \partial_{l})^2  \\
 &\frac{\Gamma}{\gamma}  \sum_{i\neq b} \left( - l_i  \partial_{l_i} + \partial_{l_il_i} \right)
\end{split}
\end{equation}
and $\mathcal{L}_1$ is such that $(\mathcal{L}_1)_{00} = 0$ and
\begin{equation}
(\mathcal{L}_1)_{bb} =  - \frac{\Gamma}{2\Gamma+\gamma}(l-l_b) \partial_x +  \partial_x(\partial_{l} -\partial_{l_b}) \frac{2 \Gamma}{\gamma + 2\Gamma}
\end{equation}
and finally
\begin{equation}
\mathcal{L}_2 = \begin{pmatrix}    
   \partial_{xx}  & 0 & 0 & ... \\
0 &   \frac{2 \Gamma}{\gamma+2\Gamma} \partial_{xx} & 0 & ... \\
0 &  0 & \frac{2 \Gamma}{\gamma+2\Gamma} \partial_{xx} & ... 
\end{pmatrix}
\end{equation}
We now seek an expanded solution $f$ of $\partial_t f = \mathcal{L} f$ as $f = f_0 + \epsilon f_1 + \epsilon^2 f_2 + ...$

At lowest order the resolution gives $f_0 = a(x,t) \begin{pmatrix} 1 \\ 1 \\  1 \\ ... \end{pmatrix}$, and the associated equilibrium distribution $\pi_0 = \pi$. At first order we need to solve $f_1$ and it is useful to seek a genuinely symmetric solution 
\begin{equation}
\mathcal{L}_0 f_1 = + \begin{pmatrix}
0 \\ (l - l_1) \\ (l - l_2) \\ ... 
\end{pmatrix} \frac{\Gamma}{\gamma + 2\Gamma} \partial_x a , \text{seeking} \,\, f_1 = \begin{pmatrix} u_0 l + u_0' l_1 + u_0' l_2 + .. \\ b_1 l + b_1' l_1 + u_1' l_2 + ... \\ b_1 l + u_1' l_1 + b_1' l_2 +... \\ ... \end{pmatrix} \partial_x a
\end{equation}
where  $u_0, u'_0, b_1, b'_1$ and $u_1'$ are constants. Notice that here $u$ and $b$ refer respectively to unbound and bound contributions, with $x$ and $x'$ corresponding respectively to leg or arm contributions, and the indices correspond to the number of bonds of the state. The constants obey the system of equations
\begin{equation}
    \begin{split}
          -  M \on u_0 + \on M b_1 - u_0 \frac{\Gamma}{\gamma} &= 0 \\
          - M \on u_0' + \on b_1' + (M-1)\on u_1' - u_0' \frac{\Gamma}{\gamma} &= 0 \\
         \off u_0 - \off b_1  - \frac{\Gamma}{2(2\Gamma + \gamma)} (b_1 - b_1') - \frac{b_1 + b_1'}{2} \frac{\Gamma}{\gamma}  &=  \frac{\Gamma}{2\Gamma + \gamma}\\
         \off u_0' - \off b_1'  + \frac{\Gamma}{2(2\Gamma +  \gamma)} (b_1 - b_1') - \frac{b_1 + b_1'}{2} \frac{\Gamma}{\gamma} &= -\frac{\Gamma}{2\Gamma + \gamma} \\
         - \off u_1' + \off u_0' - u_1' \frac{\Gamma}{\gamma} &= 0 
    \end{split}
\end{equation}
that has a unique solution. For now we do not report the coefficients here for simplicity. 

At 2nd order we need to satisfy the Fredholm alternative $\langle \partial_t f_0 - \mathcal{L}_2 f_0 - \mathcal{L}_1 f_1 , \pi_0 \rangle = 0$. We split up the terms to highlight calculation steps (discarding $Z$ terms to simplify notations, as they would cancel out eventually)
\begin{equation}
\begin{split}
\langle \mathcal{L}_1 f_1  , \pi_0 \rangle = & \sum_i \frac{\on}{\off} \frac{\Gamma}{\gamma + 2\Gamma} \langle \left( - (l-l_i)(b_1 l - b_1' l_i)   + 2 (b_1-b_1')\right) e^{-l^2/2} e^{-l_i^2/2} \rangle \partial_{xx} a \\
%= & - \left( - 2 \frac{\Gamma}{\gamma + 2\Gamma}  + 2 \frac{2\Gamma}{\gamma+2\Gamma}\right) \frac{ \partial_{xx} a}{1 + \frac{q_{\rm off}}{1 + q_{\rm on}}\frac{\gamma+2\Gamma}{\Gamma}} \\
= &  M \frac{2 \Gamma}{\gamma+2\Gamma} \frac{(b_1-b_1')}{2} \partial_{xx} a \\ 
%= & - N \frac{2 \Gamma}{\gamma+2\Gamma} \frac{1 - \frac{\frac{N-1}{2} \frac{\gamma^2}{\Gamma^2} \on \off}{(1 + N \on \frac{\gamma}{\Gamma} ) (1+ N \off \frac{\gamma}{\Gamma})}}{1 + \frac{2\Gamma + \gamma}{\Gamma}\frac{\off}{(1 + N \on \frac{\gamma}{\Gamma} )} + \frac{\frac{\gamma}{\Gamma}(N-1)\on \off }{(1 + N \on \frac{\gamma}{\Gamma} ) (1+ N \off \frac{\gamma}{\Gamma})} }  \partial_{xx} a
\end{split}
\end{equation}
and
\begin{equation}
\langle \mathcal{L}_2 f_0 , \pi_0 \rangle = \left( 1 + \frac{2 \Gamma}{\gamma+2\Gamma}  \frac{M\on}{\off} \right) \partial_{xx} a .
\end{equation}
Gathering terms as $\langle \partial_t f_0  ,  \pi_0 \rangle = \langle \mathcal{L}_2 f_0  ,  \pi \rangle  + \langle \mathcal{L}_1 f_1  , \pi_0 \rangle $  we get
\begin{equation}
\left(1 + M\frac{\on}{\off}\right)\partial_t a  =  \left( 1+ \frac{M\on}{\off} \frac{2\Gamma}{\gamma+2\Gamma} (1 + \frac{(b_1-b_1')}{2} )\right) \partial_{xx} a .
\end{equation}
Reorganizing terms slightly we arrive at (in dimensional scales)
\begin{equation}
\partial_t a = k_B T\left( \frac{p_{0,M}}{\Gamma} + \frac{p_{1,M}}{\Gamma_{1,M}} \right)\partial_{xx} a.
\end{equation}
with $p_{0,M} = \frac{\off}{\off + M \on}$ and $p_{1,M} = 1 - p_{0,M}$ and 
\begin{equation}
    \Gamma_{1,M} = \frac{\Gamma + \gamma/2}{1 - (b_1- b_1')/2}.
\end{equation}
We can further expand $\Gamma_{1,M}$ by using the expressions for $b_1$ and $b_1'$. We find
\begin{equation}
    \Gamma_{1,M} = \Gamma + \frac{\left( \gamma + \frac{k}{\off} \right)\left( \gamma + \frac{k}{\off} + \gamma \frac{M\on}{\off}\right)}{\left( \gamma + \frac{k}{\off} \right) + \left( \gamma + \frac{k}{\off} + \gamma \frac{(M-1)\on}{\off}\right)}
\end{equation}
that is exactly Eq.~\gammaarmlegN in the main paper. Notice that, since arm and leg are interchangeable, a similar effect would be observed for a particle with $M$ legs allowed to bind to 1 arm. 

Notice that when $M$ is large, we obtain that the above expression simplifies to
\begin{equation}
    \Gamma_{1,M} = \Gamma + \gamma_{\rm eff, 1, M} \,\, \text{with} \,\, \frac{1}{\gamma_{\rm eff, 1, M}} =  \frac{1}{ \gamma_{\rm eff, M, 1}} + \frac{1}{ \gamma_{\rm eff, 1, 1}} , \,\,
\end{equation}
with $\gamma_{\rm eff, M, 1} =  k \left( \frac{1}{\off} + \frac{\gamma}{k} \frac{(M-1)\on + \off}{\off} \right)$ the effective friction due to the leg $\gamma_{\rm eff,1,1} =  k \left(\frac{1}{\off} + \frac{\gamma}{k} \right)$ due to arms. %Here we see that the characteristic re-binding time for the leg is $\tau_{\rm on} = 1/(M-1)\on $, due to the increased number of other available arms to bind to. Specifically, once the leg is bound to a given arm, it can unbind and rebind to another $(M-1)$ arms. For the arms $\tau_{\rm on} = \infty$ as there are no other legs to bind to once the only leg is bound.  
%This highlights the striking interplay between top and bottom legs. 

\subsubsection{N legs facing M potential arms}

\paragraph{2 legs facing 2 arms}

To understand the dynamics at play in the case of $N$ legs facing $M$ potential arms we first investigate the more specialized $2$ for $2$ scenario.  We number legs as $1$ and $2$ and arms as $3$ and $4$.
We write the generator for this system directly in the non-dimensional scales. It is a bit lengthy as there are now 7 possible states. We arrange the states as state $\# 1$ is the unbound state, states $\#2 - 5$ correspond to 1 bond states, and states $\#6 - 7$ to 2 bond states. We write $\mathcal{L} = \mathcal{Q} + \mathcal{U}$. The transition rate matrix for the generator is simply
\begin{equation}
\mathcal{Q} = \frac{1}{\epsilon^2}  \begin{pmatrix}
- 4 q_{\rm on} & q_{\rm on} & q_{\rm on} & q_{\rm on} & q_{\rm on} & .& . \\
q_{\rm off} & - q_{\rm off} -q_{\rm on} &. & . & . &  q_{\rm on} & . \\
q_{\rm off} &. & - q_{\rm off} -q_{\rm on} & . & . &  . & .  q_{\rm on} \\
q_{\rm off} &. & . & - q_{\rm off} -q_{\rm on} & . & . q_{\rm on} & .  \\
q_{\rm off} &. & . &. & - q_{\rm off} -q_{\rm on} & . & . q_{\rm on}  \\
. & q_{\rm off} & . & q_{\rm off} & . & - 2 q_{\rm off} & .   \\
. & . & q_{\rm off} & . & q_{\rm off} & . & - 2 q_{\rm off}   \\
\end{pmatrix}
\end{equation}
Then we write the diagonal (only non zero) components of $\mathcal{U}$. For the unbound state we have
\begin{equation}
\mathcal{U}_{11} = \frac{1}{\epsilon^2} \frac{\Gamma}{\gamma} \sum_{i=1}^4 D_{l_i} + \partial_{xx}
\end{equation}
where $D_{l_i} =  - l_i \partial_{l_i}  + \partial_{l_il_i}$ is the unbound relaxation operator.  
Then in the $2 .. 5$ states where just one tether is bound we have (for example for the 2 state where tethers say 1 and 3 are bound)
\begin{equation}
\begin{split}
\mathcal{U}_{22} =  &\frac{1}{\epsilon^2} \left(  - \frac{\Gamma}{2(2\Gamma+\gamma)}(l_1-l_3) (\partial_{l_1}-\partial_{l_3} ) - \frac{\Gamma}{2\gamma}(l_1+l_3) (\partial_{l_3} + \partial_{l_1}) + \frac{ \Gamma}{2(2\Gamma+\gamma)}(\partial_{l_3} - \partial_{l_1})^2 + \frac{\Gamma}{2\gamma} (\partial_{l_3} + \partial_{l_1})^2 \right) \\
& + \frac{1}{\epsilon^2} \frac{\Gamma}{\gamma} \left( D_{l_2} + D_{l_4} \right) +  \frac{1}{\epsilon} \frac{\Gamma}{2\Gamma+\gamma} \left( - (l_1-l_3) \partial_x +  2 \partial_x(\partial_{l_1} -\partial_{l_3})\right) + 1 \left( \frac{2\Gamma}{2\Gamma + \gamma} \partial_{xx} \right)
\end{split}
\end{equation}
and similarly for the other 1 bond states. 
Finally for the $6$ and $7$ states, 2 bonds are formed. In these state we have, for example for state $\#6$ that contains the bonds $1-3$ and $2-4$ 
\begin{equation}
\begin{split}
\mathcal{U}_{66} =  & - \frac{\Gamma}{2\Gamma+2\gamma}(l_1- l_3 + l_2 - l_4) (\frac{1}{\epsilon}\partial_x - \frac{1}{\epsilon^2} \frac{1}{2}\partial_{l_3} +\frac{1}{\epsilon^2} \frac{1}{2}\partial_{l_1}  -\frac{1}{\epsilon^2} \frac{1}{2}\partial_{l_4} + \frac{1}{\epsilon^2}\frac{1}{2}\partial_{l_2} ) \\
&- \frac{1}{\epsilon^2}\frac{\Gamma}{2\gamma}(l_1+l_3) (\partial_{l_3} + \partial_{l_1})  - \frac{1}{\epsilon^2}\frac{\Gamma}{2\gamma}(l_2+l_4) (\partial_{l_4} + \partial_{l_2})   \\
& + \frac{\Gamma}{2(2\Gamma + 2\gamma)} \left( 2\partial_x + \frac{1}{\epsilon}\left( \partial_{l_1} +  \partial_{l_2}-  \partial_{l_3}-  \partial_{l_4}\right)  \right)^2 + \frac{\Gamma}{2\gamma} \frac{1}{\epsilon^2} \left[ \left( \partial_{l_1} + \partial_{l_3}\right)^2 + \left( \partial_{l_2} +  \partial_{l_4}\right)^2 \right]
\end{split}
\end{equation}
and reordering this last expression as a function of the scales in $\epsilon$
\begin{equation}
\begin{split}
\mathcal{U}_{66} =  &\frac{1}{\epsilon^2} \left( - \frac{\Gamma(l_1-l_3 + l_2 - l_4) }{2(2\Gamma+2\gamma)}(\partial_{l_1} - \partial_{l_3} + \partial_{l_2} - \partial_{l_4})- \frac{(l_1+l_3)\Gamma}{2\gamma} (\partial_{l_3} + \partial_{l_1}) - \frac{(l_2+l_4)\Gamma}{2\gamma} (\partial_{l_2} + \partial_{l_4}) \right)  \\
&   + \frac{1}{\epsilon^2} \left( \frac{\Gamma}{2(2\Gamma + 2\gamma)} \left( \partial_{l_1} +  \partial_{l_2}-  \partial_{l_3}-  \partial_{l_4}\right)^2 + \frac{\Gamma}{2\gamma} \left[ \left( \partial_{l_1} +  \partial_{l_3}\right)^2 + \left( \partial_{l_2} +  \partial_{l_4}\right)^2 \right] \right) \\
& + \frac{1}{\epsilon} \left( \frac{2\Gamma}{(2\Gamma + 2\gamma)} \partial_x  \left( \partial_{l_1} +  \partial_{l_2}-  \partial_{l_3}-  \partial_{l_4}\right)   - \frac{\Gamma (l_1-l_3 + l_2 - l_4)}{2\Gamma+2\gamma} \partial_x  \right) \\
& +1 \left( \frac{2\Gamma}{2\Gamma + 2 \gamma} \partial_{xx}\right) 
 \end{split}
\end{equation}

Here the equilibrium distribution is
\begin{equation}
\pi = \frac{1}{\sqrt{2\pi}\left( 2 + 4\frac{q_{\rm off}}{q_{\rm on}} + \frac{q_{\rm off}^2}{q_{\rm on}^2}\right)} \begin{pmatrix}
q_{\rm off}^2/q_{\rm on}^2, & q_{\rm off}/q_{\rm on}, & q_{\rm off}/q_{\rm on}, & q_{\rm off}/q_{\rm on}, & q_{\rm off}/q_{\rm on},& 1, & 1 
\end{pmatrix}^T.
\end{equation}

We seek a solution to the expanded generator $\mathcal{L} = \frac{1}{\epsilon^2}\mathcal{L}_0 +  \frac{1}{\epsilon}\mathcal{L}_1 + \mathcal{L}_2$ as $f = f_0 + \epsilon f_1 + \epsilon^2 f_2...$. 
No steps change in the resolution compared to previous calculations but for finding the solution at order $1$. Here we seek a solution 
\begin{equation}
\mathcal{L}_0  f_1 = - \mathcal{L}_1 f_0 = \begin{pmatrix}
0 \\ \frac{1}{2\Gamma + \gamma} (l_1 - l_3) \\ \frac{1}{2\Gamma + \gamma} (l_1 - l_4) \\ \frac{1}{2\Gamma + \gamma} (l_2 - l_4)  \\ \frac{1}{2\Gamma + \gamma} (l_2 - l_3) \\ \frac{1}{2\Gamma + 2 \gamma} (l_1 - l_3 + l_2 - l_4) \\ \frac{1}{2\Gamma + 2 \gamma} (l_1 - l_4 + l_2 - l_3)
\end{pmatrix} \Gamma \partial_x a
\end{equation}
The solution is expected to preserve the symmetries of the problem and therefore we may seek 
\begin{equation}
    f_1 = \begin{pmatrix} u_0 l_1 + u_0 l_2 + u'_0 l_3 + u'_0 l_4 \\
     b_1 l_1 + u_1 l_2 + b'_1 l_3 + u'_1 l_4 \\
      ...\\
     b_2 l_1 + b_2 l_2 + b'_2 l_3 + b'_2 l_4 \\
     ...
     \end{pmatrix} \partial_x a
\end{equation}
where $b_n, b_n', u_n, u_n'$ are constants that refer to bound and unbound configurations of the leg or the arms. They solve a linear system of equations that possesses a single solution that we will report below. 

The Fredholm alternative at the next order requires $\langle \partial_t f_0 - \mathcal{L}_2 f_0 - \mathcal{L}_1 f_1 , \pi \rangle$. We obtain, splitting the relevant contributions
\begin{equation}
    \begin{split}
        \langle \mathcal{L}_1 f_1 , \pi \rangle &= 0 + \frac{(b_1 - b_1')\Gamma}{2\Gamma + \gamma} p_1 + \frac{2 (b_2 -  b_2') \Gamma}{2\Gamma + 2\gamma} p_2
    \end{split}
\end{equation}
and assembling all terms allows to get an effective long time diffusion equation $\partial_t a = \frac{\Gamma}{\Gamma^{2,2}_{\rm eff}} \partial_{xx} a $ where
\begin{equation}
    \frac{1}{\Gamma^{2,2}_{\rm eff}} = \frac{1}{\Gamma} p_0 + \frac{1}{\frac{\Gamma + \gamma/2}{\left(1 - \frac{b_1 - b_1'}{2}\right)}} p_1 + \frac{1}{\frac{\Gamma + 2 \gamma/2}{\left(1 - 2 \frac{b_2 - b_2'}{2}\right)}}   p_2
\end{equation}
which is similarly as in all cases a weighted harmonic sum of friction coefficients, with $p_n$ the probability to have $n$ bonds.

The system of equations that the constants satisfy is 
\begin{equation}
    \begin{split}
        &- 4 q_{\rm on} u_0 - u_0 + 2 \on b_1 + 2 \on u_1 = 0 \\
        &- 4 q_{\rm on} u'_0 - u'_0 + 2 \on b'_1 + 2 \on u'_1 = 0 \\
        &\off u_0 - \off b_1 - \on b_1 + \on b_2 - \frac{\gamma}{2(2\Gamma + \gamma)} (b_1 - b'_1) - \frac{b_1 + b'_1}{2} = \frac{\gamma}{2\Gamma + \gamma} \\
        &\off u'_0 - \off b'_1 - \on b'_1 + \on b'_2 + \frac{\gamma}{2(2\Gamma + \gamma)} (b_1 - b'_1) - \frac{b_1 + b'_1}{2} = -\frac{\gamma}{2\Gamma + \gamma} \\
        &\off u_0 - \off u_1 - \on u_1 + \on b_2 - u_1 = 0 \\
        &\off u'_0 - \off u'_1 - \on u'_1 + \on b'_2 - u'_1 = 0 \\
        & \off b_1 + \off u_1 - 2 \off b_2  - \frac{2\gamma}{2(2\Gamma + 2\gamma)} (b_2 - b'_2) - \frac{b_2 + b'_2}{2} = \frac{\gamma}{2\Gamma + 2\gamma} \\
        & \off b'_1 + \off u'_1 - 2 \off b'_2  + \frac{2\gamma}{2(2\Gamma + 2\gamma)} (b_2 - b'_2) - \frac{b_2 + b'_2}{2} = -\frac{\gamma}{2\Gamma + 2\gamma} 
    \end{split}
\end{equation}
Since the bottom and top tethers don't exactly have the same positions it's not possible to simplify further, typically $b_n \neq -b'_n$. This really shows the structure of the equations.

\paragraph{N legs facing M arms}

The above system of equations allows to generalize the derivation for $n$ bonds in some $N$ legs for $M$ arms structure.  For each possible number of bonds say $n$, the tethers are either unbound ($u_n, u'_n$) or bound ($b_n, b'_n$) and can exchange with their counterparts. 

When there are $n$ bonds, focusing on a connected pair, one can still bind more pairs, and there are $(M-n)(N-n)$ possible ways to do so. In any case the connected pair will remain connected during that transformation.  When there are $n$ bonds, one can unbind $n$ pairs. $n-1$ possibilities lead to the given pair still being bound. 

For an unconnected leg (resp. arm), there are only $M-n$ possibilities (resp. $(N-n)$)  to form a bond that will connect the unconnected one.

We obtain the general system of equations as
\begin{equation}
    \begin{split}
         n \off u_{n-1}  - (n \off + (N-n)(M-n)\on) u_n + \on (M-n) b_{n+1} + \on (N-n-1)(M-n) u_{n+1} - u_n \frac{\Gamma}{\gamma}&= 0 \\
         n \off u'_{n-1}  - (n \off + (N-n)(M-n)\on) u'_n + \on (N-n) b'_{n+1} + \on (M-n-1)(N-n) u'_{n+1} - u'_n \frac{\Gamma}{\gamma}&= 0 \\
         \off u_{n-1} + (n-1) \off b_{n-1}  - (n \off + (N-n)(M-n)\on) b_n + \on (N-n)(M-n) b_{n+1} & ...\\   - \frac{n \Gamma}{2(2\Gamma + n \gamma)} (b_n - b'_n) - \frac{b_n + b'_n}{2}\frac{\Gamma}{\gamma} &= - \frac{\Gamma}{2\Gamma + n\gamma}\\
        \off u'_{n-1} + (n-1) \off b'_{n-1}  - (n \off + (N-n)(M-n)\on) b'_n + \on (N-n)(M-n) b'_{n+1} & ...\\  + \frac{n \Gamma}{2(2\Gamma + n \gamma)} (b_n - b'_n) - \frac{b_n + b'_n}{2} \frac{\Gamma}{\gamma}&= \frac{\Gamma}{2\Gamma + n\gamma}\\
          \frac{\Gamma + n \gamma/2}{ 1 - n (b_n - b'_n)/2} &= \Gamma_n 
    \end{split}
    \label{eq:systemMNlegs}
\end{equation}

Looking for the average number of bonds contribution, we may assume similarly that for $n = N_b$ we have $u_n \simeq u_{n-1} \simeq u_n = \bar{u}$, and similarly for other quantities. We thus obtain the closed system of equations
\begin{equation}
    \begin{split}
          -  (M-N_b) \on \bar{u} + \on (M-N_b) \bar{b} - \bar{u} \frac{\Gamma}{\gamma}&= 0 \\
          -  (N-N_b) \on \bar{u}' + \on (N-N_b) \bar{b}' - \bar{u}'\frac{\Gamma}{\gamma} &= 0 \\
         \off \bar{u}  - \off \bar{b}  - \frac{N_b\Gamma}{2(2\Gamma + N_b \gamma)} (\bar{b} - \bar{b}') - \frac{\bar{b} + \bar{b}'}{2} \frac{\Gamma}{\gamma}  &= \frac{\Gamma}{2\Gamma + N_b\gamma}\\
         \off \bar{u}'  - \off \bar{b}'  + \frac{N_b\Gamma}{2(2\Gamma + N_b \gamma)} (\bar{b} - \bar{b}') - \frac{\bar{b} + \bar{b}'}{2} \frac{\Gamma}{\gamma} &= - \frac{\Gamma}{2\Gamma + N_b\gamma}\\
          \frac{\Gamma +N_b \gamma/2}{ 1 - N_b (\bar{b} - \bar{b}')/2} &= \Gamma_{N_b} 
    \end{split}
\end{equation}

This finally yields
\begin{equation}
    \Gamma_{N_b} = \Gamma + N_b \gamma_{\rm eff, N, M}, \,\, \text{with} \,\, \frac{1}{\gamma_{\rm eff, N, M}} =  \frac{1}{ \gamma^{\rm leg}_{\rm eff}} + \frac{1}{ \gamma^{\rm arm}_{\rm eff}} , \,\,
\end{equation}
with the effective friction due to legs as $\gamma^{\rm leg}_{\rm eff} = \gamma + k \left( \frac{1}{\off}  + \frac{\gamma}{k} \frac{(M-N_b)\on}{\off}\right)$ and that due to arms $\gamma^{\rm arm}_{\rm eff} = \gamma + k \left( \frac{1}{\off}  + \frac{\gamma}{k} \frac{(N-N_b)\on}{\off}\right)$. Here we see that the characteristic binding time (for example for the leg) is $\tau_{\rm on} = 1/(M-N_b)\on $, due to the increased number of possibilities $(M-N_b)$ due to multiple available arms.

\begin{figure}[h!]
    \centering
    \includegraphics[width = 0.5\textwidth]{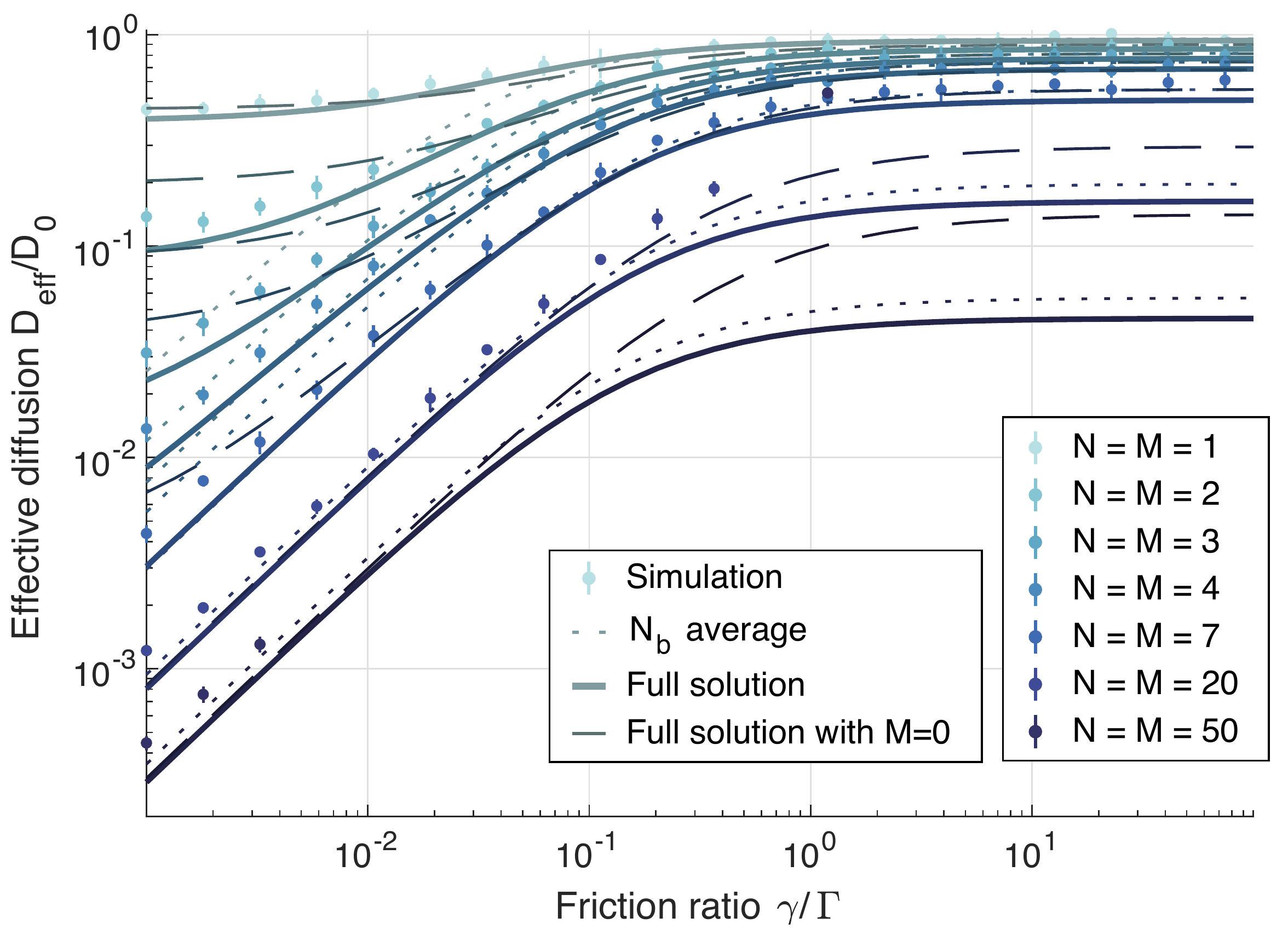}
    \caption{Effective diffusion for systems with an equal number of arms and legs all interacting with one another ($N = M$) from stochastic simulations. We overlay the predictions using Eq.~\eqref{eq:GammaNb2} ($N_b$ average) and fully solving for the system of equations Eq.~\eqref{eq:systemMNlegs} (Full solution). For reference we also show the result of the full system in the case of $N$ legs with $M = 0$ arms, from solving Eq.~\eqref{eq:systemNlegs}. Here the values of other parameters  are $ \frac{q_{\rm on} \Gamma}{k} = 1.0$ and $\frac{q_{\rm off} \Gamma}{k} = 0.8$}
    \label{fig:NMlegs}
\end{figure}

Here we can explore limiting regimes. If there are as many legs as there are arms we have $M = N$, then the effective friction simplifies to 
\begin{equation}
    \Gamma_{N_b} = \Gamma + \frac{N_b}{2} \left[ \gamma + k \left( \frac{1}{\off}  + \frac{\gamma}{k} \frac{(N-N_b)\on}{\off}\right) \right],
    \label{eq:GammaNb2}
\end{equation}
where we see that the additional friction is divided by 2, as expected from the leg and arm case. One notable difference is that here we see that the characteristic binding time $\tau_{\rm on} = 1/(N-N_b)\on $, due to the increased number of possibilities due to multiple arms and legs. According to the large $N$ limit investigated, $(N-N_b)\on$ does not necessarily diverge. In particular for very sticky systems $(N-N_b) \simeq 0$ and therefore this part does not contribute significantly to the dynamics. 

If there are a large number of arms, say $M \gg N_b$, then we find $\gamma_{\rm eff, M, N} \rightarrow \gamma^{\rm arm}_{\rm eff}$ is dominated by the arm contributions to the effective friction. 

These effective results capture well stochastic simulation results, as shown in Fig.~\ref{fig:NMlegs}.